\documentclass[accepted]{aastex62}

\usepackage{amsmath}
\usepackage{float}
\usepackage{graphicx}
\usepackage{natbib}
\usepackage{rotating}
\usepackage[caption=false]{subfig}

\begin{document}
\title{A Comparison of the Chemical Composition of Main Sequence and \\
Giant Stars in the Open Cluster NCC 752}
\author{Michael~G.~Lum\thanks{lummi@hawaii.edu}}
\author{Ann~Merchant~Boesgaard\thanks{boes@ifa.hawaii.edu}}
\affil{Institute for Astronomy, University of Hawai'i\\
2680 Woodlawn Drive, Honolulu HI 96822}

\begin{abstract}
The chemical composition of stars in open clusters provides the best information on the chemical evolution of stars via comparison of the main sequence stars with the evolved giants.  This is a case study of the abundances in the dwarfs and giants in the old, open cluster, NGC 752.  It is also a pilot program for automated abundance determinations, including equivalent width measurements, stellar parameter determinations, and abundance analysis.  We have found abundances of 31 element-ion combinations in 23 dwarfs and 6 giants.  The mean cluster abundance of Fe is solar with [Fe/H]$=-0.01\pm0.06$ with no significant difference between the dwarfs and giants.  We find that the cluster abundances of other elements, including alpha-elements, to be at, or slightly above solar levels.  We find some evidence for CNO processing in the spectra of the giants.  The enhancement of Na in giants indicates that the NeNa cycle has occurred.  The abundances of Mg and Al are similar in the dwarfs and giants indicating that the hotter MgAl cycle has not occurred.  We find no evidence of s-process enhancements in the abundances of heavy elements in the giants.
\end{abstract}

\section{Introduction}
Cluster stars, with their common origin and shared evolutionary environment, provide laboratories for analyzing atmospheric processes and for exploring compositional variances between stars with differing initial masses. Studies of cluster stars are able to utilize a large, homogeneous sample to represent a cross-section of the stellar population of a given age and origin. A cluster represents conditions at a given time and galactic location. Therefore, comparative studies between cluster member stars have distinct advantages over single star or randomly selected field star studies.\par
Prior cluster studies have leveraged the material and temporal common origin of cluster stars to establish trends with various cluster characteristics. Sample previous studies have examined the relationship between metallicity and age \citep[e.g.][]{Phelps:94}, or metallicity, age, and galactic location \citep[e.g.][]{Friel:02}. Other cluster studies have produced or supplemented catalogs of elemental abundances for large numbers of elements, stars and clusters  \citep[e.g:][]{Pancino:09}. The aforementioned cluster studies take advantage of the common origin assumption, and determine cluster abundances and metallicity measurements from small numbers of either giant or dwarf stars.\par
The primary goal of this study is to provide a comprehensive elemental abundance analysis of the stars in NGC 752. We have included both giant and dwarf stars in our sample. With the exception of some light elements (Li, Be, and B, q.v.:\cite{Boesgaard:16}), we can assume that the atmospheric elemental abundances, as measured in our dwarf sample are representative of the unaltered stellar nebula from which they formed. We then compare the abundances as measured in our giant sample, which have undergone a ``dredge-up'' of core material into their atmospheres, to those from our dwarf sample. We evaluate the differences to gain insights into the core nuclear processes which have occurred over the main-sequence lifetime of the stars.\par
Within the past decade, large, easily accessible, digitally archived spectral data have become widely available. These public archives provide an opportunity to analyze data and publish discoveries, without the necessity of exclusive telescope time. Our observational data has been drawn exclusively from publicly available archive data. As our chosen spectra have been taken by mutiple investigators, with different science goals, our resulting data set has a wide variance in wavelength coverage and signal-to-noise.\par
A second goal of this study is to establish methods for automated cluster abundance evaluation, using previously collected, archived spectral data. By using a rudimentary artificial intelligence, we implemented an objective measurement algorithm which removed several repetitive, error-prone, or subjective processes from the abundance and stellar parameter determinations.\par
For this study, we selected data from the HIRES spectrograph \citep{Vogt:94} on the 10m W.M.Keck I telescope. We also use a solar reference spectrum from the National Solar Observatory \citep{Wallace:11}, and solar calibration spectra from the ESO's HARPS \citep{Mayor:03} instrument.\par
We measured the abundances of a comprehensive list of 25 elements, 6 in both singly-ionized and neutral states, across the 23 dwarfs and 6 giants. Our abundance measurements were determined through spectral synthesis and equivalent width measurements of absorption features drawn from a list of approximately 2700 lines. While our chosen spectra contain observations spanning from the near-UV (3360~\AA) to the near-IR (9140~\AA), we restrict our evaluation to the (mostly-) optical range ranging from 4500~\AA~to 8500~\AA.\par

\section{NGC 752}
We have selected NGC 752 as the first cluster in our study due to a number of factors. It is a member a relatively small number of ``solar twin'' clusters, having a metallicity approximately solar ([Fe/H]$=+0.04\pm0.01$ \citet{Blanco-Cuaresma:14}, [Fe/H]$=-0.063\pm0.013$ \citet{Maderak:13}, [Fe/H]$=+0.08\pm0.04$ \citet{Carrera:11}). For calibration purposes, using near-solar metallicity targets permits us to use the Sun as a reference spectrum to confirm laboratory-determined absorption line parameters with observational measurements. NGC 752 is relatively old open cluster, with an age between 1.5 and 1.8 Gyr (1.45 Gyr \citet{Anthony-Twarog:09}, ~1.6 Gyr \citet{Carrera:11}, 1.78 Gyr \citet{Daniel:94}). NGC 752 shows evidence of either dynamic or tidal dispersal particularly in the lower main-sequence \citep{Carraro:14}. While we do not utilize spectra from the late-K and M region affected by this dispersal, the loss of these lower mass stars is possible evidence that NGC 752 is well into the process of disappearing into the galactic field.\par

Figure~\ref{fig:752HR} places each of our target stars on a color-color diagram, based on photometry from the GAIA ``Data Release 2'' (DR2) \citep{GAIADR2:18}. Also noted on the figure are the stars denoted as members by \citet{Platais:91}. We have also included stars from the GAIA dataset which appear to be members, based on analysis of position, proper motion, and parallax from DR2 (Lum 2019, in prep).\par
At NGC 752's age, the main-sequence turn off stars are F-type stars. These stars lie on the blue side of the Li ``gap'' \citep{Boesgaard:86b}, implying that sufficient mixing occurs during the main sequence lifetime of the star to have carried surface material to a depth where Li can be destroyed by (p,$\alpha$) reactions. \citeauthor{Hobbs:86} documented both the Li destruction in NGC 752 giants \citep{Pilachowski:88} and the presence of the ``Li gap'' in the F-stars \citep{Hobbs:86, Pilachowski:88a}. Deep mixing mechanisms may also affect atmospheric elemental abundances, as measured at the surface, after the first dredge up portion of the red giant branch.\par
We selected 29 stars, 23 dwarfs and 6 giants, within the cluster to perform a complete, precise accounting of the material composition of the cluster. We analyzed differences between the evolved and unevolved members, with the aim of quantifying the compositional changes between the evolutionary stages of our sample sets. Our sample spectra uses archived observational data from the Keck Observatory Archive (KOA).\par

\section{Stellar Spectra Data}
\subsection{NGC 752 spectra from Keck HIRES}

Table~\ref{tab:752Obs} lists the J2000 coordinates (RA/Dec), spectral wavelength range, and the signal-to-noise for the composite spectra of each star in our survey. We chose the \citet{Platais:91} identifiers for our reference, but have included cross-references for \citet{Heinemann:26} and \citet{Rohlfs:62} (WebDA) as reported by the SIMBAD astronomical database \citep{Wenger:00}. The wavelength ranges cited reflect the minimum and maximum values for each spectrum. The spectral coverage for a given range is generally not complete, due to inter-order gaps and gaps between coverage of individual observation configurations.\par

All NGC 752 spectra were obtained from the Keck Observatory Archive (KOA) and were measured using the HIRES spectrograph, with a resolution of \textasciitilde{48,000}, as calculated from the instrument setup parameters (KOA ``SPECRES'' keyword). We utilize observations from both the ``original'' (prior to August 2004) and ``upgraded'' (post-2004) HIRES. The upgraded detector added two CCDs, for a total of three, increasing the wavelength coverage while reducing pixel size from $24\mu$ to $15\mu$~.The difference between the two detectors is significant, and manifests in our data as a difference in the signal-to-noise ratio with respect to exposure time, and increase in wavelength coverage for a given spectrum.\par

Table~\ref{tab:752Obs} lists the signal-to-noise (S/N) ratio at the specified wavelength; there are one or two measurements, depending on the wavelength coverage of the spectrum. The reported S/N values are an average, rounded to the nearest ten, for the region as reported by the  \emph{MAKEE} \citep{Barlow:makee} software, during the reduction process.  For the representative measures, we selected two regions, the first near 6200~\AA , the region containing the $T_{eff}$ calibration lines, and the second near 7800~\AA , where we measured the O\,\small{I} triplet.\par

Our minimum measured S/N ratios of 40-60 for the dwarf sample are similar to those of comparable studies - eg: \citet{Maderak:13} had a minimum S/N of 60. The S/N ratio of the giant star spectra in this study are significantly higher than those of other studies, topping out at well over 400. The S/N range in the \citet{Topcu:15} giant study was 140-290, while \citet{Reddy:12} had S/N ratios in the 100-120 range. \par

We opted to use the spectra as reduced by the KOA process, heeding the warning that ``the quality and content of the data...may not necessarily be suitable for publishable science.''~(\url{https://koa.ipac.caltech.edu/UserGuide/HIRES/extracted.html}). Our reasoning being that, should we choose to do our own reduction, we would use the same tools- specifically \emph{MAKEE}, choice of calibration frames (as suggested by the archive search), and process as the automated pipeline. We do not have sufficient knowledge of the observation session to override the recommendations of the archive. However, we performed several calibration processes in addition to the archive calibrations. We applied wavelength correction to the pipeline's wavelength solution, using a selection of Fraunhofer and other strong iron lines. We also normalized each spectrum to a continuum level of 1.0, using a $5^{th}$ order polynomial fit for each spectral order, which we found as the highest-order polynomial we could use to fit the relatively smooth regions of the longer wavelength orders of the spectra, without over-smoothing the broad lines, and ``noisier'' regions. In the cases where multiple spectra were available for a single star, we co-added each spectrum using a weighted mean, using the STScI's IRAF\footnote{IRAF is distributed by The Association of Universities for Research in Astronomy, Inc., under cooperative agreement with the National Science Foundation} function \emph{scombine}, with each weighted by its signal-to-noise ratio.\par

As a final spectra quality control process, we performed a visual inspection of the individual orders, as provided in the Keck Observatory Archives image previews. We excluded two orders (out of several hundred) which showed evidence of reduction or observational faults. In both cases, the missing orders were conveniently duplicated in other spectra, without the flaws.\par

During our analysis process, we elected to exclude spectra for four stars from our final results. Although high resolution and high S/N spectra were available for PLA-413, PLA-552, PLA-859, and PLA-1284, they were excluded for the following reasons: For PLA-859, the parallax data from GAIA DR2 ($0.28\pm0.04$mas) places it beyond the cluster (mean parallax from GAIA DR2 data of $2.2\pm0.1$, Lum (2019, in prep)). For PLA-1284, our automated process for measuring absorption lines in this (SB2) binary system did not properly account for the shifted components. After analysis, we found that the spectroscopic binary, PLA-552 displayed evidence of line broadening, which affected abundance calculations (particularly of Ti and Cr). While PLA-413 is not catalogued as a spectroscopic binary, it lies above NGC 752's main sequence, near other binaries, and displays identical broadened metal lines. For this reason, we have also excluded both PLA-552 and PLA-413 from our final analysis.\par

\subsection{Data Processing} \label{sec:DataProcessing}
Due to the relatively large number of spectra (47) and large number of lines to measure ($\sim2700$ per spectrum), we created a series of automated processes for analyzing our spectra. The code consists of mostly Python scripts, with some calls to external C and Fortran functions. Intermediate data products, such as spectral line physical parameters (e.g. log\,$g_f$, excitation potential), equivalent width measurements, spectra file data, ``astrophysical'' log$g_f$ corrections, etc. are stored in a SQLite3 database. The code is freely available on github at: \url{https://github.com/mikelum/ClusterAnalysis}.\par

Our software is responsible for: a) management of the line list database and associated references, b) absorption line measurement and equivalent width calculation, c) stellar atmosphere parameter determination, and d) elemental abundance determination. It also produces plots and tables used in data analysis and in this work. In addition, we used \citep[DAOSpec]{Stetson:08} software routines to assist with and calibrate our Doppler corrections. We also used \emph{IRAF} and \emph{STDAS} \citep{STDAS} for spectra stacking and continuum normalization. Our elemental abundance calculations were obtained through use of \emph{MOOG} \citep{Sneden:73, Sneden:12} \emph{synth}, \emph{blends}, and \emph{abfind} ``drivers''. Our analysis spanned several \emph{MOOG} version upgrades, between 2014 and 2017. We see no significant difference in results between the earliest (July 2014) and latest (February 2017) versions. Note that we have made superficial and cosmetic modifications to the standard MOOG program and make files to allow compilation under the \emph{GNU Compiler Collection's GFortran} (Free Software Foundation, \url{https://gcc.gnu.org/}) compiler, and removed status messages to allow the \emph{MOOGSILENT} function to run truly silently.\par

As mentioned previously, our spectra cover a wide range of S/N ratios and resolutions from two different versions of the HIRES instrument. We account for lower resolution and/or lower S/N ratios by establishing a minimum EQuivalent Width measurement (EQW) for any given spectra of five (5) times the product of the resolution (in m\AA/pixel) and inverse of the signal-to-noise, with a minimum of 2.0 m\AA. The calculated minimum EQW limit for each star is listed in Table~\ref{tab:atmParms} (see additional discussion of Table~\ref{tab:atmParms} in section~\ref{sec:LineQuality}). Similarly, with lower S/N ratios, we expect higher uncertanties and lower sensitivity to weaker absorption lines. Figure~\ref{fig:FeSpecSN} shows the region near 6200~\AA~which contains several Fe\,\small{I} lines. Temperature effects on the Fe lines are evident, and we have used them to assist with parameter determination. Over the 300K range of the three spectra in the figure, the Fe absorption lines strengthen noticeably between the hottest and coolest spectra. Figure~\ref{fig:OSpecSN} compares the Oxygen ``triplet'' region between 7770 and 7782~\AA~for three dwarf spectra with S/N of approximately 140. At this high S/N, the triplet is clearly visible, with temperature effects visible to a discerning eye as broader line widths and increased absorption in the line wings.\par

In addition to setting a minimum EQW measurement threshold, we also compensate for low S/N by measuring a larger sample of absorption features, where available. As an example, our list of potential Fe I lines contains over 400 entries. Our complete line list is available as an online database, with a sample format shown in Table \ref{tab:LineList}.\par

We include one additional criterium for accepting a measured line from lower S/N spectra. We measure the line profile in terms of the ratio of the Full Width at Half Maximum (FWHM) of the Gaussian portion of the line core (see section \ref{sec:EQWMeasures}, below) to the resultant EQW measurement ($FWHM/EQW$). When our automated measurement process finds a potential absorption line, which was incorrectly measured from random noise fluctuation in the spectrum, the $FWHM/EQW$ ratio is almost always below 1.0 ($>99.8$\% for the lowest S/N spectra). We determined this ratio limit of 1.0 through Monte Carlo (MC) measurement simulations on 250 random noise spectra. Each noise spectrum point consists of a (normalized) flux value, generated around a continuum value of 1.0, with a variance determined by the S/N being modeled. We then insert randomly selected regions from our actual data, containing a single absorption feature, into the random noise spectra. We run each noise spectrum through our line measurement process, and evaluate the resulting line measures. We apply the same criteria for line acceptability (within the linear portion of the curve of growth, within the expected wavelength range, and above the detection threshold) to these line candidates, eliminating approximately 80\% of the (false) lines from a noise spectrum prior to applying the line profile criterium. We are able to recover 100\% of ``real'' lines inserted into our test spectra, using the 1.0 (FWHM/EQW) ratio limit, and incorrectly classify less than 0.2\% of the ``false'' lines as legitimate measurements. Of the incorrectly measured lines, an average of 3 per Monte Carlo run of 2700 lines in 250 simulated spectra passed both the 1.0 ratio limit and the line acceptability criteria, as listed above.\par

\subsection{Absorption Line Evaluation}
\label{sec:LineQuality}
One of our goals of this project was to produce a method for selecting target lines, using objective criteria, from all available measured transitions within our spectral coverage range. Most comparable abundance studies will use a substantially smaller set of ``trusted'' lines, adopted from previous, similar studies, and/or  lines which have been reliable for the particular researcher in their prior work. In our own work (e.g. \citet{Boesgaard:15, Boesgaard:13}), we have opted for a combination of the two selection processes.\par

For small-scale, high-resolution, high S/N studies, this process works well. However, for larger-scale and potentially lower S/N studies, a much larger sample of lines are needed. When compiling all the available line resources, including large online databases such as those run by VALD \citep{VALD}, and NIST \citep{Kramida:14}, we can easily find a massive number of available lines. The challenge is identifying which to use.\par

For this project, we categorized each of the lines in our database by measuring the line in a high-resolution solar spectrum \citep{Wallace:11}, calculating the abundance of the respective element from that line width, and then comparing the result to the calculated solar abundances from Table 1 in \citet{Asplund:09}. We then categorized the lines as ``excellent'', where the difference ($\Delta$) in calculated abundance from that in \citeauthor{Asplund:09} is less than 0.05 dex, ``good'' ($0.05\leq\Delta\leq0.10$), ``fair'' ($0.10\leq\Delta\leq0.20$), ``poor'' ($0.20\leq\Delta\leq0.40$), ``bad/mis-measured'' ($0.40\leq\Delta\leq1.00$), and ``detected'' ($\Delta\geq1.00$). We assign each of these categories a numerical ``Quality'' score of 10 (``excellent''), 8, 6, 4, 2, or 1 (``detected''), and also use a score of 0 for lines which were either not detected in the solar spectrum, or which were measured using spectral synthesis.\par

Table~\ref{tab:SolarAbs} shows abundances as calculated using each of the five above categories, plus the abundance as calculated using all measured lines. Using these calculations, we determined that our line choices for Pr\,\small{II} and Y\,\small{I} were not accurate enough to use for further analysis. Similarly, we were unable to measure the small number of S\,\small{I} and Eu\,\small{II} lines in our spectral sample to provide analysis of these elements.\par

Our choice to base our astrophysical log\,$g_f$ values on the solar spectrum is a reasonable assumption due to the metallicity of our sample stars falling into that of solar ([Fe/H]$_{NGC752}$\textasciitilde{}+0.00). Furthermore, the atmospheric parameters for our dwarf sample is similar to solar (F,G, and K-stars). For future work in stars and clusters with metallicities and atmospheric parameters farther from solar, we will not have independently verifiable abundance values (as we do from meteoric and solar samples). For these future astrophysical log\,$g_f$ values, we expect that proper calibration of a given line's correction  requiring an iterative, statistical process. The first-order correction being the (constant) solar correction we use in this work. Future corrections would be calculated using a function of atmospheric parameters and metallicities from the abundances of a small sample of stars (such as in this work). The resulting function would be expanded through the inclusion of data from stars with increasing variance in parameters and metallicities. The eventual goal being a functional solution for the astrophysical log\,$g_f$ of any given line, based on a large sample of stellar parameters and metallicities. However, examining such a project is well beyond the scope of this work.\par

Within the wavelength range of our study, the absorption lines for both C and N are weak or are blended with other nearby lines. While our automated line measurement process can evaluate blended lines of comparable strength, we chose to supplement our C and N abundance measures with additional measurements from synthesized spectra. We used MOOG's \emph{synth} driver and line lists which combined elemental data from the VALD \citep{VALD} database, with CN molecular data from \citet{Sneden:14}. We discuss the specific regions in Section~\ref{sec:CNO}. Figure~\ref{fig:Synth} shows a sample synthesis of the region near 7115~\AA~for the Sun and for PLA-350, a red giant in NGC-752. We used this region to determine C\,\small{I} abundance.\par

\subsection{Absorption Line Selection}
\label{sec:LineSelection}
When we measured abundances of elements with large numbers of lines in a particular spectrum, we have the option of using a subset of the lines. While the immediate tendency might be to only use the lines with the highest quality scores (as described above in Section~\ref{sec:LineQuality}), we must also consider the effect of adding more samples to a given population in increasing the accuracy of a measurement. In general, adding more measurements of \emph{equal quality} will result in a more accurate measurement of a given quantity. However, since we know the quality of our population, our process would begin by selecting lines of the highest quality, and subsequent additions to our measurement subset would be of lower quality. For many elements, we have a relatively large number of quality-rated lines, and feel the need to address the question of accuracy vs. precision.\par

By hand-selecting a small number of lines, we can guarantee a precise result - simply put, we can select for small error bars. Conversely, a set of lines exists which would provide the most accurate measurement of the abundance of a given element in a given star. Unfortunately, we cannot know the latter set, without knowing the actual abundance(s) from another independant measure. For our purposes, we are interested in the most accurate measurement of atmospheric abundances. However, we also wish to have at least a ``reasonable'' level of precision. To answer the question of accuracy vs. precision, we created a Monte-Carlo simulation of the line measurement process.\par

Our simulation began with an assumption of an ``actual'' elemental abundance of $[X/Fe]=0.00$. We randomly generated a set of between 1 and 250 lines, distributed among the 6 ``quality'' categories as indicated by the population in our solar measurements (Table~\ref{tab:SolarAbs}) - Approximately 16\% in the top (``10'') category, 13\% in  the next (``8''), and 20\%, 21\%, 18\%, and 12\% in the remaining four categories, respectively. For each line, we generated an experimental abundance measurement, using a randomized Gaussian distribution around the solar value. The standard deviation ($\sigma$) used was the average $\sigma$ of all the lines in the corresponding category in our solar measurements, added in quadriture with an atmospheric term of 0.10, representative of a typical atmospheric parameter error term (see Table~\ref{tab:atmErrs}). By comparing the experimental abundance, calculated as the mean of all lines of a given quality or better, with the actual value of 0.00, we determined the quality-number relationship used herein.\par

Our simulation ran 10,000 trials, using N=1,\,3,\,5,\,10,\,20,\,50,\,100, and 250 lines to measure the simulated abundance of our simulated element. We then evaluated both the accuracy and precision of abundances calculated from lines of varying ``quality'' ratings. Unsurprisingly, the most precise abundances were calculated when only the lines of the highest available quality ranking were used. However, as the number of available lines for a given element increase, we found that including more lines of lower quality does increase the accuracy of the measurement.\par

In summary, our simulation found, for elements where we measure 4 or fewer lines, using only the lines in the highest available quality category provided the most accurate abundance measurement. For example, if we measured 3 lines with quality scores of 10, 8, and 6, we would calculate the abundance, using only the single line of the highest quality (10). Should none of the measured lines for a given element have the highest quality rating, we would take the abundance as measured by lines in the next lower category, reducing our acceptable quality category until at least one line was used.\par

In the case where we measured between 5 and 10 lines for an element, we would not discriminate between lines of the highest two categories (10, 8), and would use all available lines in these two quality categories. As in the cases with small numbers of lines, should none of our measured lines rank in the top two quality categories, we calculated the elemental abundance using the highest quality lines available from the measured set.\par

For $N\geq20$, the most accurate abundace values are achieved by using all measured lines in the top 3 categories. In all spectra for this project, when 20 or more lines were measured for a given element, a large fraction (generally the majority) of lines fell within the top three quality categories. Therefore the abundances calculated for large-N elements always used the ``best'' lines.\par

Finally, we note that for any given line, we apply the $\Delta$ value (from the solar calculation, Table~\ref{tab:SolarAbs}) to our resulting abundance calculation, commonly referred to as an ``Astrophysical log\,$g_f$ correction''.\par

\subsection{Equivalent Width Measurements}
\label{sec:EQWMeasures}
Using our continuum-fit and Doppler-corrected spectra, we measured approximately 2400 absorption line features per spectrum from a list of 31 elements, 8 of which had two ionization states. We were able to obtain usable line measurements for 24 elements (7 with both neutral and singly-ionized states). A complete list of our line list (2430 lines), reference list (68) and equivalent width measurements (17000+ measurements) for all stars are available as supplemental online data.\par

Needless to say, with such a massive line list and large collection of spectra, we did not perform the EQW measurements manually. We utilized a combination of Python scripts and \emph{AstroPy} \citep{Astropy:13} functions to obtain our EQW measurements (using a simple linear combination of Lorenzian and Gaussian curves). However, we found that, after trying multiple line profiles (pure Gaussian, Voight, and rudimentary linear and trapezoidal interpolations), our chosen line profile had less effect on the calculation of the equivalent width of a given line than individual data point errors from signal-to-noise, and spectral dispersion.\par

Our line profiles, and the resulting EQW values, are based from a localized continuum, using the values of the local maxima in the 2\AA~window containing the center of our target line. This process is similar to that used in \citet{Ramirez:11}, except instead of a visual inspection of the local window, our process automates the continuum placement by adding a small factor (0.02 on a 1.0-normalized continuum) to the mean of the highest local maxima. However, as with the specific form of the line fit, above, the exact placement of the local continuum has a very small effect (\textasciitilde{}1m\AA) as per \citet{Carrera:11}.\par

Figure \ref{fig:RelEQWs} compares the resulting equivalent width measurements taken by our automated process against over 1200 IRAF \emph{splot} measurements taken by \citet{Reddy:12} and \citet{Topcu:15}. We find no systematic difference between the techniques ($\Delta_{EQW}= 1.2\pm5.9$~m\AA), nor is there a significant difference between the variance between our method and either of the prior works, or between the two prior works.\par

As an additional check, we manually examined the five outliers (circled in the figure), and the fitting procedure for the questionable lines. In order of increasing EQW, as measured by our automated process: 1) We measured the EQW of the Ce\,\small{II} line at 5274.21~\AA~as 37.4~m\AA~while \citet{Reddy:12} record a measurement of 12.2~m\AA. Our manual re-evaluation did not yield a significantly different measurement from our automated process, and we suspect that the difference is due to either a mis-measurement or a typographical error with \citeauthor{Reddy:12}, as their EQW measures for the same line in their other giants is similar to ours. In PLA-687 they measure 39.9~m\AA~vs. our measurement of 35.6~m\AA and in PLA-1172, 39.0~m\AA~(42.1~m\AA~from our process). 2) For the Ti\,\small{II} line at 4911.19~\AA, we determined that our measurement (of 55.8~m\AA) was affected by a form of signal clipping, probably introduced during the cosmic ray removal process, which caused the measurement process to underestimate the line width. Since this is an unusual occurance, we elected to retain our measurement as taken, and account for it as an expected random variation. 3) We elected to treat the Co\,\small{I} line at 5212.69~\AA~(our EQW=78.9~m\AA) in a similar fashion. This line is blended with a Ti\,\small{I} line at 5212.28~\AA, and our automated process fits the Co\,\small{I} and Ti\,\small{I} lines as a two-component blend. Manual fitting did not yield a significant difference. The final two selected outliers; 4) a Zn\,\small{I} line at 4722.15~\AA~in PLA-687 (our EQW=89.2~m\AA, \citeauthor{Reddy:12}'s EQW=36.2~m\AA) and 5) a Fe\,\small{II} line at 6246.32~\AA~in PLA-1089 (our EQW=134.9~m\AA, \citeauthor{Topcu:15}'s EQW=35.0~m\AA), also fall into the ``probable error in the prior work'' (likely typographical or transposition error) category, as those measurements are also outliers in their respective work, relative to measurements of the same lines in other stars. \par

\subsection{Atmospheric Models and Parameters}
We derived the stellar atmosphere conditions used to calculate abundances with an iterative process, based on our spectroscopic data. We began with an initial effective temperature ($T_{\rm{eff}}$), determined by either spectroscopic (preferred) or photometric methods. We calculated initial spectroscopic $T_{\rm{eff}}$s using the Line-Depth Ratio (LDR) technique \citep{Gray:91} with the selection of lines and the polynomial relations in \citet{Biazzo:07}. We took the weighted mean of all $T_{\rm{eff}}$s from all available line combinations, weighted by their sensitivity as stated in \citeauthor{Biazzo:07}, Table 2. For spectra which did not contain sufficient data in the 6199-6275~\AA~region containing the lines in \citeauthor{Biazzo:07}, we used photometry from the \emph{SIMBAD Astronomical Database} \citep{Wenger:00} to determine our initial $T_{\rm{eff}}$. The majority of the NGC 752 photometry contained in the SIMBAD database was taken from the Tycho \citep{Hog:00} and 2MASS \citep{Skrutskie:06} catalogs. We calculated photometric starting temperatures using the polynomial color-temperature-metallicity relation \citep[ex:][]{Ramirez:05,Casagrande:10}, and polynomial coefficients from \citet{Huang:15}. The initial photometric $T_{\rm{eff}}$ for a given star was the mean of all \citeauthor{Huang:15} temperatures, determined from all available SIMBAD color combinations, weighted by their error. Typically, the SIMBAD database contained B,\,V,\,R,\,J,\,H, and K photometry for our stars, yielding 5 separate color-temperature relations in \citeauthor{Huang:15}. We adopt the reddening value for E(B-V) of $0.034\pm0.004$ from \citet{Twarog:15}, and adjust that values for other colors using the relations from \citet{Rieke:85}. For both spectroscopic and photometric temperatures, when available color or line combinations resulted in more than 4 $T_{\rm{eff}}$ results, the highest and lowest $T_{\rm{eff}}$ were omitted before calculating a weighted mean.\par

We then adopt the gravity value (log\,g), which fufills the requirement that the Fe abundance, as measured using absorption features from the two ionization states (Fe\,\small{I}, Fe\,\small{II}), is the same for both states. We then refine our $T_{\rm{eff}}$ values to fufill the requirement that the measurement of iron abundance (Fe\,\small{I}) should show no trend when compared to the excitation potential of each line. We iterated between the Fe\,\small{I}/Fe\,\small{II} ``balance'' and the Fe\,\small{I} ``slope'' processes until we arrived at a $T_{\rm{eff}}$ - log\,g combination which fufilled both the balance and the slope requirements. This spectroscopic parameter determination process is well detailed in \citet{Takeda:02} and used by other automated processes like iSpec's ``Equivalent Width Method'' \citep{Blanco-Cuaresma:14}.\par

To incorporate the variances of our abundance measurements, we then performed an additional iterative process. We allowed the Fe\,\small{I} ``slope'' parameter to vary by up to $\pm0.05$ dex/eV, and the Fe abundances to ``match'' as long as their 1-$\sigma$ error bars overlapped. This produced a range of acceptable $T_{\rm{eff}}$-$log\,g$ values, from which we selected the $T_{\rm{eff}}$-log\,g pair which minimized the Fe\,\small{I}/Fe\,\small{II} abundance difference and the distance to a 1.5Gyr \emph{PARSEC} \citep{Bressan:12} isochrone at solar metallicity. We generated the \emph{PARSEC} isochrones with the \emph{CMD} tool \url{http://stev.oapd.inaf.it/cgi-bin/cmd}.\par

Figure~\ref{fig:ParmComps} plots our final $T_{\rm{eff}}$/log\,g combinations against other published values for both dwarfs (Figure~\ref{fig:ParmComps:Dwarfs}) and giants (Figure~\ref{fig:ParmComps:Giants}). We also include the 1.5 Gyr \emph{PARSEC} isochrone at solar metallicity. The difference between selecting a 1.45 Gyr \citep{Anthony-Twarog:09}, 1.6 Gyr \citep{Carrera:11}, or our plotted 1.5 Gyr isochrone is less than the typical point size on the plots, and results in log\,g and $T_{\rm{eff}}$ differences of less than 0.01 and 10K, respectively. It is worth noting that the close fit of the \citet{Maderak:13} and \citet{Castro:16} parameter choices to the PARSEC isochrone is due to their choice of using a photometrically-determined $T_{\rm{eff}}$, and selecting their log\,g value from the corresponding point on an isochrone. Table~\ref{tab:TComps} provides a $T_{\rm{eff}}$ comparison between this work and other literature values. The ``Delta" column compares the difference between this work and the literature value for a given star while the ``Lit. Std." column shows the standard deviation of the literature values (where available).\par

To determine the microturbulent velocity, we calculate the Ca abundance from selected absorption lines, using our previously-determined $T_{\rm{eff}}$ and log\,g, for a range of $\xi$ values. Our final $\xi$ value for each model is the one where the calculated Ca abundance variance is minimized. Figure~\ref{fig:vturbDet} illustrates the process for an example star. Note that while this is essentially the same process used in \citet{Reddy:12}, we selected Ca to determine the minimum variance point, as opposed to th combination of Fe, V, Cr, and Ti in the prior work. The choice of Ca led to the $\xi$ value with the least ambiguity, and Ca absorption lines were consistently available in both giant and main-sequence populations. We also use Fe\,\small{I]/Fe\,\small{II} and Ti\,\small{I}/Ti\,\small{II} measurements for $T_{\rm{eff}}$ and log\,g determination.\par

Our parameters were used to create atmospheric models using the \emph{ATLAS9} grids \citep{Castelli:04}. We used a linear interpolation algorithm between the existing grid points, based on our $T_{\rm{eff}}$, log\,g, and $\xi$ values. Since stellar metallicity is a parameter in determining the proper model, we used an iterative process to select the appropriate value for a given model. Starting with a solar metallicity, we calculated the Fe/H value for a given star, and then repeated the process, using the new metallicity, until we found a stable value. When a general metallicity value was required, notably in isochrone generation, we used a weighted average of our Fe\,\small{I} and Fe\,\small{II} measurements from both giant and dwarf stars. For Fe abundances, we use the notation of $log_{10}({N(Fe)}/{N(H)})+12.00$ where $log_{10}N(H)$ is set at 12.00. We measured the cluster (Fe) metallicity value at $7.49\pm0.06$. Using the solar value of 7.50 from \citet{Asplund:09}, our cluster metallicity is $[Fe/H]=-0.01\pm0.06$. \par
Our EQW measurements and interpolated atmospheric models were used as inputs to the \emph{MOOG} ``abfind'', ``blends'',  and ``synth'' drivers, resulting in our final abundance calculations.\par

\subsection{GAIA Reference Spectra Calibration}
To verify our abundance calculation methodology, we obtained spectra from The Gaia FGK Benchmark Star archive\footnote{\url{https://www.blancocuaresma.com/s/benchmarkstars}} \citep{Blanco-Cuaresma:14b}, supplemented by the low-metallicity sample from \citet{Hawkins:16}. To best match the wavelength coverage and resolution of our NGC 752 spectra from the Keck Archive, we preferentially selected the spectra from the NARVAL \citep{Auriere:03} (24 spectra) and UVES-POP \citep{Bagnulo:03} (9 spectra) instruments, with a small number of spectra from the HARPS \citep{Mayor:03} (4 spectra) and ESPaDOnS \citep{Donati:03} (2 spectra) instruments.\par

In order to minimize abundance variances due to atmospheric parameter variation, we adopted the parameters as determined in \citet{Heiter:15}. Using our measurement and calculation techniques described earlier, we derived abundances for our element suite in the 39 reference stars. We then compared our results with those compiled in \citet{Jofre:15}\footnote{Note that several measures from \citet{Pavlenko:12} appear to be anomalous ($[X/Fe]\approx{-7.0}$) in \citeauthor{Jofre:15}, possibly typographic errors. Therefore, we excluded them from our literature abundance(s).}. Figure~\ref{fig:GAIA_Ab_Comp} presents the comparison in graphical form. Each point plots  the ``literature'' abundance (X-Axis) against our measured abundance (Y-axis). In cases where an element was not explicitly measured in \citeauthor{Jofre:15}, the literature value is calculated by using the individual star's [Fe/H], and assuming a solar ratio for [X/Fe] to produce a [X/H]\footnote{Where [X/H] is shorthand for: $log_{10}({N(X)}/{N(H)})+12.00$} value. The red points in Figure~\ref{fig:GAIA_Ab_Comp} represent the comparison to actual literature values while the blue points are those which use the extrapolated abundances.\par

Although our software is intended for use with stars of (approximately) solar metallicity, we also include the low metallicity stars of the GAIA reference spectra set with points in a lighter shade (pink/cyan) of their corresponding solar metallicity counterparts.\par

When comparing individual elemental abundances on a star-to-star basis, our calculated abundances for all stars and elements in the GAIA set (noted as:[X/H]) differed by $-0.03\pm0.13$. When comparing solar metallicity star abundances that were ``directly" measured in the \citeauthor{Jofre:15} compilation, our measurements were also essentially the same, at: $-0.04\pm0.12$. The same can be said if we analyze the comparison between our measurements and ``extrapolated" literature values ($-0.06\pm0.14$). For the low-metallicity sample, our measured results were also the same as both the directly-measured and extrapolated values: $-0.01\pm0.12$ and $+0.02\pm0.14$, respectively. It is important to note that the variance among the literature values when multiple values are given in \citeauthor{Jofre:15} is $\pm0.10$.\par

\section{Abundance Results}
Our abundance results are shown in Table~\ref{tab:752Abs}, which contains the cluster averages, with dwarf and giants averaged separately, along with the standard deviation of the individual star measurements, and the count of stars in which a given element was measured. While Fe is measured on the scale of $log_{10}({N(Fe)}/{N(H)})+12.00$, all other elements are measured relative to Fe, as denoted by [X/Fe].\par 

We have chosen to list the standard deviation of the individual line measurements as our uncertainty for the abundance measurement for a given star. While this method produces a larger uncertainty (by a factor of $\sqrt{n-1}$) when compared to the varaince of the mean(s), we find the larger uncertainties are more representative of our measurement techniques. We also note that individual line measurements do not represent truly independant measurements of a star's elemental abundance, as they can be non-uniformly affected by factors dependant on wavelength (through the reduction process), physical line parameters (excitation potential and log\,gf, and their associated uncertainties), and equivalent width measurements. The ``Q.'' score column in is the average of all line quality scores (see Section~\ref{sec:LineQuality}) used in calculating the listed abundance.\par

Individual star data is shown in the extended abundance table, available in the digital version of Table~\ref{tab:752Abs}. In order to properly characterize the uncertainty in individual star abundance measurements, caused by varying model atmosphere parameters, we re-ran the full abundance determination process, varying each of the three parameters, as shown in Table~\ref{tab:atmErrs}. We then added the largest error of each parameter (either the ``+'', or the ``-'' error) in quadrature, to produce an atmospheric parameter error. Uncertainty, due to variance in line measurements, is designated as $\sigma_{var}$ in Table~\ref{tab:752Abs}. Abundance uncertainty, due to uncertainty in atmospheric parameters, is designated as $\sigma_{atm}$ in the same table.\par

\subsection{Iron and Iron-Peak Elements}
Our measured Fe abundance was based on over 400 absorption features. As discussed in Section~\ref{sec:EQWMeasures}, the large number of measurements was pared to a smaller subset to eliminate data which were likely affected by measurement errors. Using a weighted average of our calculated Fe\,\small{I}/H (75\%) and Fe\,\small{II}/H (25\%) for both dwarf and giant members, we derive a cluster (Fe) metallicity of $7.49\pm0.06$. We use this value for all atmospheric models and isochrones during the abundance and atmospheric parameter determination process.\par
In addition to Fe\,\small{I} and Fe\,\small{II}, we also measured the four other ``Iron Peak'' elements, Cr (\small{I} and \small{II}), Mn\,\small{I}, Co\,\small{I}, Ni\,\small{I}, and two ``nearby'' elements V\,\small{I}, and Cu\,\small{I}. Composite cluster values are shown in Table~\ref{tab:752Abs}, and individual line measurements in the online version of Table~\ref{tab:LineList}. As with Fe abundances, we find no significant differences between abundance measures of these elements in the dwarf and giant population. We accounted for the hyperfine structure of V, Mn, and Cu lines by using the MOOG \emph{blends} driver, with the line component parameters taken from the line sources listed in Table~\ref{tab:LineList}. As with Sc\,\small{I} below, we expect the odd-Z element Co to also have hyperfine structure in its line profiles. However, we have no appropriate source to implement these characteristics, and must rely upon an assumed similarity between the NGC-752 stars, and the solar reference counterpart to adjust for this structure.\par
Our measured abundances for the five elements (V\,\small{I}, Cr\,\small{I}, Fe\,\small{I}, Co\,\small{I}, and Ni\,\small{I}) commonly measured in our study and in the giant studies of \citet{Carrera:11} and \citet{Reddy:12}, match well. Table~\ref{tab:FeGroupComps} provides a summary of the five elements over the three studies.\par

\subsection{Carbon, Nitrogen and Oxygen}
\label{sec:CNO}
With the assumed age of 1.45\,Gyr, and a Main Sequence Turn Off (MSTO) mass of $1.5M_\odot$ \citep{Bartasiute:07}, we expect the evolved members of NGC 752 to show evidence of CNO-cycle processing. Even though the CNO processing is ongoing throughout a star's lifetime, conditions at the core-radiative zone boundary isolate core materials until core hydrogen fusion ends. Convective processes during the H shell fusion, and core He fusion phases allow the altered core materials to mix into the stars' atmosphere, becoming detectable in our giant star spectra. \citet{Iben:64} and \citet{Iben:91} summarize the atmospheric differences we expect to see as a result of this ``first dredge-up''. Specifically, we should expect increases in $^{13}$C (relative to $^{12}$C) and $^{14}$N.\par

While all three light elements are common in stellar atmospheres, measurement is difficult, even with high S/N spectra, due to the lack of low-excitation transitions in optical spectra. Our measurements consist of abundances as calculated from both synthesis and from absorption line feature measurements. Unfortunately, we were unable differentiate between isotope states, particularly $^{12}$C and $^{13}$C, so our analysis is based on relative elemental (instead of isotope) measures.\par

For our giant population, we utilized a combination of elemental and molecular absorption features to calculate C and N abundances. As C-, N-, and O-molecule abundances are inter-related, we first determined the cluster O abundances using equivalent width measurements of several features, including the O\,\small{I} ``triplet'' at 7771-7776~\AA. Figure~\ref{fig:OSpecSN} illustrates the spectral region encompasing the triplet region. While MOOG abundance calculations assume Local Thermal Equilibrium (LTE), we adjusted our final abundances for the expected non-LTE (NLTE) conditions of the high-excitation potentials ($\geq{9.14eV}$), as suggested in \citet{Takeda:03a}. We find [O\,\small{I}/Fe]$=-0.11\pm0.08$, with similar abundance measurements in the dwarf and giant populations. The cluster, dwarf, and giant (individual star measures in the digital version of the table) [O/Fe] measurements are compiled in Table-\ref{tab:752Abs}. Our measured [O/Fe] of $-0.15\pm0.07$ in our giant sample is lower than that of the giants measured by \citet{Carrera:11} of $+0.03\pm0.04$, possibly due to differences in the measurement methods. \citeauthor{Carrera:11} calculated their O abundances using synthesis around the [O\,\small{I}] forbidden line near 6300\AA.\par

Our abundance determination for C is a composite of equivalent width measurements of the features at 5382~\AA, 6381~\AA, and 8335~\AA~and spectral synthesis of the region near 7115~\AA, which contains several high-excitation C\,\small{I} absorption features (see Figure~\ref{fig:CSpecSN}). The synthesis process was partially iterative between C and N abundances, as there are CN features in the same region. In determining the C abundance, we gave higher priority to fitting the lines at 7111~\AA~and 7113~\AA, since they were more isolated from CN features. Table-\ref{tab:752Abs} lists our [C/Fe] results; [C/Fe]$=-0.10\pm0.13$ (dwarfs) and [C/Fe]$=-0.22\pm0.08$ (giants).\par

Individual EQW measurements were given equal weight, with the abundance calculated from the synthesis given five (5) times the weight of a single EQW measurement (appropriate for the 5 blended lines in that region). While all of our measured C lines have (relatively) high excitation potentials ($>{7.68eV}$), at the effective temperature of our sample stars,  we do not expect NLTE effects to be significant (as per \citet{RentzschHolm:96}).\par

Nitrogen abundance was exclusively calculated using spectral synthesis. As none of our spectra had coverage of the strong CN bandheads in the near-UV and near-IR regions, we used a combination of the N abundance derived during synthesis of the 7115~\AA~region, and an additional synthesis around the 7442~\AA~region. Based on the measurement of a single N absorption line in one dwarf spectrum, and sysntesis in the five giant spectra, we find [N/Fe]$=0.12$(1 dwarf) and $0.28\pm0.07$(giants)\par

Our measurements of the giant sample show a decrease of C to below that of solar, and N significantly higher. The general pattern of higher N and lower C abundance corresponds to the results of the giants measured by \citet{Topcu:15}, although our measured C abundance is significantly higher, at $[C/Fe]=-0.22\pm0.08$ vs. their value of$-0.46\pm0.03$~probably due to the difference in synthesis region selection. Likewise, our measured N abundance of $+0.27\pm0.06$ is also higher than that measured by \citeauthor{Topcu:15}, which we partially attribute to the unavailability of the violet and red bands of the CN spectrum.\par

\citeauthor{Topcu:15} explain the relative N enhancement and C depletion to main-sequence evolutionary processes, namely He production in the core through the CNO cycle. The relevant portion of the CNO cycle is that which occurs at the lowest temperature: 
\begin{gather*}
 ^{12}C(p,\gamma)^{13}N(,e^{+}\gamma_{e})^{13}C(p,\gamma) \rightarrow \\
 ^{14}N(p,\gamma)^{15}O(, e^{+}\gamma_{e})^{15}N(p,\alpha)^{12}C
\end{gather*}
The slow step, or ``bottleneck'', is the $^{14}$N proton capture. Over the course of the main-sequence lifetime of a star, this results in an increase of N abundance at the expense of the $^{12}$C abundance. With the MSTO mass of at least $1.5M_\odot$, evidence of processing from the higher temperature branch of the CNO-I-cycle (CNO-II) might also be present. Specifically, the high temperature branch has two steps which require proton capture by O. Like the N increase in the CNO-I cycle, increased O abundance might be explained through the capture steps in the higher temperature CNO-II cycle. Since we do not measure an increased O abundance, we conclude that these higher temperature branches are not significant contributors to the elemental alteration of NGC 752's stars.\par

Admittedly, measuring the CNO abundances in giant stars only shows half the picture. The critical comparison should occur between main-sequence and evolved members, not between a solar reference and the evolved members. To that end, we have measured C and O abundances using the aformentioned absorption features at 7115~\AA~and 7775~\AA, respectively. These abundance measurements do hint at the expected trend for CNO processing, but measurement uncertainty, particularly due to the lack of strong CN features in our dwarf sample, prevents us from making a definitive statement to that effect.\par

The third ``piece'' - N abundance - also shows the expected trend for CNO processing, but again, we would need more, higher resolution spectra to make this claim definite. N abundances are the result of molecular CN feature synthesis, so we were largely unable to determine N abundances for the dwarf population. Our sole dwarf N abundance comes from a single EQW measurement of the 7442.29~\AA~N\,\small{I} feature in PLA-520. We also recorded similar measurements from PLA-859 and PLA-889, but disqualified them due to their EQW measurements falling below the acceptable minimum width threshold for the respective spectra's S/N ratio. We also note that the [C/N] ratio of -0.50, as measured in our giant sample, agrees well with the expected [C/N] of -0.54 for a 1.5Gyr giant just after its first dredge-up ([C/N]$_{\rm{FDU}}$) at solar metallicity given in \citet{Salaris:15}.\par
Figure~\ref{fig:CNOAbs} shows the summary results of our C and O light-element measurements. Individual star measurements are compiled in the digital version of Table~\ref{tab:752Abs}. \par

\subsection{Light Elements}
The ``light'' elements, with $11\leq{Z}\leq22$, provide viable targets for measurement, even with our lowest S/N spectra. Na, Mg, Al, Si, Ca, and Ti all have either a large number of measureable transitions, or strong, well documented features. We find that, within our error margins, the abundance of these elements follow their solar counterparts. We chose to separate the elements into two groups for discussion. Our odd-Z light element sample of Na\,\small{I} and Al\,\small{I} are shown in Figure~\ref{fig:OddZAbs}. While both light elements show significant star-to-star variation, Na appears to be enhanced in the giants, while Al is likely unchanged. The ``$\alpha$-'', or even-Z elements (Mg\,\small{I}, Si\,\small{I}, Ca\,\small{I}, and Ti\,\small{I}), as shown in Figure~\ref{fig:AlphaAbs}, potentially show evidence of evolution. Silicon abundance appears to increase, from $0.05\pm0.05$ in the dwarf population to $0.11\pm0.02$ in the giants, while Ti appears to decrease ($0.13\pm0.05$ to $0.06\pm0.04$). However, both of these changes are within the $1-\sigma$ error bars, and would need more accurate measurements to show a definite trend.\par

When comparing Na abundances between the dwarf and giant population, we find an increase from [Na/Fe]$=-0.18\pm0.12$ to $0.13\pm0.05$. For both dwarf and giant samples, we applied the NLTE Na corrections from \citet{Lind:11} using the calculator at: \url{http://inspect.coolstars19.com/cp/application.py/A_from_e?element_name=Na} Our measured Na increase in the giants is possibly due to the same mechanism as \citet{Boesgaard:13} noted in comparing their Praesepe dwarf population with the giants studied by \citet{Carrera:11}. Specifically, that the enrichment is a result of the NeNa cycle as detailed in \citet{Arnould:95}. As with the O and N enhancement during CNO processing, the $^{23}$Na proton capture has the smallest cross-section, which provides an explanation for our measured enhancement.\par

\citeauthor{Arnould:95} and \citet{Ventura:11} also discuss Al enhancement through the MgAl cycle. While \citeauthor{Ventura:11} specifically address the MgAl cycle in AGB stars, the conditions in the core during the ``He-flash'' can easily reach the necessary conditions for the process to occur. When comparing our dwarf and giant populations, the measured Al abundances differ by less than the star-to-star variance, but possibly show an increase (from dwarfs $0.17\pm0.11$ to giants $0.32\pm0.09$). However, we do not measure significant decrease in Mg, the depleted element in the MgAl cycle, between our two populations. As with \citeauthor{Boesgaard:13}, we may even measure an \emph{increase} in Mg in the giants. This is not entirely unexpected, as the $^{23}Na(p,\gamma)^{24}Mg$ step within the NeNa cycle can provide a ``leakage'' of Mg, resulting in an abundance increase. \citeauthor{Ventura:11} also state that increases in Si as a result of the MgAl cycle are possible. More than with our Al and Mg measurements, the giant Si abundance shows an increase, corresponding to that expected in \citeauthor{Ventura:11}, but star-to-star variances and the lack of evidence for Mg depletion prevent us from making the definitive statement that the Si increase is a result of such cycling. \par

Our abundance measures for both Sc\,\small{II}, V\,\small{I} and V\,\small{II} are included in Table~\ref{tab:752Abs}, and show consistent measures between the two populations.\par

\subsection{Heavy Elements}
We have also included measurements of several elements with $Z\geq30$. Generally speaking, high-Z elements track slightly above solar ratios. While s-process enhancement of elements with N=50 (Y) and N=80 (Ba, and Ce to a lesser extent) is possible, the availability of neutron flux relies on higher temperature reactions. These reactions, $^{13}C(\alpha,n)^{16}O$ and $^{22}Ne(\alpha,n)^{25}Mg$, occur at approximately $1.5x10^{8}$ and $3x10^{8}K$, respectively \citep{Kaeppeler:90} - temperatures which we do not expect our target star cores to reach. While our sample giant stars have not yet reached the Assymptotic Giant Branch (AGB) phase where these reactions could proceed, the He degeneracy fraction prior to the He-flash may allow core temperatures to reach the threshold for these n-generating processes to proceed. However, since we do not measure a significant abundance increase in the end products, O and Mg, of the neutron generating process, we would not expect to see the results from s-processing. Correpondingly, we did not find a significant change in the abundances of the heavy, s-process elements, Y and Ba, and to a lesser extent, Ce, which are included in Table~\ref{tab:752Abs}. Our Sm\,\small{II} abundances, at first glance, would indicate a significant decrease in the abundance of the element. However, this does not hold up to scrutiny - the four values from the dwarf population are based on EQW measurements barely above our measurement threshold of 2m\AA. Similar to the weak N\,\small{I} feature, we did produce five additional measurements of Sm abundances, but did not include them due to the EQW measures falling below our 2m\AA~minimum threshold for EQW measurements. If we do include these abundances in our dwarf population, the apparent difference between the dwarf and giant population disappears. Therefore, when measuring Sm in this cluster, we would only rely on EQW measures from the giant population, arriving at a cluster value of [Sm/H]=$+0.05\pm0.03$. Individual star abundance measurements on an element-by-element basis are in the digital version of Table~\ref{tab:752Abs}, and the even more detailed individual EQW measures for each line in every star are listed in the included digital material.\par

\section{Summary and Conclusions}
Using spectra obtained from public archive data, we have computed atmospheric elemental composition of 23 main sequence stars and 5 red-giant ``clump'' stars in the open cluster NGC 752. Using an automated process, we determined atmospheric parameters, measured absorption line equivalent widths, and calculated elemental abundances for 25 elements, 7 with multiple ionization states. We enhanced the accuracy of our abundance calculations by using a substantially larger set of line measurements, and an automated measurement process which removed the subjective element from the equivalent width measurement process. This process also performs a line quality assessment, by comparing the abundance calculated from our automated process in a high-resolution solar spectrum, with the ``known'' values in \citet{Asplund:09}. \par

Overall, the cluster abundances track closely with their solar counterparts, with very slight enhancement ([X/Fe]$<+0.05$~dex) of most elements, relative to Fe. We find a cluster Fe abundance of $7.49\pm0.06$ ($[Fe/H]=-0.01\pm0.06$), with no differences between giant and dwarf, and/or Fe\,\small{I} and Fe\,\small{II} measurements. Similarly, the abundances of the other iron-peak elements follow their solar counterparts, and do not differ between the dwarf and giant populations ([Cr/Fe]$=0.04\pm0.06$, [Mn/Fe]$=-0.01\pm0.04$, [Co/Fe]$=0.11\pm0.07$, and [Ni/Fe]$=-0.02\pm0.04$).\par

Since our sample population contained both dwarf and giant stars, we looked for evidence of nucleosynthetic processing, including CNO cycling. We find that C\,\small{I} is slightly depleted ([C/Fe]$=-0.10\pm0.13$) in the dwarfs, moreso in the giants ($-0.22\pm0.08$). We measured a N\,\small{I} enhancement in the giant population ($+0.28\pm0.07$), whereas O\,\small{I} is consistent between the dwarfs and giants ($-0.10\pm0.08$, and $-0.15\pm0.07$, respectively). The combination of C depletion with N enhancement, is evidence of CNO cycling during the main-sequence lifetime of the evolved stars.\par

We also found evidence of the NeNa light element ``cycle''. We measured a Na enhancement in the giants ($+0.13\pm0.05$), and slight depletion in the dwarfs ($-0.18\pm0.12$). We did not find evidence of the hotter MgAl cycle, as the dwarf and giant populations measured the same Mg and Al abundances (within errors), with only a slight increase in Si (a possible secondary product of the MgAl cycle) abundance.\par

We find no evidence for enhancement of $\alpha$-elements (O, Mg, Si, Ca, Ti). The group show only slight to no enhancement ($<0.05$ dex) over either solar ratios, or between dwarfs and giants.\par

We also sampled a number of heavier (Z$\geq29$) elements. While several had slight abundance differences between dwarfs and giants, we do not attribute it to the ``slow'' neutron capture process (s-process), due to the lack of enhancement of the two high-Z elements with low cross-sections, Y and Ba, as well as lack of evidence of the requisite neutron-producing reactions.\par

\section{Acknowledgements}
We would like to thank Fabio Bresolin, Roberto Mendez, and Rolf-Peter Kudritzki for their questions, feedback and suggestions, which greatly improved the quality of this work.

This research has made use of the Keck Observatory Archive (KOA), which is operated by the W. M. Keck Observatory and the NASA Exoplanet Science Institute (NExScI), under contract with the National Aeronautics and Space Administration.\par

We are extremely grateful to the principal investigators (PIs) of the original observations of the archived data used in our research: J. Cohen, B.Jones, G. Marcy, and S. Schuler. We also wish to thank the maintainers of the digital online archives for their respective instruments, and the observatories and staff for the archives and instruments. When available, we include any requested attributions, below.\par

This work has made use of the VALD database, operated at Uppsala University, the Institute of Astronomy RAS in Moscow, and the University of Vienna.\par

This publication makes use of data products from the Two Micron All Sky Survey, which is a joint project of the University of Massachusetts and the Infrared Processing and Analysis Center/California Institute of Technology, funded by the National Aeronautics and Space Administration and the National Science Foundation.\par

We acknowledge the ESO's HARPS and UVES staff and archives for providing our solar calibration spectra.\par

The National Solar Observatory provided our solar baseline spectrum.\par

\clearpage
\bibliographystyle{apj}
\bibliography{752Refs_Bib}

\clearpage

\setlength{\tabcolsep}{4pt} 
\renewcommand{\arraystretch}{0.85} 
\begin{table}[H]
\scriptsize
\caption{NGC 752 Archive Observations}
\begin{tabular}{llllllllrc}
\multicolumn{3}{l}{Reference Number} & \multicolumn{1}{l}{RA (J2000)} &\multicolumn{1}{l}{Dec (J2000)} & \multicolumn{1}{l}{Wavelength} & \multicolumn{2}{c}{S/N} & \multicolumn{1}{l}{Date of} & \multicolumn{1}{l}{Observer} \\
 \multicolumn{1}{l}{Pla.} & \multicolumn{1}{l}{Hen.} & \multicolumn{1}{l}{Rohlfs} & \multicolumn{1}{l}{(hh:mm:ss)} &\multicolumn{1}{l}{(Deg:mm:ss)} & \multicolumn{1}{c}{Range (\AA)} & \multicolumn{1}{c}{6200~\AA} & \multicolumn{1}{c}{7800~\AA} & \multicolumn{1}{l}{Obs.} & \multicolumn{1}{l}{Inits.}\\
\hline
\multicolumn{8}{c}{Dwarfs} \\
\hline
300  & --- & 487 & 01:55:27 & +38:08:32 & 4350-6860 &  60 & --- & 2000-10-08 & BJ \\
361  & 29  & 310 & 01:55:44 & +37:54:31 & 4350-6860 &  50 & --- & 2000-10-08 & BJ \\
391  & 38  & 313 & 01:55:53 & +37:49:26 & 6240-8680 & --- &  70 & 2000-10-09 & BJ \\
413$^{\ddagger}$  & 48  & 294 & 01:55:59 & +37:40:49 & 6240-8680 & --- &  70 & 2000-10-09 & BJ \\
429  & 54  & 288 & 01:56:04 & +37:36:42 & 6240-8680 & --- &  70 & 2000-10-09 & BJ \\
520  & 80  & 290 & 01:56:23 & +37:38:14 & 5650-8090 & 150 & 130 & 2003-11-02 & AB \\
552$^{\ddagger}$ & 87  & 280 & 01:56:32 & +37:34:33 & 6240-8680 & --- &  60 & 2000-10-09 & BJ \\
575  & 94  & 303 & 01:56:36 & +37:45:35 & 6240-8680 & --- &  40 & 2000-10-09 & BJ \\
699$^{*}$ & 144 & 437 & 01:57:04 & +38:07:20 & 4350-6860 &  70 & --- & 2000-10-08 & BJ \\
& &      &     &     & 5650-8090 & 150 & 130 & 2003-11-02 & AB \\
701  & 146 & 260 & 01:57:06 & +37:50:43 & 6240-8680 & --- &  60 & 2000-10-09 & BJ \\
& &      &     &     & 5650-8090 &  90 &  80 & 2003-11-02 & AB \\
& &      &     &     & 4000-8490 &  90 & 130 & 2009-08-27 & JC \\
& &      &     &     & 4000-8490 &  90 & 120 & 2009-08-27 & JC \\
786  & 183 & 270 & 01:57:22 & +37:38:21 & 4350-6860 &  70 & --- & 2000-10-09 & BJ \\
         & &      &     &     & 4000-8490 &  90 & 130 & 2009-08-27 & JC \\
         & &      &     &     & 4000-8490 &  80 & 120 & 2009-08-27 & JC \\
790  & 185 & --- & 01:57:24 & +37:52:12 & 5650-8090 & 160 & 140 & 2003-11-02 & AB \\
791  & 184 & 434 & 01:57:24 & +38:06:10 & 5650-8090 & 150 & 130 & 2003-11-02 & AB \\
828  & 199 & 266 & 01:57:34 & +37:42:01 & 6240-8680 &  70 & --- & 2000-10-09 & BJ \\
859$^{**}$  & 207 & 252 & 01:57:38 & +37:49:44 & 6240-8680 &  70 & --- & 2000-10-09 & BJ \\
         & &      &     &     & 4000-8490 & 110 & 140 & 2009-08-27 & JC \\
         & &      &     &     & 4000-8490 &  70 & 100 & 2009-08-27 & JC \\
864  & 211 & 430 & 01:57:39 & +38:08:39 & 6240-8680 & --- &  40 & 2000-10-09 & BJ \\
         & &      &     &     & 5650-8090 & 150 & 130 & 2003-11-02 & AB \\
889  & 216 & 427 & 01:57:45 & +38:11:07 & 6240-8680 & --- &  30 & 2000-10-09 & BJ \\
         & &      &     &     & 5650-8090 & 150 & 130 & 2003-11-02 & AB \\
921  & 229 & 103 & 01:57:52 & +37:27:46 & 5650-8090 & 160 & 140 & 2003-11-02 & AB \\
964  & 244 & 388 & 01:58:03 & +38:02:30 & 5650-8090 & 130 & 120 & 2003-11-02 & AB \\
993  & 256 & 101 & 01:58:09 & +37:28:36 & 6240-8680 & --- &  60 & 2000-10-09 & BJ \\
999  & --- & 107 & 01:58:11 & +37:23:53 & 6240-8680 & --- &  70 & 2000-10-09 & BJ \\
1012 & --- & 118 & 01:58:13 & +37:15:20 & 5650-8090 & 150 & 120 & 2003-11-02 & AB \\
1017 & 265 & 240 & 01:58:16 & +37:33:26 & 6240-8680 & --- &  70 & 2000-10-09 & BJ \\
1107 & 298 & 232 & 01:58:34 & +37:40:19 & 6240-8680 & --- &  70 & 2000-10-09 & BJ \\
1270 & --- & --- & 01:59:19 & +37:49:50 & 6240-8680 & --- &  70 & 2000-10-09 & BJ \\
1284$^{\ddagger}$ & --- & --- & 01:59:20 & +37:23:23 & 6240-8680 & --- &  70 & 2000-10-09 & BJ \\
         &           &      &     &     & 4000-8490 & 100 & 140 & 2009-08-27 & JC \\
         &           &      &     &     & 4000-8490 & 110 & 150 & 2009-08-27 & JC \\
1365 & --- & 187 & 01:59:47 & +37:49:48 & 4350-6860 &  35 & --- & 2000-10-08 & BJ \\
\hline
\multicolumn{8}{c}{Giants} \\
\hline
350  & 24  & 25  & 01:55:40 & +37:52:28 & 3360-8100 &  55 &  70 & 2008-06-20 & GM \\
         &           &      &     &     & 3360-8100 & 160 & 220 & 2008-06-22$^{\dagger}$ & GM \\
356  & 20  & 27  & 01:55:43 & +37:37:39 & 3360-8100 &  55 & 60 & 2008-06-20$^{\dagger}$ & GM \\
         &           &      &     &     & 3360-8100 & 160 & 220 & 2008-06-22 & GM \\
506  & 29  & 75  & 01:56:19 & +37:58:02 & 4690-9140 & 290 & 400 & 2009-10-08 & SS\\
687  & --- & 137 & 01:57:04 & +38:07:57 & 3360-8100 &  50 &  70 & 2008-06-20 & GM \\
         &           &      &     &     & 3360-8100 & 160 & 220 & 2008-06-22 & GM \\
         &           &      &     &     & 4690-9140 & 350 & 450 & 2009-10-08 & SS\\
1089 & 67  & 295 & 01:58:30 & +37:51:37 & 3360-8100 &  50 &  70 & 2008-06-20 & GM \\
         &           &      &     &     & 4690-9140 & 290 & 400 & 2009-10-08 & SS\\
1172 & 65  & 311 & 01:58:53 & +37:48:57 & 3360-8100 &  50 &  80 & 2008-06-20 & GM \\
\hline
\multicolumn{8}{l|}{* - Possible single-line binary} & \multicolumn{2}{l}{AB- Boesgaard} \\
\multicolumn{8}{l|}{$\dagger$ - Possible mis-aligned order fit (see text)} & \multicolumn{2}{l}{JC- Cohen} \\
\multicolumn{8}{l|}{$\ddagger$ - Spectroscopic Binary (unused)} & \multicolumn{2}{l}{BJ- Jones} \\
\multicolumn{8}{l|}{** - Parallax differential (unused)} & \multicolumn{2}{l}{GM- Marcy} \\
\multicolumn{8}{l|}{} & \multicolumn{2}{l}{SS- Schuler} \\
\hline
\hline
\label{tab:752Obs}
\end{tabular}
\end{table}
\setlength{\tabcolsep}{6pt} 
\renewcommand{\arraystretch}{1.0} 

\begin{table}[H]
\caption{NGC 752 Atmospheric Parameters}
\begin{tabular}{lrrrr}
\multicolumn{1}{l}{PLA \#} & \multicolumn{1}{c}{$T_{eff}$} & \multicolumn{1}{c}{log\,g} & \multicolumn{1}{c}{$\xi$} & \multicolumn{1}{l}{$min. EQW$}\\
\multicolumn{1}{c}{} & \multicolumn{1}{c}{(K)} & \multicolumn{1}{l}{} & \multicolumn{1}{c}{(km s$^{-1}$)} & \multicolumn{1}{c}{(m\AA)}\\
\hline
\multicolumn{5}{c}{Dwarfs} \\
\hline
300 & 5650 & 4.50 & 2.1 & 12.5 \\
361 & 5550 & 4.55 & 1.8 & 8.3 \\
391 & 5450 & 4.55 & 1.8 & 6.3 \\
429 & 5250 & 4.55 & 1.8 & 6.5 \\
520 & 6100 & 4.40 & 1.6 & 2.6 \\
575 & 5525 & 4.55 & 1.8 & 8.8 \\
699 & 5875 & 4.50 & 1.6 & 2.7 \\
701 & 5725 & 4.45 & 1.7 & 1.5* \\
786 & 5575 & 4.50 & 2.0 & 1.6* \\
790 & 6250 & 4.35 & 2.1 & 3.0 \\
791 & 6125 & 4.40 & 1.8 & 3.0 \\
828 & 5425 & 4.50 & 1.9 & 6.3 \\
864 & 6050 & 4.45 & 1.7 & 3.0 \\
889 & 6175 & 4.35 & 1.7 & 3.0 \\
921 & 6100 & 4.45 & 1.6 & 2.7 \\
964 & 6100 & 4.45 & 1.6 & 3.6 \\
993 & 5675 & 4.55 & 1.4 & 6.5 \\
999 & 5750 & 4.45 & 1.4 & 5.8 \\
1012 & 6275 & 4.35 & 1.7 & 3.3 \\
1017 & 5775 & 4.55 & 1.6 & 6.0 \\
1107 & 5750 & 4.55 & 1.7 & 5.6 \\
1270 & 5575 & 4.55 & 1.8 & 5.6 \\
1365 & 5825 & 4.50 & 1.4 & 11.2 \\
\hline
\multicolumn{5}{c}{Giants} \\
\hline
350 & 5000 & 2.65 & 1.6 & 1.9* \\
356 & 4950 & 2.60 & 1.5 & 1.9* \\
506 & 4975 & 2.65 & 1.5 & 1.1* \\
687 & 5000 & 2.65 & 1.5 & 0.7* \\
1089 & 5100 & 2.60 & 1.5 & 1.0* \\
1172 & 5000 & 2.55 & 1.6 & 4.0 \\
\hline
\multicolumn{5}{l}{*-Adjusted to minimum of 2.0~m\AA} \\
\hline
\hline
\label{tab:atmParms}
\end{tabular}
\end{table}

\begin{table}[H]
\caption{Effective Temperature Comparison}
\begin{tabular}{lrrrrrrrrrr}
{PLA ID} & {This Work} & {Hobbs} & {Bala.} & {Sest.} & {Biaz.} & {Made.} & {Phot.} & {Lit. Mean} & {Delta} & {Lit. Std}\\
{300} & {5650} & {} & {} & {} & {} & {} & {} & {n/a} & {n/a} & {n/a}\\
{350} & {5000} & {} & {} & {} & {5006} & {} & {} & {5006} & {-6} & {n/a}\\
{356} & {4950} & {} & {} & {} & {5017} & {} & {} & {5017} & {-67} & {n/a}\\
{361} & {5550} & {} & {} & {} & {} & {5537} & {} & {5537} & {13} & {n/a}\\
{391} & {5450} & {} & {} & {} & {} & {5517} & {} & {5517} & {-67} & {n/a}\\
{429} & {5250} & {} & {} & {} & {} & {5124} & {} & {5124} & {126} & {n/a}\\
{506} & {4975} & {} & {} & {} & {5009} & {} & {} & {5009} & {-34} & {n/a}\\
{520} & {6100} & {} & {} & {6151} & {5999} & {6102} & {6102} & {6089} & {12} & {64}\\
{575} & {5525} & {5350} & {} & {5425} & {} & {} & {} & {5388} & {138} & {53}\\
{687} & {5000} & {} & {} & {} & {5010} & {} & {} & {5010} & {-10} & {n/a}\\
{699} & {5875} & {} & {} & {5970} & {5754} & {} & {5899} & {5874} & {1} & {110}\\
{701} & {5725} & {5690} & {5693} & {5677} & {5722} & {5656} & {5674} & {5685} & {40} & {22}\\
{786} & {5575} & {} & {} & {5531} & {5653} & {5517} & {} & {5567} & {8} & {75}\\
{790} & {6250} & {6220} & {} & {6325} & {5992} & {6264} & {6248} & {6210} & {40} & {128}\\
{791} & {6125} & {} & {} & {} & {6076} & {6208} & {6198} & {6161} & {-36} & {73}\\
{828} & {5425} & {} & {} & {} & {} & {} & {} & {n/a} & {n/a} & {n/a}\\
{864} & {6050} & {} & {} & {6060} & {6021} & {6071} & {6073} & {6056} & {-6} & {24}\\
{889} & {6175} & {} & {} & {6231} & {6091} & {6177} & {6169} & {6167} & {8} & {58}\\
{921} & {6100} & {5870} & {6160} & {6223} & {6081} & {6169} & {6162} & {6111} & {-11} & {126}\\
{964} & {6100} & {} & {} & {6102} & {6048} & {} & {6059} & {6070} & {30} & {29}\\
{993} & {5675} & {} & {} & {5611} & {} & {5593} & {} & {5602} & {73} & {13}\\
{999} & {5750} & {} & {} & {} & {} & {5728} & {} & {5728} & {22} & {n/a}\\
{1012} & {6275} & {} & {} & {6282} & {6217} & {} & {6212} & {6237} & {38} & {39}\\
{1017} & {5775} & {5740} & {5760} & {5791} & {} & {5764} & {} & {5764} & {11} & {21}\\
{1089} & {5100} & {} & {} & {} & {5081} & {} & {} & {5081} & {19} & {n/a}\\
{1107} & {5750} & {} & {} & {5791} & {} & {5764} & {} & {5778} & {-28} & {19}\\
{1172} & {5000} & {} & {} & {} & {4975} & {} & {} & {4975} & {25} & {n/a}\\
{1270} & {5575} & {} & {} & {} & {} & {5621} & {} & {5621} & {-46} & {n/a}\\
{1365} & {5825} & {} & {} & {5603} & {5912} & {5586} & {} & {5700} & {125} & {184}\\
\hline
\multicolumn{11}{l}{References:}\\
\multicolumn{11}{l}{Hobbs: \citet{Hobbs:86}}\\
\multicolumn{11}{l}{Bala.: \citet{Balachandran:95}}\\
\multicolumn{11}{l}{Sest.: \citet{Sestito:04}}\\
\multicolumn{11}{l}{Biaz.: \citet{Biazzo:07}}\\
\multicolumn{11}{l}{Made.: \citet{Maderak:13}}\\
\multicolumn{11}{l}{Phot.: Photometric value as calculated by \citet{Cummings:17},}\\
\multicolumn{1}{l}{} & \multicolumn{10}{l}{using photometry from \citet{Daniel:94}.}\\

\label{tab:TComps}
\end{tabular}
\end{table}

\begin{table}[H]
\caption{Absorption Line List}
\begin{tabular}{lllll}
\multicolumn{1}{l}{} & \multicolumn{1}{l}{Wavelength} & \multicolumn{1}{l}{Ex. Pot.} & \multicolumn{1}{l}{} & \multicolumn{1}{l}{}\\
\multicolumn{1}{l}{Ion} & \multicolumn{1}{c}{(\AA)} & \multicolumn{1}{c}{(eV)} & \multicolumn{1}{l}{$Log(gf)$} & \multicolumn{1}{r}{Ref.$^{*}$}\\
\hline
C\,\small{I} & 5380.340 & \multicolumn{1}{r}{ 7.68} & \multicolumn{1}{r}{-1.62} & \multicolumn{1}{r}{58}\\
C\,\small{I} & 6587.620 & \multicolumn{1}{r}{ 8.53} & \multicolumn{1}{r}{-1.00} & \multicolumn{1}{r}{23}\\
C\,\small{I} & 7100.130 & \multicolumn{1}{r}{ 8.64} & \multicolumn{1}{r}{-1.02} & \multicolumn{1}{r}{23}\\
C\,\small{I} & 7111.450 & \multicolumn{1}{r}{ 8.64} & \multicolumn{1}{r}{-1.08} & \multicolumn{1}{r}{23}\\
C\,\small{I} & 7113.170 & \multicolumn{1}{r}{ 8.65} & \multicolumn{1}{r}{-0.77} & \multicolumn{1}{r}{23}\\
C\,\small{I} & 7115.170 & \multicolumn{1}{r}{ 8.64} & \multicolumn{1}{r}{-1.47} & \multicolumn{1}{r}{23, 59}\\
\multicolumn{5}{c}{...}\\
\hline
\multicolumn{5}{l}{*-See Table:\ref{tab:LineRefs} for reference key}\\
\hline
\label{tab:LineList}
\end{tabular}
\end{table}

\clearpage
\begin{table}[H]
\caption{Absorption Line References}
\begin{tabular}{llll} 
\multicolumn{1}{l}{Ref.} & \multicolumn{1}{l|}{} & \multicolumn{1}{l}{Ref.} & \multicolumn{1}{l}{}\\
\multicolumn{1}{l}{No.} & \multicolumn{1}{l|}{Reference} & \multicolumn{1}{l}{No.} & \multicolumn{1}{l|}{Reference}\\
\hline
\multicolumn{1}{r}{1} & \multicolumn{1}{l|}{\citet{Allen:11}} & \multicolumn{1}{r}{35} & \multicolumn{1}{l|}{\citet{Garz:73}} \\
\multicolumn{1}{r}{2} & \multicolumn{1}{l|}{\citet{AllendePrieto:98}} & \multicolumn{1}{r}{36} & \multicolumn{1}{l|}{\citet{Giridhar:95}} \\
\multicolumn{1}{r}{3} & \multicolumn{1}{l|}{\citet{Bard:91}} & \multicolumn{1}{r}{37} & \multicolumn{1}{l|}{\citet{Hannaford:82}} \\
\multicolumn{1}{r}{4} & \multicolumn{1}{l|}{\citet{Bard:94}} & \multicolumn{1}{r}{38} & \multicolumn{1}{l|}{\citet{Kock:68}} \\
\multicolumn{1}{r}{5} & \multicolumn{1}{l|}{\citet{Beveridge:94}} & \multicolumn{1}{r}{39} & \multicolumn{1}{l|}{\citet{Kramida:14}} \\
\multicolumn{1}{r}{6} & \multicolumn{1}{l|}{\citet{Biemont:91}} & \multicolumn{1}{r}{40} & \multicolumn{1}{l|}{\citet{Kroll:87}} \\
\multicolumn{1}{r}{7} & \multicolumn{1}{l|}{\citet{Biemont:80}} & \multicolumn{1}{r}{41} & \multicolumn{1}{l|}{\citet{Lambert:68}} \\
\multicolumn{1}{r}{8} & \multicolumn{1}{l|}{\citet{Bizzarri:93}} & \multicolumn{1}{r}{42} & \multicolumn{1}{l|}{\citet{Lambert:78}} \\
\multicolumn{1}{r}{9} & \multicolumn{1}{l|}{\citet{Blackwell:80a}a} & \multicolumn{1}{r}{43} & \multicolumn{1}{l|}{\citet{Lawler:89}} \\
\multicolumn{1}{r}{10} & \multicolumn{1}{l|}{\citet{Blackwell:86a}a} & \multicolumn{1}{r}{44} & \multicolumn{1}{l|}{\citet{Lawler:13}} \\
\multicolumn{1}{r}{11} & \multicolumn{1}{l|}{\citet{Blackwell:86b}b} & \multicolumn{1}{r}{45} & \multicolumn{1}{l|}{\citet{Lawler:08}} \\
\multicolumn{1}{r}{12} & \multicolumn{1}{l|}{\citet{Blackwell:86c}c} & \multicolumn{1}{r}{46} & \multicolumn{1}{l|}{\citet{Lawler:15}} \\
\multicolumn{1}{r}{13} & \multicolumn{1}{l|}{\citet{Blackwell:83}} & \multicolumn{1}{r}{47} & \multicolumn{1}{l|}{\citet{Lawler:09}} \\
\multicolumn{1}{r}{14} & \multicolumn{1}{l|}{\citet{Blackwell:82a}a} & \multicolumn{1}{r}{48} & \multicolumn{1}{l|}{\citet{Lawler:14}} \\
\multicolumn{1}{r}{15} & \multicolumn{1}{l|}{\citet{Blackwell:82d}d} & \multicolumn{1}{r}{49} & \multicolumn{1}{l|}{\citet{Martin:88}} \\
\multicolumn{1}{r}{16} & \multicolumn{1}{l|}{\citet{Blackwell:76}} & \multicolumn{1}{r}{50} & \multicolumn{1}{l|}{\citet{McWilliam:94}} \\
\multicolumn{1}{r}{17} & \multicolumn{1}{l|}{\citet{Blackwell:79a}a} & \multicolumn{1}{r}{51} & \multicolumn{1}{l|}{\citet{McWilliam:95}} \\
\multicolumn{1}{r}{18} & \multicolumn{1}{l|}{\citet{Blackwell:79b}b} & \multicolumn{1}{r}{52} & \multicolumn{1}{l|}{\citet{Meylan:93}} \\
\multicolumn{1}{r}{19} & \multicolumn{1}{l|}{\citet{Blackwell:80b}b} & \multicolumn{1}{r}{53} & \multicolumn{1}{l|}{\citet{Moity:83}} \\
\multicolumn{1}{r}{20} & \multicolumn{1}{l|}{\citet{Blackwell:82b}b} & \multicolumn{1}{r}{54} & \multicolumn{1}{l|}{\citet{Nissen:97}} \\
\multicolumn{1}{r}{21} & \multicolumn{1}{l|}{\citet{Blackwell:82c}c} & \multicolumn{1}{r}{55} & \multicolumn{1}{l|}{\citet{OBrian:91}} \\
\multicolumn{1}{r}{22} & \multicolumn{1}{l|}{\citet{Blackwell:84}} & \multicolumn{1}{r}{56} & \multicolumn{1}{l|}{\citet{Prochaska:00}} \\
\multicolumn{1}{r}{23} & \multicolumn{1}{l|}{\citet{Boesgaard:13}} & \multicolumn{1}{r}{57} & \multicolumn{1}{l|}{\citet{Prochaska:00}} \\
\multicolumn{1}{r}{24} & \multicolumn{1}{l|}{\citet{Booth:84}} & \multicolumn{1}{r}{58} & \multicolumn{1}{l|}{\citet{Ramirez:11}} \\
\multicolumn{1}{r}{25} & \multicolumn{1}{l|}{\citet{Buurman:86}} & \multicolumn{1}{r}{59} & \multicolumn{1}{l|}{\citet{Reddy:13}} \\
\multicolumn{1}{r}{26} & \multicolumn{1}{l|}{\citet{Cardon:82}} & \multicolumn{1}{r}{60} & \multicolumn{1}{l|}{\citet{Savanov:90}} \\
\multicolumn{1}{r}{27} & \multicolumn{1}{l|}{\citet{Cohen:03}} & \multicolumn{1}{r}{61} & \multicolumn{1}{l|}{\citet{Schnabel:99}} \\
\multicolumn{1}{r}{28} & \multicolumn{1}{l|}{\citet{DenHartog:14}} & \multicolumn{1}{r}{62} & \multicolumn{1}{l|}{\citet{Smith:81}} \\
\multicolumn{1}{r}{29} & \multicolumn{1}{l|}{\citet{DenHartog:11}} & \multicolumn{1}{r}{63} & \multicolumn{1}{l|}{\citet{Sobeck:07}} \\
\multicolumn{1}{r}{30} & \multicolumn{1}{l|}{\citet{Edvardsson:93}} & \multicolumn{1}{r}{64} & \multicolumn{1}{l|}{\citet{Stephens:99}} \\
\multicolumn{1}{r}{31} & \multicolumn{1}{l|}{\citet{Francois:88}} & \multicolumn{1}{r}{65} & \multicolumn{1}{l|}{\citet{Wickliffe:97}} \\
\multicolumn{1}{r}{32} & \multicolumn{1}{l|}{\citet{Fry:97}} & \multicolumn{1}{r}{66} & \multicolumn{1}{l|}{\citet{Wiese:80}} \\
\multicolumn{1}{r}{33} & \multicolumn{1}{l|}{\citet{Fuhr:88}} & \multicolumn{1}{r}{67} & \multicolumn{1}{l|}{\citet{Wood:13}} \\
\multicolumn{1}{r}{34} & \multicolumn{1}{l|}{\citet{Fuhrmann:95}} & \multicolumn{1}{r}{68} & \multicolumn{1}{l|}{\citet{Wood:14}} \\

\hline
\label{tab:LineRefs}
\end{tabular}
\end{table}

\begin{sidewaystable}
\caption{Solar absorption line quality evaluation}
\begin{tabular}{|l|l|l|l|l|l|l|l|l|l|}
\multicolumn{1}{l}{Ion} & \multicolumn{1}{l}{Asplund (2009)} & \multicolumn{1}{l}{$\Delta\leq0.05$} & \multicolumn{1}{l}{$\Delta\leq0.10$} & \multicolumn{1}{l}{$\Delta\leq0.20$} & \multicolumn{1}{l}{$\Delta\leq0.40$} & \multicolumn{1}{l}{$\Delta\leq1.00$} & \multicolumn{1}{l}{All lines}\\
\hline
C\,\small{I} & 8.43$\pm$0.05 & 8.44$\pm$0.02 (3) & 8.42$\pm$0.04 (4) & 8.41$\pm$0.09 (6) & 8.47$\pm$0.13 (8) & 8.55$\pm$0.25 (9) & 8.55$\pm$0.25 (9)\\
\hline
O\,\small{I} & 8.69$\pm$0.05 & 8.69$\pm$0.02 (2) & 8.72$\pm$0.05 (3) & 8.77$\pm$0.06 (6) & 8.77$\pm$0.06 (6) & 8.84$\pm$0.18 (7) & 8.84$\pm$0.18 (7)\\
\hline
Na\,\small{I} & 6.24$\pm$0.04 & 6.26$\pm$0.01 (2) & 6.26$\pm$0.01 (2) & 6.27$\pm$0.08 (5) & 6.23$\pm$0.12 (6) & 6.36$\pm$0.32 (7) & 6.36$\pm$0.32 (7)\\
\hline
Mg\,\small{I} & 7.60$\pm$0.04 & 7.62$\pm$0.02 (6) & 7.61$\pm$0.05 (10) & 7.60$\pm$0.07 (12) & 7.60$\pm$0.07 (12) & 7.48$\pm$0.27 (15) & 7.20$\pm$0.91 (17)\\
\hline
Al\,\small{I} & 6.45$\pm$0.03 & 6.42$\pm$0.00 (1) & 6.41$\pm$0.01 (2) & 6.35$\pm$0.06 (4) & 6.31$\pm$0.08 (6) & 6.31$\pm$0.08 (6) & 6.31$\pm$0.08 (6)\\
\hline
Si\,\small{I} & 7.51$\pm$0.03 & 7.53$\pm$0.03 (8) & 7.52$\pm$0.05 (16) & 7.57$\pm$0.10 (33) & 7.59$\pm$0.11 (37) & 7.62$\pm$0.15 (40) & 7.71$\pm$0.41 (42)\\
\hline
Si\,\small{II} & 7.51$\pm$0.03 & \multicolumn{1}{c|}{---}  & \multicolumn{1}{c|}{---}  & 7.70$\pm$0.00 (2) & 7.70$\pm$0.00 (2) & 7.70$\pm$0.00 (2) & 7.70$\pm$0.00 (2)\\
\hline
S\,\small{I} & 7.12$\pm$0.03 & \multicolumn{1}{c|}{---}  & \multicolumn{1}{c|}{---}  & 7.27$\pm$0.00 (1) & 7.37$\pm$0.10 (2) & 7.37$\pm$0.10 (2) & 7.37$\pm$0.10 (2)\\
\hline
Ca\,\small{I} & 6.34$\pm$0.04 & 6.31$\pm$0.01 (3) & 6.32$\pm$0.05 (5) & 6.32$\pm$0.14 (14) & 6.23$\pm$0.21 (29) & 6.15$\pm$0.27 (35) & 6.15$\pm$0.27 (35)\\
\hline
Sc\,\small{I} & 3.15$\pm$0.04 & \multicolumn{1}{c|}{---}  & 3.22$\pm$0.00 (1) & 3.19$\pm$0.13 (5) & 3.18$\pm$0.19 (7) & 3.18$\pm$0.19 (7) & 3.18$\pm$0.19 (7)\\
\hline
Sc\,\small{II} & 3.15$\pm$0.04 & 3.14$\pm$0.01 (4) & 3.15$\pm$0.05 (7) & 3.15$\pm$0.09 (11) & 3.16$\pm$0.15 (13) & 3.19$\pm$0.18 (14) & 3.19$\pm$0.18 (14)\\
\hline
Ti\,\small{I} & 4.95$\pm$0.05 & 4.95$\pm$0.03 (60) & 4.94$\pm$0.05 (103) & 4.94$\pm$0.10 (166) & 4.97$\pm$0.19 (255) & 5.09$\pm$0.36 (344) & 5.55$\pm$0.96 (471)\\
\hline
Ti\,\small{II} & 4.95$\pm$0.05 & 4.94$\pm$0.03 (9) & 4.96$\pm$0.06 (16) & 4.97$\pm$0.10 (25) & 4.95$\pm$0.21 (41) & 5.03$\pm$0.36 (54) & 5.13$\pm$0.69 (59)\\
\hline
V\,\small{I} & 3.93$\pm$0.08 & 3.92$\pm$0.02 (24) & 3.94$\pm$0.05 (40) & 3.95$\pm$0.09 (58) & 3.94$\pm$0.12 (64) & 3.96$\pm$0.15 (66) & 3.99$\pm$0.28 (67)\\
\hline
V\,\small{II} & 3.93$\pm$0.08 & 3.92$\pm$0.00 (1) & 3.88$\pm$0.04 (2) & 3.86$\pm$0.04 (3) & 3.79$\pm$0.13 (4) & 3.79$\pm$0.13 (4) & 3.79$\pm$0.13 (4)\\
\hline
Cr\,\small{I} & 5.64$\pm$0.04 & 5.63$\pm$0.03 (20) & 5.62$\pm$0.06 (46) & 5.63$\pm$0.11 (78) & 5.63$\pm$0.16 (100) & 5.68$\pm$0.33 (130) & 5.79$\pm$0.58 (141)\\
\hline
Cr\,\small{II} & 5.64$\pm$0.04 & 5.63$\pm$0.03 (2) & 5.64$\pm$0.06 (4) & 5.67$\pm$0.10 (7) & 5.62$\pm$0.20 (12) & 5.58$\pm$0.29 (15) & 5.58$\pm$0.29 (15)\\
\hline
Mn\,\small{I} & 5.43$\pm$0.05 & 5.43$\pm$0.02 (5) & 5.41$\pm$0.06 (11) & 5.36$\pm$0.08 (20) & 5.34$\pm$0.09 (23) & 5.34$\pm$0.09 (23) & 5.41$\pm$0.34 (24)\\
\hline
Fe\,\small{I} & 7.50$\pm$0.04 & 7.49$\pm$0.03 (49) & 7.49$\pm$0.05 (79) & 7.47$\pm$0.10 (148) & 7.39$\pm$0.19 (265) & 7.32$\pm$0.30 (355) & 7.32$\pm$0.39 (368)\\
\hline
Fe\,\small{II} & 7.50$\pm$0.04 & 7.50$\pm$0.04 (12) & 7.49$\pm$0.05 (22) & 7.48$\pm$0.09 (35) & 7.43$\pm$0.15 (46) & 7.37$\pm$0.27 (56) & 7.34$\pm$0.32 (58)\\
\hline
Co\,\small{I} & 4.99$\pm$0.07 & 4.99$\pm$0.03 (19) & 4.96$\pm$0.05 (33) & 4.96$\pm$0.09 (50) & 4.99$\pm$0.17 (66) & 5.08$\pm$0.37 (92) & 5.69$\pm$1.10 (130)\\
\hline
Ni\,\small{I} & 6.22$\pm$0.04 & 6.22$\pm$0.03 (29) & 6.23$\pm$0.06 (62) & 6.24$\pm$0.10 (97) & 6.23$\pm$0.16 (125) & 6.22$\pm$0.29 (150) & 6.29$\pm$0.51 (158)\\
\hline
Cu\,\small{I} & 4.19$\pm$0.04 & 4.16$\pm$0.00 (1) & 4.18$\pm$0.08 (3) & 4.18$\pm$0.08 (3) & 4.23$\pm$0.12 (4) & 4.31$\pm$0.19 (5) & 4.31$\pm$0.19 (5)\\
\hline
Zn\,\small{I} & 4.56$\pm$0.05 & \multicolumn{1}{c|}{---}  & 4.65$\pm$0.00 (2) & 4.67$\pm$0.03 (3) & 4.67$\pm$0.03 (3) & 4.67$\pm$0.03 (3) & 4.67$\pm$0.03 (3)\\
\hline
Y\,\small{I} & 2.21$\pm$0.05 & \multicolumn{1}{c|}{---}  & \multicolumn{1}{c|}{---}  & \multicolumn{1}{c|}{---}  & 2.43$\pm$0.00 (1) & 2.63$\pm$0.21 (2) & 2.63$\pm$0.21 (2)\\
\hline
Y\,\small{II} & 2.21$\pm$0.05 & 2.20$\pm$0.01 (2) & 2.17$\pm$0.04 (3) & 2.26$\pm$0.09 (7) & 2.34$\pm$0.19 (14) & 2.25$\pm$0.29 (16) & 2.25$\pm$0.29 (16)\\
\hline
Zr\,\small{I} & 2.58$\pm$0.04 & 2.55$\pm$0.00 (1) & 2.59$\pm$0.05 (2) & 2.64$\pm$0.11 (6) & 2.66$\pm$0.11 (7) & 2.66$\pm$0.11 (7) & 2.66$\pm$0.11 (7)\\
\hline
Zr\,\small{II} & 2.58$\pm$0.04 & 2.56$\pm$0.00 (1) & 2.56$\pm$0.00 (1) & 2.56$\pm$0.00 (1) & 2.56$\pm$0.00 (1) & 2.56$\pm$0.00 (1) & 2.56$\pm$0.00 (1)\\
\hline
Ba\,\small{II} & 2.18$\pm$0.09 & \multicolumn{1}{c|}{---}  & \multicolumn{1}{c|}{---}  & 2.03$\pm$0.04 (2) & 2.15$\pm$0.17 (3) & 2.04$\pm$0.24 (4) & 1.84$\pm$0.46 (5)\\
\hline
Ce\,\small{II} & 1.58$\pm$0.04 & 1.61$\pm$0.01 (3) & 1.59$\pm$0.06 (8) & 1.61$\pm$0.08 (9) & 1.65$\pm$0.15 (14) & 1.68$\pm$0.17 (15) & 1.93$\pm$1.00 (16)\\
\hline
Pr\,\small{II} & 0.72$\pm$0.04 & \multicolumn{1}{c|}{---}  & \multicolumn{1}{c|}{---}  & \multicolumn{1}{c|}{---}  & \multicolumn{1}{c|}{---}  & 1.60$\pm$0.00 (1) & 1.60$\pm$0.00 (1)\\
\hline
Nd\,\small{II} & 1.42$\pm$0.04 & 1.42$\pm$0.02 (8) & 1.42$\pm$0.03 (10) & 1.45$\pm$0.07 (12) & 1.47$\pm$0.11 (13) & 1.52$\pm$0.20 (14) & 1.52$\pm$0.20 (14)\\
\hline
Sm\,\small{II} & 0.96$\pm$0.04 & 0.96$\pm$0.00 (1) & 1.02$\pm$0.04 (4) & 1.02$\pm$0.04 (4) & 1.09$\pm$0.11 (6) & 1.15$\pm$0.17 (7) & 1.15$\pm$0.17 (7)\\
\hline
Eu\,\small{II} & 0.52$\pm$0.04 & \multicolumn{1}{c|}{---}  & 0.45$\pm$0.00 (1) & 0.45$\pm$0.00 (1) & 0.45$\pm$0.00 (1) & 0.71$\pm$0.26 (2) & 0.71$\pm$0.26 (2)\\
\hline
\label{tab:SolarAbs}
\end{tabular}
\end{sidewaystable}

\setlength{\tabcolsep}{4pt} 
\renewcommand{\arraystretch}{1.0} 
\begin{table}[H]
\caption{Elemental Abundances for NGC 752:\\Fe\,\small{I}~and Fe\,\small{II}~abundances are listed on the scale $log_{10}({N(Fe)}/{N(H)})+12.00$ where $log_{10}N(H)$ is set at 12.00, as stated in the text. Listed abundance errors are the star-to-star variations ($\sigma_{var}$) and the atmospheric errors ($\sigma_{atm}$) from Table~\ref{tab:atmErrs}. The atmospheric errors for ``All Stars" abundances are the mean of the dwarf and giant stars, weighted by population. The ``Q.'' (quality) score is the average of all measured lines for a given element (see section \ref{sec:LineSelection}). Quality scores for synthesized measurements are listed as ``n/a''.}
\begin{tabular}{llllllllllllllll}
\multicolumn{1}{l}{} & \multicolumn{5}{c}{All Stars} & \multicolumn{5}{c}{Dwarfs} & \multicolumn{5}{c}{Giants} \\
\hline
\multicolumn{1}{|l||}{Ion} & \multicolumn{1}{l|}{Ab.} & \multicolumn{1}{r}{$\sigma_{var}$} & \multicolumn{1}{r|}{$\sigma_{atm}$} & \multicolumn{1}{l|}{\#} & \multicolumn{1}{l||}{Q.} & \multicolumn{1}{l|}{Ab.} & \multicolumn{1}{r}{$\sigma_{var}$} & \multicolumn{1}{r|}{$\sigma_{atm}$} & \multicolumn{1}{l|}{\#} & \multicolumn{1}{l||}{Q.} & \multicolumn{1}{l|}{Ab.} & \multicolumn{1}{r}{$\sigma_{var}$} & \multicolumn{1}{r|}{$\sigma_{atm}$} & \multicolumn{1}{l|}{\#} & \multicolumn{1}{l|}{Q.} \\
\hline
\multicolumn{1}{|l||}{Fe\,\small{I}}   & \multicolumn{1}{r|}{  7.50}& \multicolumn{1}{r}{0.06} & \multicolumn{1}{r|}{0.08} & \multicolumn{1}{r|}{ 29} & \multicolumn{1}{r||}{  8.0} & \multicolumn{1}{r|}{  7.50}& \multicolumn{1}{r}{0.07} & \multicolumn{1}{r|}{0.08} & \multicolumn{1}{r|}{ 23} & \multicolumn{1}{r||}{  8.1} & \multicolumn{1}{r|}{  7.51}& \multicolumn{1}{r}{0.03} & \multicolumn{1}{r|}{0.11} & \multicolumn{1}{r|}{ 6} & \multicolumn{1}{r|}{ 7.9}  \\
\hline
\multicolumn{1}{|l||}{Fe\,\small{II}}  & \multicolumn{1}{r|}{  7.45}& \multicolumn{1}{r}{0.06} & \multicolumn{1}{r|}{0.06} & \multicolumn{1}{r|}{ 29} & \multicolumn{1}{r||}{  8.5} & \multicolumn{1}{r|}{  7.45}& \multicolumn{1}{r}{0.07} & \multicolumn{1}{r|}{0.05} & \multicolumn{1}{r|}{ 23} & \multicolumn{1}{r||}{  8.6} & \multicolumn{1}{r|}{  7.45}& \multicolumn{1}{r}{0.05} & \multicolumn{1}{r|}{0.10} & \multicolumn{1}{r|}{ 6} & \multicolumn{1}{r|}{  7.8}  \\
\hline
\hline
\multicolumn{1}{|l||}{[C\,\small{I}/Fe]}   & \multicolumn{1}{r|}{ -0.13}& \multicolumn{1}{r}{0.13} & \multicolumn{1}{r|}{0.13} & \multicolumn{1}{r|}{ 27} & \multicolumn{1}{r||}{  n/a} & \multicolumn{1}{r|}{ -0.10}& \multicolumn{1}{r}{0.13} & \multicolumn{1}{r|}{0.02} & \multicolumn{1}{r|}{ 21} & \multicolumn{1}{r||}{  0.5} & \multicolumn{1}{r|}{ -0.22}& \multicolumn{1}{r}{0.08} & \multicolumn{1}{r|}{0.02} & \multicolumn{1}{r|}{ 6} & \multicolumn{1}{r|}{  n/a}  \\
\hline
\multicolumn{1}{|l||}{[N\,\small{I}/Fe]}   & \multicolumn{1}{r|}{  0.25}& \multicolumn{1}{r}{0.08} & \multicolumn{1}{r|}{0.01} & \multicolumn{1}{r|}{  7} & \multicolumn{1}{r||}{  n/a} & \multicolumn{1}{r|}{  0.12}& \multicolumn{1}{r}{n/a} & \multicolumn{1}{r|}{0.08 } & \multicolumn{1}{r|}{  1} & \multicolumn{1}{r||}{  1.0} & \multicolumn{1}{r|}{  0.28}& \multicolumn{1}{r}{0.07} & \multicolumn{1}{r|}{0.00} & \multicolumn{1}{r|}{ 6} & \multicolumn{1}{r|}{  n/a}  \\
\hline
\multicolumn{1}{|l||}{[O\,\small{I}/Fe]}   & \multicolumn{1}{r|}{ -0.11}& \multicolumn{1}{r}{0.08} & \multicolumn{1}{r|}{0.10} & \multicolumn{1}{r|}{ 26} & \multicolumn{1}{r||}{  8.3} & \multicolumn{1}{r|}{ -0.10}& \multicolumn{1}{r}{0.08} & \multicolumn{1}{r|}{0.10} & \multicolumn{1}{r|}{ 20} & \multicolumn{1}{r||}{  8.0} & \multicolumn{1}{r|}{ -0.15}& \multicolumn{1}{r}{0.07} & \multicolumn{1}{r|}{0.11} & \multicolumn{1}{r|}{ 6} & \multicolumn{1}{r|}{  9.1}  \\
\hline
\multicolumn{1}{|l||}{[Na\,\small{I}/Fe]}  & \multicolumn{1}{r|}{ -0.09}& \multicolumn{1}{r}{0.18} & \multicolumn{1}{r|}{0.07} & \multicolumn{1}{r|}{ 20} & \multicolumn{1}{r||}{  8.6} & \multicolumn{1}{r|}{  -0.18}& \multicolumn{1}{r}{0.12} & \multicolumn{1}{r|}{0.06} & \multicolumn{1}{r|}{ 14} & \multicolumn{1}{r||}{  8.0} & \multicolumn{1}{r|}{  0.13}& \multicolumn{1}{r}{0.05} & \multicolumn{1}{r|}{0.09} & \multicolumn{1}{r|}{ 6} & \multicolumn{1}{r|}{ 10.0}  \\
\hline
\multicolumn{1}{|l||}{[Mg\,\small{I}/Fe]}  & \multicolumn{1}{r|}{  0.04}& \multicolumn{1}{r}{0.07} & \multicolumn{1}{r|}{0.04} & \multicolumn{1}{r|}{ 29} & \multicolumn{1}{r||}{  9.2} & \multicolumn{1}{r|}{  0.03}& \multicolumn{1}{r}{0.07} & \multicolumn{1}{r|}{0.04} & \multicolumn{1}{r|}{ 23} & \multicolumn{1}{r||}{  9.0} & \multicolumn{1}{r|}{  0.05}& \multicolumn{1}{r}{0.07} & \multicolumn{1}{r|}{0.05} & \multicolumn{1}{r|}{ 6} & \multicolumn{1}{r|}{  9.8}  \\
\hline
\multicolumn{1}{|l||}{[Al\,\small{I}/Fe]}  & \multicolumn{1}{r|}{  0.20}& \multicolumn{1}{r}{0.12} & \multicolumn{1}{r|}{0.06} & \multicolumn{1}{r|}{ 28} & \multicolumn{1}{r||}{  9.9} & \multicolumn{1}{r|}{  0.17}& \multicolumn{1}{r}{0.11} & \multicolumn{1}{r|}{0.05} & \multicolumn{1}{r|}{ 22} & \multicolumn{1}{r||}{  9.8} & \multicolumn{1}{r|}{  0.32}& \multicolumn{1}{r}{0.09} & \multicolumn{1}{r|}{0.08} & \multicolumn{1}{r|}{ 6} & \multicolumn{1}{r|}{  10.0}  \\
\hline
\multicolumn{1}{|l||}{[Si\,\small{I}/Fe]}  & \multicolumn{1}{r|}{  0.07}& \multicolumn{1}{r}{0.05} & \multicolumn{1}{r|}{0.03} & \multicolumn{1}{r|}{ 29} & \multicolumn{1}{r||}{  7.6} & \multicolumn{1}{r|}{  0.05}& \multicolumn{1}{r}{0.05} & \multicolumn{1}{r|}{0.02} & \multicolumn{1}{r|}{ 23} & \multicolumn{1}{r||}{  7.6} & \multicolumn{1}{r|}{  0.11}& \multicolumn{1}{r}{0.02} & \multicolumn{1}{r|}{0.06} & \multicolumn{1}{r|}{ 6} & \multicolumn{1}{r|}{  7.7}  \\
\hline
\multicolumn{1}{|l||}{[Ca\,\small{I}/Fe]}  & \multicolumn{1}{r|}{  0.02}& \multicolumn{1}{r}{0.04} & \multicolumn{1}{r|}{0.10} & \multicolumn{1}{r|}{ 29} & \multicolumn{1}{r||}{  6.6} & \multicolumn{1}{r|}{  0.02}& \multicolumn{1}{r}{0.04} & \multicolumn{1}{r|}{0.09} & \multicolumn{1}{r|}{ 23} & \multicolumn{1}{r||}{  6.4} & \multicolumn{1}{r|}{  0.05}& \multicolumn{1}{r}{0.03} & \multicolumn{1}{r|}{0.12} & \multicolumn{1}{r|}{ 6} & \multicolumn{1}{r|}{  7.2}  \\
\hline
\multicolumn{1}{|l||}{[Sc\,\small{II}/Fe]} & \multicolumn{1}{r|}{  0.00}& \multicolumn{1}{r}{0.07} & \multicolumn{1}{r|}{0.03} & \multicolumn{1}{r|}{ 27} & \multicolumn{1}{r||}{  8.7} & \multicolumn{1}{r|}{ -0.02}& \multicolumn{1}{r}{0.07} & \multicolumn{1}{r|}{0.02} & \multicolumn{1}{r|}{ 21} & \multicolumn{1}{r||}{  8.9} & \multicolumn{1}{r|}{  0.04}& \multicolumn{1}{r}{0.03} & \multicolumn{1}{r|}{0.08} & \multicolumn{1}{r|}{ 6} & \multicolumn{1}{r|}{  8.0}  \\
\hline
\multicolumn{1}{|l||}{[Ti\,\small{I}/Fe]}  & \multicolumn{1}{r|}{  0.11}& \multicolumn{1}{r}{0.05} & \multicolumn{1}{r|}{0.11} & \multicolumn{1}{r|}{ 29} & \multicolumn{1}{r||}{  8.2} & \multicolumn{1}{r|}{  0.13}& \multicolumn{1}{r}{0.05} & \multicolumn{1}{r|}{0.10} & \multicolumn{1}{r|}{ 23} & \multicolumn{1}{r||}{  8.2} & \multicolumn{1}{r|}{  0.06}& \multicolumn{1}{r}{0.04} & \multicolumn{1}{r|}{0.14} & \multicolumn{1}{r|}{ 6} & \multicolumn{1}{r|}{ 8.0}  \\
\hline
\multicolumn{1}{|l||}{[Ti\,\small{II}/Fe]} & \multicolumn{1}{r|}{  0.09}& \multicolumn{1}{r}{0.10} & \multicolumn{1}{r|}{0.03} & \multicolumn{1}{r|}{ 27} & \multicolumn{1}{r||}{  8.5} & \multicolumn{1}{r|}{  0.11}& \multicolumn{1}{r}{0.11} & \multicolumn{1}{r|}{0.02} & \multicolumn{1}{r|}{ 21} & \multicolumn{1}{r||}{  8.6} & \multicolumn{1}{r|}{  0.04}& \multicolumn{1}{r}{0.05} & \multicolumn{1}{r|}{0.09} & \multicolumn{1}{r|}{ 6} & \multicolumn{1}{r|}{  7.9}  \\
\hline
\multicolumn{1}{|l||}{[V\,\small{I}/Fe]}   & \multicolumn{1}{r|}{  0.12}& \multicolumn{1}{r}{0.09} & \multicolumn{1}{r|}{0.12} & \multicolumn{1}{r|}{ 20} & \multicolumn{1}{r||}{  8.6} & \multicolumn{1}{r|}{  0.12}& \multicolumn{1}{r}{0.10} & \multicolumn{1}{r|}{0.11} & \multicolumn{1}{r|}{ 14} & \multicolumn{1}{r||}{  8.7} & \multicolumn{1}{r|}{  0.13}& \multicolumn{1}{r}{0.05} & \multicolumn{1}{r|}{0.16} & \multicolumn{1}{r|}{ 6} & \multicolumn{1}{r|}{  8.5}  \\
\hline
\multicolumn{1}{|l||}{[V\,\small{II}/Fe]}  & \multicolumn{1}{r|}{  0.19}& \multicolumn{1}{r}{0.12} & \multicolumn{1}{r|}{0.12} & \multicolumn{1}{r|}{ 17} & \multicolumn{1}{r||}{  8.1} & \multicolumn{1}{r|}{  0.21}& \multicolumn{1}{r}{0.12} & \multicolumn{1}{r|}{0.02} & \multicolumn{1}{r|}{ 11} & \multicolumn{1}{r||}{  7.1} & \multicolumn{1}{r|}{  0.15}& \multicolumn{1}{r}{0.10} & \multicolumn{1}{r|}{0.04} & \multicolumn{1}{r|}{ 6} & \multicolumn{1}{r|}{  9.8}  \\
\hline
\multicolumn{1}{|l||}{[Cr\,\small{I}/Fe]}  & \multicolumn{1}{r|}{  0.06}& \multicolumn{1}{r}{0.06} & \multicolumn{1}{r|}{0.08} & \multicolumn{1}{r|}{ 29} & \multicolumn{1}{r||}{  8.3} & \multicolumn{1}{r|}{  0.06}& \multicolumn{1}{r}{0.07} & \multicolumn{1}{r|}{0.08} & \multicolumn{1}{r|}{ 23} & \multicolumn{1}{r||}{  8.4} & \multicolumn{1}{r|}{ 0.05}& \multicolumn{1}{r}{0.02} & \multicolumn{1}{r|}{0.09} & \multicolumn{1}{r|}{ 6} & \multicolumn{1}{r|}{  7.9}  \\
\hline
\multicolumn{1}{|l||}{[Cr\,\small{II}/Fe]} & \multicolumn{1}{r|}{  -0.06}& \multicolumn{1}{r}{0.07} & \multicolumn{1}{r|}{0.09} & \multicolumn{1}{r|}{ 13} & \multicolumn{1}{r||}{  7.9} & \multicolumn{1}{r|}{  -0.03}& \multicolumn{1}{r}{0.07} & \multicolumn{1}{r|}{0.05} & \multicolumn{1}{r|}{  7} & \multicolumn{1}{r||}{  8.0} & \multicolumn{1}{r|}{  -0.10}& \multicolumn{1}{r}{0.06} & \multicolumn{1}{r|}{0.10} & \multicolumn{1}{r|}{ 6} & \multicolumn{1}{r|}{  7.7}  \\
\hline
\multicolumn{1}{|l||}{[Mn\,\small{I}/Fe]}  & \multicolumn{1}{r|}{ -0.01}& \multicolumn{1}{r}{0.05} & \multicolumn{1}{r|}{0.09} & \multicolumn{1}{r|}{ 18} & \multicolumn{1}{r||}{  7.1} & \multicolumn{1}{r|}{ -0.03}& \multicolumn{1}{r}{0.04} & \multicolumn{1}{r|}{0.08} & \multicolumn{1}{r|}{ 12} & \multicolumn{1}{r||}{  6.7} & \multicolumn{1}{r|}{ 0.04}& \multicolumn{1}{r}{0.02} & \multicolumn{1}{r|}{0.11} & \multicolumn{1}{r|}{ 6} & \multicolumn{1}{r|}{  7.8}  \\
\hline
\multicolumn{1}{|l||}{[Co\,\small{I}/Fe]}  & \multicolumn{1}{r|}{  0.11}& \multicolumn{1}{r}{0.07} & \multicolumn{1}{r|}{0.09} & \multicolumn{1}{r|}{ 29} & \multicolumn{1}{r||}{  8.5} & \multicolumn{1}{r|}{  0.11}& \multicolumn{1}{r}{0.08} & \multicolumn{1}{r|}{0.08} & \multicolumn{1}{r|}{ 23} & \multicolumn{1}{r||}{  8.5} & \multicolumn{1}{r|}{  0.11}& \multicolumn{1}{r}{0.04} & \multicolumn{1}{r|}{0.11} & \multicolumn{1}{r|}{ 6} & \multicolumn{1}{r|}{  8.4}  \\
\hline
\multicolumn{1}{|l||}{[Ni\,\small{I}/Fe]}  & \multicolumn{1}{r|}{ -0.02}& \multicolumn{1}{r}{0.04} & \multicolumn{1}{r|}{0.07} & \multicolumn{1}{r|}{ 29} & \multicolumn{1}{r||}{  7.8} & \multicolumn{1}{r|}{ -0.03}& \multicolumn{1}{r}{0.04} & \multicolumn{1}{r|}{0.07} & \multicolumn{1}{r|}{ 23} & \multicolumn{1}{r||}{  7.8} & \multicolumn{1}{r|}{ 0.00}& \multicolumn{1}{r}{0.01} & \multicolumn{1}{r|}{0.08} & \multicolumn{1}{r|}{ 6} & \multicolumn{1}{r|}{ 7.9}  \\
\hline
\multicolumn{1}{|l||}{[Cu\,\small{I}/Fe]}  & \multicolumn{1}{r|}{ -0.06}& \multicolumn{1}{r}{0.06} & \multicolumn{1}{r|}{0.06} & \multicolumn{1}{r|}{ 22} & \multicolumn{1}{r||}{  7.7} & \multicolumn{1}{r|}{ -0.06}& \multicolumn{1}{r}{0.06} & \multicolumn{1}{r|}{0.06} & \multicolumn{1}{r|}{ 22} & \multicolumn{1}{r||}{  7.7} & \multicolumn{1}{r|}{  ---}& \multicolumn{1}{r}{---} & \multicolumn{1}{r|}{---} & \multicolumn{1}{r|}{ 0} & \multicolumn{1}{r|}{  ---}  \\
\hline
\multicolumn{1}{|l||}{[Zn\,\small{I}/Fe]}  & \multicolumn{1}{r|}{ -0.16}& \multicolumn{1}{r}{0.11} & \multicolumn{1}{r|}{0.05} & \multicolumn{1}{r|}{ 20} & \multicolumn{1}{r||}{  8.0} & \multicolumn{1}{r|}{ -0.21}& \multicolumn{1}{r}{0.09} & \multicolumn{1}{r|}{0.03} & \multicolumn{1}{r|}{ 14} & \multicolumn{1}{r||}{  8.0} & \multicolumn{1}{r|}{ -0.06}& \multicolumn{1}{r}{0.08} & \multicolumn{1}{r|}{0.14} & \multicolumn{1}{r|}{ 6} & \multicolumn{1}{r|}{  8.0}  \\
\hline
\multicolumn{1}{|l||}{[Y\,\small{II}/Fe]}  & \multicolumn{1}{r|}{  0.02}& \multicolumn{1}{r}{0.13} & \multicolumn{1}{r|}{0.05} & \multicolumn{1}{r|}{ 20} & \multicolumn{1}{r||}{  7.0} & \multicolumn{1}{r|}{  0.01}& \multicolumn{1}{r}{0.15} & \multicolumn{1}{r|}{0.03} & \multicolumn{1}{r|}{ 14} & \multicolumn{1}{r||}{  6.1} & \multicolumn{1}{r|}{ 0.05}& \multicolumn{1}{r}{0.07} & \multicolumn{1}{r|}{0.09} & \multicolumn{1}{r|}{ 6} & \multicolumn{1}{r|}{  9.1}  \\
\hline
\multicolumn{1}{|l||}{[Zr\,\small{II}/Fe]} & \multicolumn{1}{r|}{ -0.01}& \multicolumn{1}{r}{0.15} & \multicolumn{1}{r|}{0.02} & \multicolumn{1}{r|}{ 10} & \multicolumn{1}{r||}{  10.0} & \multicolumn{1}{r|}{  0.09}& \multicolumn{1}{r}{0.17} & \multicolumn{1}{r|}{0.01} & \multicolumn{1}{r|}{  4} & \multicolumn{1}{r||}{ 10.0} & \multicolumn{1}{r|}{ -0.08}& \multicolumn{1}{r}{0.09} & \multicolumn{1}{r|}{0.04} & \multicolumn{1}{r|}{ 6} & \multicolumn{1}{r|}{  10.0}  \\
\hline
\multicolumn{1}{|l||}{[Ba\,\small{II}/Fe]} & \multicolumn{1}{r|}{ -0.14}& \multicolumn{1}{r}{0.06} & \multicolumn{1}{r|}{0.09} & \multicolumn{1}{r|}{ 18} & \multicolumn{1}{r||}{  6.0} & \multicolumn{1}{r|}{ -0.17}& \multicolumn{1}{r}{0.03} & \multicolumn{1}{r|}{0.09} & \multicolumn{1}{r|}{ 12} & \multicolumn{1}{r||}{  6.0} & \multicolumn{1}{r|}{ -0.08}& \multicolumn{1}{r}{0.07} & \multicolumn{1}{r|}{0.08} & \multicolumn{1}{r|}{ 6} & \multicolumn{1}{r|}{  6.0}  \\
\hline
\multicolumn{1}{|l||}{[Ce\,\small{II}/Fe]} & \multicolumn{1}{r|}{  0.06}& \multicolumn{1}{r}{0.12} & \multicolumn{1}{r|}{0.04} & \multicolumn{1}{r|}{ 13} & \multicolumn{1}{r||}{  9.2} & \multicolumn{1}{r|}{  0.06}& \multicolumn{1}{r}{0.16} & \multicolumn{1}{r|}{0.02} & \multicolumn{1}{r|}{  7} & \multicolumn{1}{r||}{  9.5} & \multicolumn{1}{r|}{  0.07}& \multicolumn{1}{r}{0.03} & \multicolumn{1}{r|}{0.06} & \multicolumn{1}{r|}{ 6} & \multicolumn{1}{r|}{  8.8}  \\
\hline
\multicolumn{1}{|l||}{[Nd\,\small{II}/Fe]} & \multicolumn{1}{r|}{  0.22}& \multicolumn{1}{r}{0.10} & \multicolumn{1}{r|}{0.03} & \multicolumn{1}{r|}{ 13} & \multicolumn{1}{r||}{  9.4} & \multicolumn{1}{r|}{  0.27}& \multicolumn{1}{r}{0.11} & \multicolumn{1}{r|}{0.03} & \multicolumn{1}{r|}{  7} & \multicolumn{1}{r||}{  9.5} & \multicolumn{1}{r|}{  0.15}& \multicolumn{1}{r}{0.04} & \multicolumn{1}{r|}{0.04} & \multicolumn{1}{r|}{ 6} & \multicolumn{1}{r|}{  9.3}  \\
\hline
\multicolumn{1}{|l||}{[Sm\,\small{II}/Fe]} & \multicolumn{1}{r|}{  0.15}& \multicolumn{1}{r}{0.12} & \multicolumn{1}{r|}{0.04} & \multicolumn{1}{r|}{ 10} & \multicolumn{1}{r||}{  8.8} & \multicolumn{1}{r|}{  0.29}& \multicolumn{1}{r}{0.06} & \multicolumn{1}{r|}{0.02} & \multicolumn{1}{r|}{  4} & \multicolumn{1}{r||}{  9.2} & \multicolumn{1}{r|}{  0.05}& \multicolumn{1}{r}{0.03} & \multicolumn{1}{r|}{0.05} & \multicolumn{1}{r|}{ 6} & \multicolumn{1}{r|}{  8.5}  \\
\hline
\label{tab:752Abs}
\end{tabular}
\end{table}
\setlength{\tabcolsep}{6pt} 
\renewcommand{\arraystretch}{1.0} 

\begin{sidewaystable}
\caption{NGC 752 Atmospheric Errors}
\begin{tabular}{lrrrrrrrrrrrrrrrr} 
\multicolumn{1}{l}{Parameter} & \multicolumn{1}{c}{C\,\small{I}} & \multicolumn{1}{c}{N\,\small{I}} & \multicolumn{1}{c}{O\,\small{I}} & \multicolumn{1}{c}{Na\,\small{I}} & \multicolumn{1}{c}{Mg\,\small{I}} & \multicolumn{1}{c}{Al\,\small{I}} & \multicolumn{1}{c}{Si\,\small{I}} & \multicolumn{1}{c}{Si\,\small{II}} & \multicolumn{1}{c}{Ca\,\small{I}} & \multicolumn{1}{c}{Sc\,\small{I}} & \multicolumn{1}{c}{Sc\,\small{II}} & \multicolumn{1}{c}{Ti\,\small{I}} & \multicolumn{1}{c}{Ti\,\small{II}} & \multicolumn{1}{c}{V\,\small{I}} & \multicolumn{1}{c}{V\,\small{II}} & \multicolumn{1}{c}{Cr\,\small{I}} \\
\hline
\multicolumn{17}{|c|}{Dwarfs}\\
\hline
\multicolumn{1}{|l|}{$T_{\rm{eff}}+100K$} & \multicolumn{1}{r|}{-0.01} & \multicolumn{1}{r|}{-0.08} & \multicolumn{1}{r|}{-0.10} & \multicolumn{1}{r|}{0.04} & \multicolumn{1}{r|}{0.03} & \multicolumn{1}{r|}{0.05} & \multicolumn{1}{r|}{0.01} & \multicolumn{1}{r|}{-0.10} & \multicolumn{1}{r|}{0.07} & \multicolumn{1}{r|}{0.10} & \multicolumn{1}{r|}{0.00} & \multicolumn{1}{r|}{0.09} & \multicolumn{1}{r|}{-0.01} & \multicolumn{1}{r|}{0.10} & \multicolumn{1}{r|}{-0.01} & \multicolumn{1}{r|}{0.07}\\
\multicolumn{1}{|l|}{$T_{\rm{eff}}-100K$} & \multicolumn{1}{r|}{0.01} & \multicolumn{1}{r|}{0.06} & \multicolumn{1}{r|}{0.10} & \multicolumn{1}{r|}{-0.04} & \multicolumn{1}{r|}{-0.03} & \multicolumn{1}{r|}{-0.05} & \multicolumn{1}{r|}{-0.00} & \multicolumn{1}{r|}{0.10} & \multicolumn{1}{r|}{-0.07} & \multicolumn{1}{r|}{-0.10} & \multicolumn{1}{r|}{-0.00} & \multicolumn{1}{r|}{-0.10} & \multicolumn{1}{r|}{0.02} & \multicolumn{1}{r|}{-0.11} & \multicolumn{1}{r|}{0.01} & \multicolumn{1}{r|}{-0.08}\\
\multicolumn{1}{|l|}{log\,g+0.02} & \multicolumn{1}{r|}{0.00} & \multicolumn{1}{r|}{0.01} & \multicolumn{1}{r|}{0.01} & \multicolumn{1}{r|}{-0.00} & \multicolumn{1}{r|}{-0.00} & \multicolumn{1}{r|}{0.00} & \multicolumn{1}{r|}{0.00} & \multicolumn{1}{r|}{0.01} & \multicolumn{1}{r|}{-0.00} & \multicolumn{1}{r|}{0.01} & \multicolumn{1}{r|}{0.01} & \multicolumn{1}{r|}{0.00} & \multicolumn{1}{r|}{0.01} & \multicolumn{1}{r|}{0.00} & \multicolumn{1}{r|}{0.01} & \multicolumn{1}{r|}{-0.00}\\
\multicolumn{1}{|l|}{log\,g-0.02} & \multicolumn{1}{r|}{-0.00} & \multicolumn{1}{r|}{-0.01} & \multicolumn{1}{r|}{-0.01} & \multicolumn{1}{r|}{0.00} & \multicolumn{1}{r|}{0.00} & \multicolumn{1}{r|}{-0.00} & \multicolumn{1}{r|}{-0.00} & \multicolumn{1}{r|}{-0.01} & \multicolumn{1}{r|}{0.00} & \multicolumn{1}{r|}{0.00} & \multicolumn{1}{r|}{-0.01} & \multicolumn{1}{r|}{-0.00} & \multicolumn{1}{r|}{-0.01} & \multicolumn{1}{r|}{-0.00} & \multicolumn{1}{r|}{-0.01} & \multicolumn{1}{r|}{-0.00}\\
\multicolumn{1}{|l|}{$\xi+0.2\rm{km/s}$} & \multicolumn{1}{r|}{-0.00} & \multicolumn{1}{r|}{-0.00} & \multicolumn{1}{r|}{-0.02} & \multicolumn{1}{r|}{-0.01} & \multicolumn{1}{r|}{-0.02} & \multicolumn{1}{r|}{-0.01} & \multicolumn{1}{r|}{-0.02} & \multicolumn{1}{r|}{-0.03} & \multicolumn{1}{r|}{-0.04} & \multicolumn{1}{r|}{0.01} & \multicolumn{1}{r|}{-0.02} & \multicolumn{1}{r|}{-0.01} & \multicolumn{1}{r|}{-0.01} & \multicolumn{1}{r|}{-0.00} & \multicolumn{1}{r|}{-0.00} & \multicolumn{1}{r|}{-0.01}\\
\multicolumn{1}{|l|}{$\xi-0.2\rm{km/s}$} & \multicolumn{1}{r|}{0.00} & \multicolumn{1}{r|}{0.00} & \multicolumn{1}{r|}{0.02} & \multicolumn{1}{r|}{0.02} & \multicolumn{1}{r|}{0.02} & \multicolumn{1}{r|}{0.01} & \multicolumn{1}{r|}{0.02} & \multicolumn{1}{r|}{0.03} & \multicolumn{1}{r|}{0.05} & \multicolumn{1}{r|}{0.01} & \multicolumn{1}{r|}{0.02} & \multicolumn{1}{r|}{0.01} & \multicolumn{1}{r|}{0.01} & \multicolumn{1}{r|}{0.01} & \multicolumn{1}{r|}{0.00} & \multicolumn{1}{r|}{0.02}\\
\hline 
\multicolumn{1}{|l|}{Dwarf Totals} & \multicolumn{1}{r|}{0.02} & \multicolumn{1}{r|}{0.08} & \multicolumn{1}{r|}{0.10} & \multicolumn{1}{r|}{0.04} & \multicolumn{1}{r|}{0.04} & \multicolumn{1}{r|}{0.05} & \multicolumn{1}{r|}{0.02} & \multicolumn{1}{r|}{0.10} & \multicolumn{1}{r|}{0.09} & \multicolumn{1}{r|}{0.10} & \multicolumn{1}{r|}{0.02} & \multicolumn{1}{r|}{0.10} & \multicolumn{1}{r|}{0.02} & \multicolumn{1}{r|}{0.11} & \multicolumn{1}{r|}{0.02} & \multicolumn{1}{r|}{0.08}\\
\hline
\multicolumn{17}{|c|}{Giants}\\
\hline
\multicolumn{1}{|l|}{$T_{\rm{eff}}+100K$} & \multicolumn{1}{r|}{-0.02} & \multicolumn{1}{r|}{-0.00} & \multicolumn{1}{r|}{-0.09} & \multicolumn{1}{r|}{0.05} & \multicolumn{1}{r|}{0.04} & \multicolumn{1}{r|}{0.07} & \multicolumn{1}{r|}{-0.02} & \multicolumn{1}{r|}{-0.11} & \multicolumn{1}{r|}{0.08} & \multicolumn{1}{r|}{0.12} & \multicolumn{1}{r|}{-0.01} & \multicolumn{1}{r|}{0.13} & \multicolumn{1}{r|}{-0.01} & \multicolumn{1}{r|}{0.15} & \multicolumn{1}{r|}{-0.03} & \multicolumn{1}{r|}{0.08}\\
\multicolumn{1}{|l|}{$T_{\rm{eff}}-100K$} & \multicolumn{1}{r|}{0.02} & \multicolumn{1}{r|}{0.00} & \multicolumn{1}{r|}{0.11} & \multicolumn{1}{r|}{-0.05} & \multicolumn{1}{r|}{-0.03} & \multicolumn{1}{r|}{-0.06} & \multicolumn{1}{r|}{0.04} & \multicolumn{1}{r|}{0.10} & \multicolumn{1}{r|}{-0.09} & \multicolumn{1}{r|}{-0.13} & \multicolumn{1}{r|}{0.01} & \multicolumn{1}{r|}{-0.13} & \multicolumn{1}{r|}{0.02} & \multicolumn{1}{r|}{-0.15} & \multicolumn{1}{r|}{0.04} & \multicolumn{1}{r|}{-0.09}\\
\multicolumn{1}{|l|}{log\,g+0.02} & \multicolumn{1}{r|}{0.00} & \multicolumn{1}{r|}{0.00} & \multicolumn{1}{r|}{0.01} & \multicolumn{1}{r|}{-0.00} & \multicolumn{1}{r|}{0.00} & \multicolumn{1}{r|}{0.00} & \multicolumn{1}{r|}{0.00} & \multicolumn{1}{r|}{0.01} & \multicolumn{1}{r|}{-0.00} & \multicolumn{1}{r|}{-0.00} & \multicolumn{1}{r|}{0.01} & \multicolumn{1}{r|}{-0.00} & \multicolumn{1}{r|}{0.01} & \multicolumn{1}{r|}{-0.00} & \multicolumn{1}{r|}{0.01} & \multicolumn{1}{r|}{-0.00}\\
\multicolumn{1}{|l|}{log\,g-0.02} & \multicolumn{1}{r|}{-0.00} & \multicolumn{1}{r|}{-0.00} & \multicolumn{1}{r|}{-0.01} & \multicolumn{1}{r|}{-0.00} & \multicolumn{1}{r|}{-0.00} & \multicolumn{1}{r|}{0.00} & \multicolumn{1}{r|}{-0.00} & \multicolumn{1}{r|}{-0.01} & \multicolumn{1}{r|}{0.00} & \multicolumn{1}{r|}{0.00} & \multicolumn{1}{r|}{-0.01} & \multicolumn{1}{r|}{0.00} & \multicolumn{1}{r|}{-0.01} & \multicolumn{1}{r|}{0.00} & \multicolumn{1}{r|}{-0.01} & \multicolumn{1}{r|}{0.00}\\
\multicolumn{1}{|l|}{$\xi+0.2\rm{km/s}$} & \multicolumn{1}{r|}{-0.00} & \multicolumn{1}{r|}{-0.00} & \multicolumn{1}{r|}{-0.02} & \multicolumn{1}{r|}{-0.02} & \multicolumn{1}{r|}{-0.03} & \multicolumn{1}{r|}{-0.03} & \multicolumn{1}{r|}{-0.04} & \multicolumn{1}{r|}{-0.05} & \multicolumn{1}{r|}{-0.08} & \multicolumn{1}{r|}{-0.01} & \multicolumn{1}{r|}{-0.06} & \multicolumn{1}{r|}{-0.03} & \multicolumn{1}{r|}{-0.07} & \multicolumn{1}{r|}{-0.03} & \multicolumn{1}{r|}{-0.01} & \multicolumn{1}{r|}{-0.03}\\
\multicolumn{1}{|l|}{$\xi-0.2\rm{km/s}$} & \multicolumn{1}{r|}{0.00} & \multicolumn{1}{r|}{0.00} & \multicolumn{1}{r|}{0.02} & \multicolumn{1}{r|}{0.02} & \multicolumn{1}{r|}{0.04} & \multicolumn{1}{r|}{0.04} & \multicolumn{1}{r|}{0.05} & \multicolumn{1}{r|}{0.05} & \multicolumn{1}{r|}{0.08} & \multicolumn{1}{r|}{0.02} & \multicolumn{1}{r|}{0.07} & \multicolumn{1}{r|}{0.03} & \multicolumn{1}{r|}{0.09} & \multicolumn{1}{r|}{0.03} & \multicolumn{1}{r|}{0.02} & \multicolumn{1}{r|}{0.03}\\
\hline 
\multicolumn{1}{|l|}{Giant Totals} & \multicolumn{1}{r|}{0.02} & \multicolumn{1}{r|}{0.00} & \multicolumn{1}{r|}{0.11} & \multicolumn{1}{r|}{0.09} & \multicolumn{1}{r|}{0.05} & \multicolumn{1}{r|}{0.08} & \multicolumn{1}{r|}{0.06} & \multicolumn{1}{r|}{0.12} & \multicolumn{1}{r|}{0.12} & \multicolumn{1}{r|}{0.13} & \multicolumn{1}{r|}{0.08} & \multicolumn{1}{r|}{0.14} & \multicolumn{1}{r|}{0.09} & \multicolumn{1}{r|}{0.16} & \multicolumn{1}{r|}{0.04} & \multicolumn{1}{r|}{0.09}\\
\hline
\multicolumn{1}{l}{Parameter} & \multicolumn{1}{c}{Cr\,\small{II}} & \multicolumn{1}{c}{Mn\,\small{I}} & \multicolumn{1}{c}{Fe\,\small{I}} & \multicolumn{1}{c}{Fe\,\small{II}} & \multicolumn{1}{c}{Co\,\small{I}} & \multicolumn{1}{c}{Ni\,\small{I}} & \multicolumn{1}{c}{Cu\,\small{I}} & \multicolumn{1}{c}{Zn\,\small{I}} & \multicolumn{1}{c}{Y\,\small{II}} & \multicolumn{1}{c}{Zr\,\small{I}} & \multicolumn{1}{c}{Zr\,\small{II}} & \multicolumn{1}{c}{Ba\,\small{II}} & \multicolumn{1}{c}{Ce\,\small{II}} & \multicolumn{1}{c}{Nd\,\small{II}} & \multicolumn{1}{c}{Sm\,\small{II}} \\
\hline
\multicolumn{16}{|c|}{Dwarfs}\\
\hline
\multicolumn{1}{|l|}{$T_{\rm{eff}}+100K$} & \multicolumn{1}{r|}{-0.03} & \multicolumn{1}{r|}{0.08} & \multicolumn{1}{r|}{0.07} & \multicolumn{1}{r|}{-0.04} & \multicolumn{1}{r|}{0.08} & \multicolumn{1}{r|}{0.06} & \multicolumn{1}{r|}{0.06} & \multicolumn{1}{r|}{-0.00} & \multicolumn{1}{r|}{0.00} & \multicolumn{1}{r|}{0.11} & \multicolumn{1}{r|}{0.00} & \multicolumn{1}{r|}{0.02} & \multicolumn{1}{r|}{0.02} & \multicolumn{1}{r|}{0.02} & \multicolumn{1}{r|}{0.02}\\
\multicolumn{1}{|l|}{$T_{\rm{eff}}-100K$} & \multicolumn{1}{r|}{0.03} & \multicolumn{1}{r|}{-0.06} & \multicolumn{1}{r|}{-0.07} & \multicolumn{1}{r|}{0.04} & \multicolumn{1}{r|}{-0.08} & \multicolumn{1}{r|}{-0.06} & \multicolumn{1}{r|}{-0.04} & \multicolumn{1}{r|}{0.01} & \multicolumn{1}{r|}{-0.00} & \multicolumn{1}{r|}{-0.12} & \multicolumn{1}{r|}{-0.00} & \multicolumn{1}{r|}{-0.03} & \multicolumn{1}{r|}{-0.02} & \multicolumn{1}{r|}{-0.02} & \multicolumn{1}{r|}{-0.02}\\
\multicolumn{1}{|l|}{log\,g+0.02} & \multicolumn{1}{r|}{0.01} & \multicolumn{1}{r|}{-0.00} & \multicolumn{1}{r|}{0.00} & \multicolumn{1}{r|}{0.01} & \multicolumn{1}{r|}{0.00} & \multicolumn{1}{r|}{0.00} & \multicolumn{1}{r|}{0.00} & \multicolumn{1}{r|}{0.01} & \multicolumn{1}{r|}{0.01} & \multicolumn{1}{r|}{-0.00} & \multicolumn{1}{r|}{0.01} & \multicolumn{1}{r|}{0.01} & \multicolumn{1}{r|}{0.01} & \multicolumn{1}{r|}{0.01} & \multicolumn{1}{r|}{0.01}\\
\multicolumn{1}{|l|}{log\,g-0.02} & \multicolumn{1}{r|}{-0.01} & \multicolumn{1}{r|}{-0.00} & \multicolumn{1}{r|}{-0.00} & \multicolumn{1}{r|}{-0.01} & \multicolumn{1}{r|}{-0.00} & \multicolumn{1}{r|}{-0.00} & \multicolumn{1}{r|}{-0.00} & \multicolumn{1}{r|}{-0.00} & \multicolumn{1}{r|}{-0.01} & \multicolumn{1}{r|}{0.00} & \multicolumn{1}{r|}{-0.01} & \multicolumn{1}{r|}{-0.01} & \multicolumn{1}{r|}{-0.01} & \multicolumn{1}{r|}{-0.01} & \multicolumn{1}{r|}{-0.01}\\
\multicolumn{1}{|l|}{$\xi+0.2\rm{km/s}$} & \multicolumn{1}{r|}{-0.03} & \multicolumn{1}{r|}{-0.02} & \multicolumn{1}{r|}{-0.03} & \multicolumn{1}{r|}{-0.03} & \multicolumn{1}{r|}{-0.01} & \multicolumn{1}{r|}{-0.03} & \multicolumn{1}{r|}{-0.01} & \multicolumn{1}{r|}{-0.03} & \multicolumn{1}{r|}{-0.02} & \multicolumn{1}{r|}{-0.00} & \multicolumn{1}{r|}{-0.00} & \multicolumn{1}{r|}{-0.07} & \multicolumn{1}{r|}{-0.01} & \multicolumn{1}{r|}{-0.01} & \multicolumn{1}{r|}{-0.00}\\
\multicolumn{1}{|l|}{$\xi-0.2\rm{km/s}$} & \multicolumn{1}{r|}{0.04} & \multicolumn{1}{r|}{0.02} & \multicolumn{1}{r|}{0.03} & \multicolumn{1}{r|}{0.03} & \multicolumn{1}{r|}{0.01} & \multicolumn{1}{r|}{0.03} & \multicolumn{1}{r|}{0.01} & \multicolumn{1}{r|}{0.03} & \multicolumn{1}{r|}{0.03} & \multicolumn{1}{r|}{0.00} & \multicolumn{1}{r|}{0.01} & \multicolumn{1}{r|}{0.08} & \multicolumn{1}{r|}{0.01} & \multicolumn{1}{r|}{0.01} & \multicolumn{1}{r|}{0.01}\\
\hline 
\multicolumn{1}{|l|}{Dwarf Totals} & \multicolumn{1}{r|}{0.05} & \multicolumn{1}{r|}{0.08} & \multicolumn{1}{r|}{0.08} & \multicolumn{1}{r|}{0.05} & \multicolumn{1}{r|}{0.08} & \multicolumn{1}{r|}{0.07} & \multicolumn{1}{r|}{0.06} & \multicolumn{1}{r|}{0.03} & \multicolumn{1}{r|}{0.03} & \multicolumn{1}{r|}{0.12} & \multicolumn{1}{r|}{0.01} & \multicolumn{1}{r|}{0.09} & \multicolumn{1}{r|}{0.02} & \multicolumn{1}{r|}{0.03} & \multicolumn{1}{r|}{0.02}\\
\hline
\multicolumn{16}{|c|}{Giants}\\
\hline
\multicolumn{1}{|l|}{$T_{\rm{eff}}+100K$} & \multicolumn{1}{r|}{-0.06} & \multicolumn{1}{r|}{0.10} & \multicolumn{1}{r|}{0.08} & \multicolumn{1}{r|}{-0.07} & \multicolumn{1}{r|}{0.09} & \multicolumn{1}{r|}{0.05} & \multicolumn{1}{r|}{0.05} & \multicolumn{1}{r|}{-0.04} & \multicolumn{1}{r|}{0.00} & \multicolumn{1}{r|}{0.09} & \multicolumn{1}{r|}{-0.01} & \multicolumn{1}{r|}{0.03} & \multicolumn{1}{r|}{0.01} & \multicolumn{1}{r|}{0.02} & \multicolumn{1}{r|}{0.02}\\
\multicolumn{1}{|l|}{$T_{\rm{eff}}-100K$} & \multicolumn{1}{r|}{0.05} & \multicolumn{1}{r|}{-0.09} & \multicolumn{1}{r|}{-0.07} & \multicolumn{1}{r|}{0.08} & \multicolumn{1}{r|}{-0.08} & \multicolumn{1}{r|}{-0.03} & \multicolumn{1}{r|}{-0.03} & \multicolumn{1}{r|}{0.06} & \multicolumn{1}{r|}{0.01} & \multicolumn{1}{r|}{-0.09} & \multicolumn{1}{r|}{0.02} & \multicolumn{1}{r|}{-0.02} & \multicolumn{1}{r|}{-0.01} & \multicolumn{1}{r|}{-0.01} & \multicolumn{1}{r|}{-0.02}\\
\multicolumn{1}{|l|}{log\,g+0.02} & \multicolumn{1}{r|}{0.00} & \multicolumn{1}{r|}{-0.00} & \multicolumn{1}{r|}{0.00} & \multicolumn{1}{r|}{0.01} & \multicolumn{1}{r|}{0.00} & \multicolumn{1}{r|}{0.00} & \multicolumn{1}{r|}{0.00} & \multicolumn{1}{r|}{0.01} & \multicolumn{1}{r|}{0.01} & \multicolumn{1}{r|}{-0.00} & \multicolumn{1}{r|}{0.01} & \multicolumn{1}{r|}{0.01} & \multicolumn{1}{r|}{0.01} & \multicolumn{1}{r|}{0.01} & \multicolumn{1}{r|}{0.01}\\
\multicolumn{1}{|l|}{log\,g-0.02} & \multicolumn{1}{r|}{-0.01} & \multicolumn{1}{r|}{0.00} & \multicolumn{1}{r|}{-0.00} & \multicolumn{1}{r|}{-0.01} & \multicolumn{1}{r|}{-0.00} & \multicolumn{1}{r|}{-0.00} & \multicolumn{1}{r|}{-0.00} & \multicolumn{1}{r|}{-0.01} & \multicolumn{1}{r|}{-0.01} & \multicolumn{1}{r|}{0.00} & \multicolumn{1}{r|}{-0.01} & \multicolumn{1}{r|}{-0.01} & \multicolumn{1}{r|}{-0.01} & \multicolumn{1}{r|}{-0.01} & \multicolumn{1}{r|}{-0.01}\\
\multicolumn{1}{|l|}{$\xi+0.2\rm{km/s}$} & \multicolumn{1}{r|}{-0.07} & \multicolumn{1}{r|}{-0.04} & \multicolumn{1}{r|}{-0.07} & \multicolumn{1}{r|}{-0.05} & \multicolumn{1}{r|}{-0.04} & \multicolumn{1}{r|}{-0.06} & \multicolumn{1}{r|}{-0.10} & \multicolumn{1}{r|}{-0.11} & \multicolumn{1}{r|}{-0.07} & \multicolumn{1}{r|}{-0.01} & \multicolumn{1}{r|}{-0.02} & \multicolumn{1}{r|}{-0.07} & \multicolumn{1}{r|}{-0.04} & \multicolumn{1}{r|}{-0.03} & \multicolumn{1}{r|}{-0.03}\\
\multicolumn{1}{|l|}{$\xi-0.2\rm{km/s}$} & \multicolumn{1}{r|}{0.08} & \multicolumn{1}{r|}{0.04} & \multicolumn{1}{r|}{0.08} & \multicolumn{1}{r|}{0.05} & \multicolumn{1}{r|}{0.05} & \multicolumn{1}{r|}{0.07} & \multicolumn{1}{r|}{0.13} & \multicolumn{1}{r|}{0.13} & \multicolumn{1}{r|}{0.09} & \multicolumn{1}{r|}{0.02} & \multicolumn{1}{r|}{0.03} & \multicolumn{1}{r|}{0.07} & \multicolumn{1}{r|}{0.06} & \multicolumn{1}{r|}{0.04} & \multicolumn{1}{r|}{0.04}\\
\hline 
\multicolumn{1}{|l|}{Giant Totals} & \multicolumn{1}{r|}{0.10} & \multicolumn{1}{r|}{0.11} & \multicolumn{1}{r|}{0.11} & \multicolumn{1}{r|}{0.10} & \multicolumn{1}{r|}{0.10} & \multicolumn{1}{r|}{0.08} & \multicolumn{1}{r|}{0.14} & \multicolumn{1}{r|}{0.14} & \multicolumn{1}{r|}{0.09} & \multicolumn{1}{r|}{0.09} & \multicolumn{1}{r|}{0.04} & \multicolumn{1}{r|}{0.08} & \multicolumn{1}{r|}{0.06} & \multicolumn{1}{r|}{0.04} & \multicolumn{1}{r|}{0.05}\\
\hline
\hline
\label{tab:atmErrs}
\end{tabular}
\end{sidewaystable}

\setlength{\tabcolsep}{4pt} 
\renewcommand{\arraystretch}{1.0} 
\begin{table}[H]
\centering
\begin{tabular}{lccc}
\multicolumn{1}{l}{} & \multicolumn{1}{c}{} & \multicolumn{1}{c}{Carrera \&} & \multicolumn{1}{c}{Reddy, Giridhar,}\vspace{-1mm}\\
\multicolumn{1}{l}{Element/Ion} & \multicolumn{1}{c}{This work} & \multicolumn{1}{c}{Pancino} & \multicolumn{1}{c}{\& Lambert}\\
\multicolumn{1}{l}{[Fe/H]} & \multicolumn{1}{r}{$-0.01\pm0.06$} & \multicolumn{1}{r}{$+0.08\pm0.04$} & \multicolumn{1}{r}{$-0.04\pm0.03$}\\
\multicolumn{1}{l}{[V/H]} & \multicolumn{1}{r}{$+0.08\pm0.03$} & \multicolumn{1}{r}{$+0.09\pm0.09$} & \multicolumn{1}{r}{$-0.01\pm0.05$}\\
\multicolumn{1}{l}{[Cr/H]} & \multicolumn{1}{r}{$+0.01\pm0.02$} & \multicolumn{1}{r}{$+0.08\pm0.01$} & \multicolumn{1}{r}{$-0.06\pm0.04$}\\
\multicolumn{1}{l}{[Co/H]} & \multicolumn{1}{r}{$+0.09\pm0.02$} & \multicolumn{1}{r}{$+0.09\pm0.04$} & \multicolumn{1}{r}{$-0.05\pm0.02$}\\
\multicolumn{1}{l}{[Ni/H]} & \multicolumn{1}{r}{$-0.01\pm0.01$} & \multicolumn{1}{r}{$+0.07\pm0.03$} & \multicolumn{1}{r}{$-0.05\pm0.03$}\\
\hline
\label{tab:FeGroupComps}
\end{tabular}
\caption{Comparison of Fe-group abundance measurements between \citet{Carrera:11} and \citet{Reddy:12} and this work.}
\end{table}
\setlength{\tabcolsep}{6pt} 
\renewcommand{\arraystretch}{1.0} 

\begin{figure}[H]
    \centering
    \includegraphics[width=0.90\textwidth]{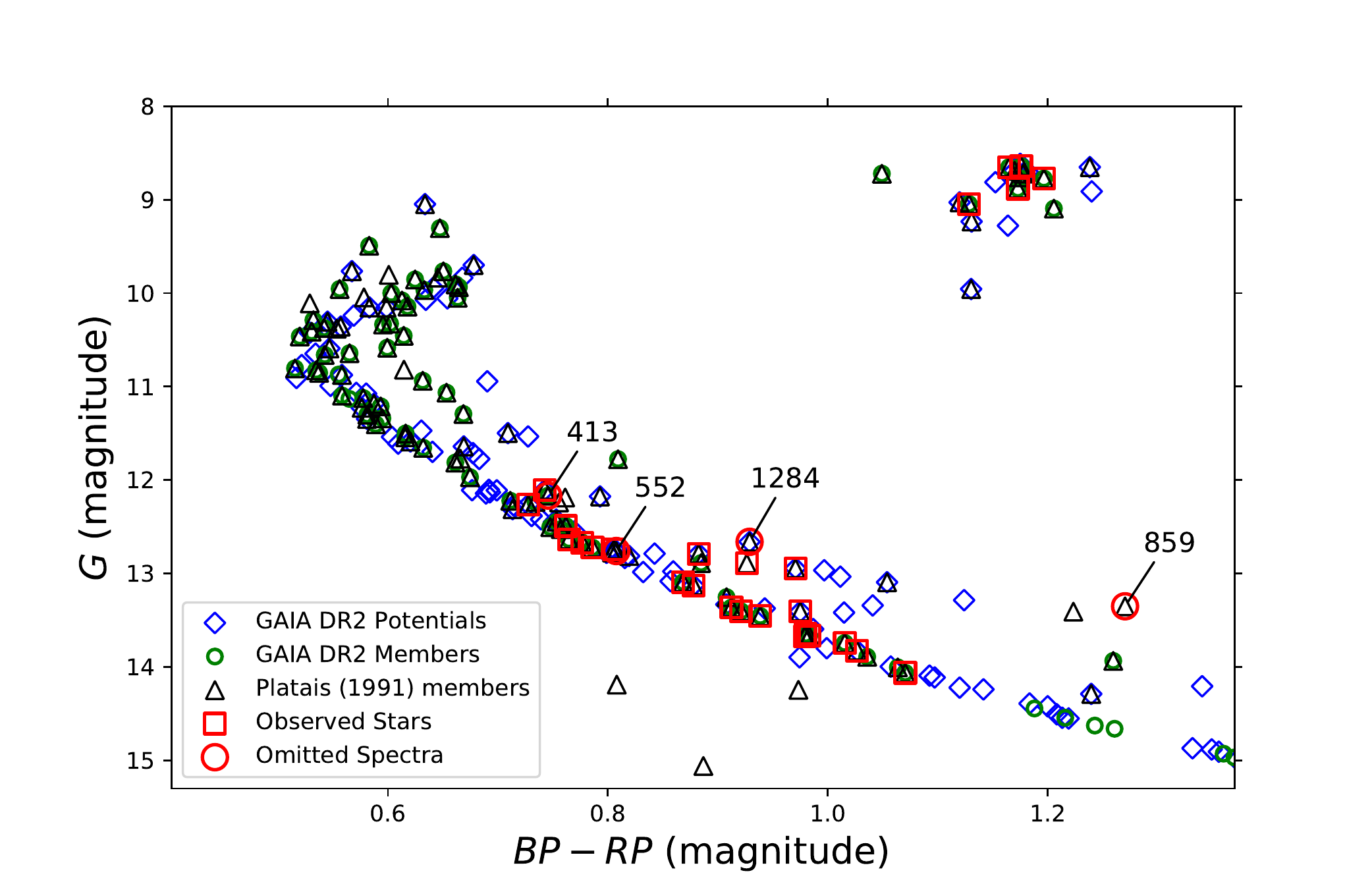}
    \caption{The color-color digagram for NGC 752. Photometry from GAIA's Data Release 2 (DR2) \citep{GAIADR2:18}. Target stars for this work are designated by red squares; stars designated as cluster members by \citet{Platais:91} are black triangles, potential and probable additional members, as determined from the GAIA DR2 dataset (Lum 2019, in prep), are designated by blue diamonds and green circles, respectively. We also note the locations of the four Platais member spectra which we do not include in our analysis. Stars which are listed as both members in Platais, and which we consider (potential) members from the GAIA photometry/astrometry, have both the black triangle, and the appropriate GAIA marking.}
 \label{fig:752HR}
\end{figure}

\begin{figure}[H]
    \centering
    \includegraphics[width=0.80\textwidth]{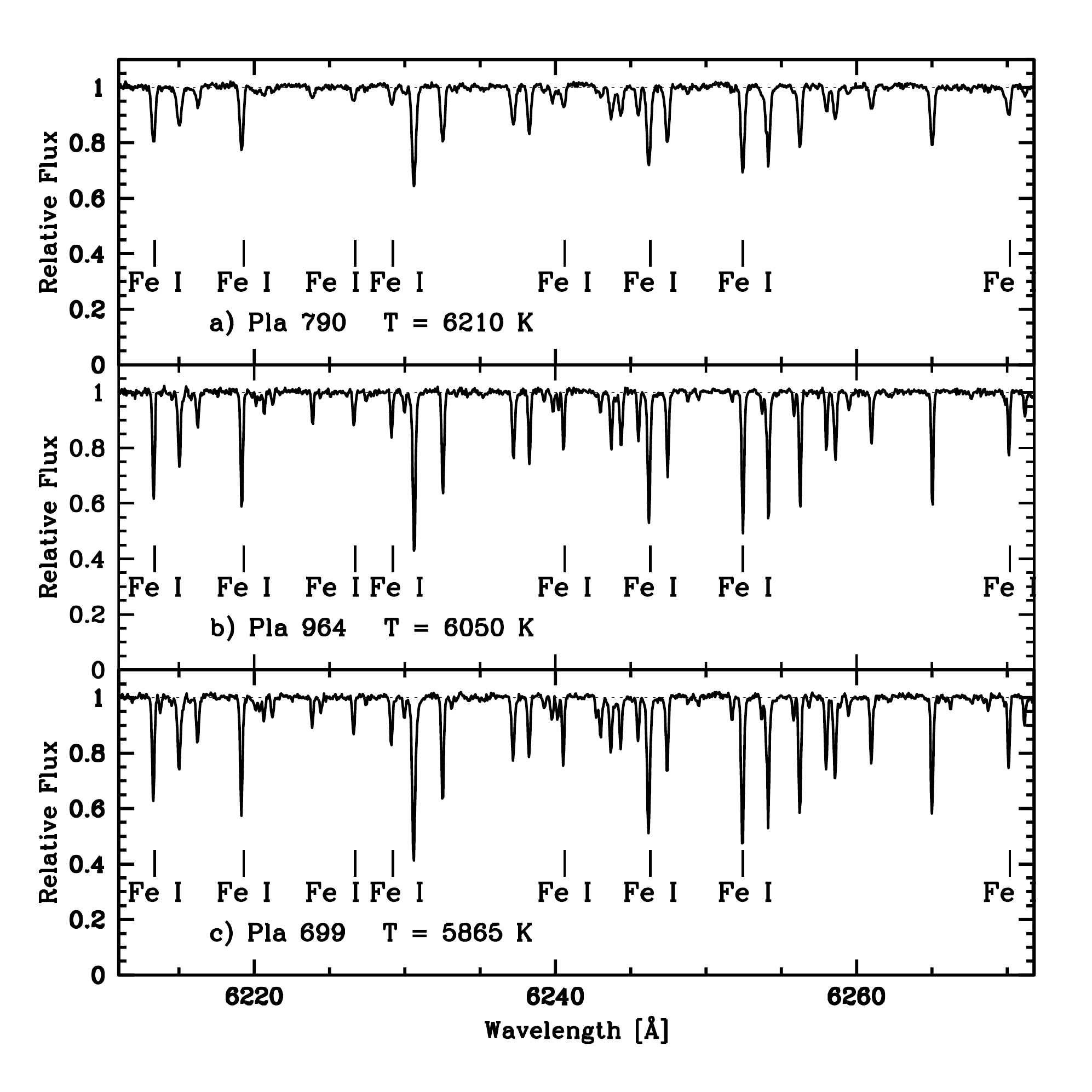}
    \caption{A example of three spectra spanning the $T_{\rm{eff}}$ range for our sample showing the 6200-6270~\AA~range used for atmospheric parameter determination. Strong Fe lines are noted.}    \label{fig:FeSpecSN}
\end{figure}

\begin{figure}[H]
    \centering
    \includegraphics[width=0.80\textwidth]{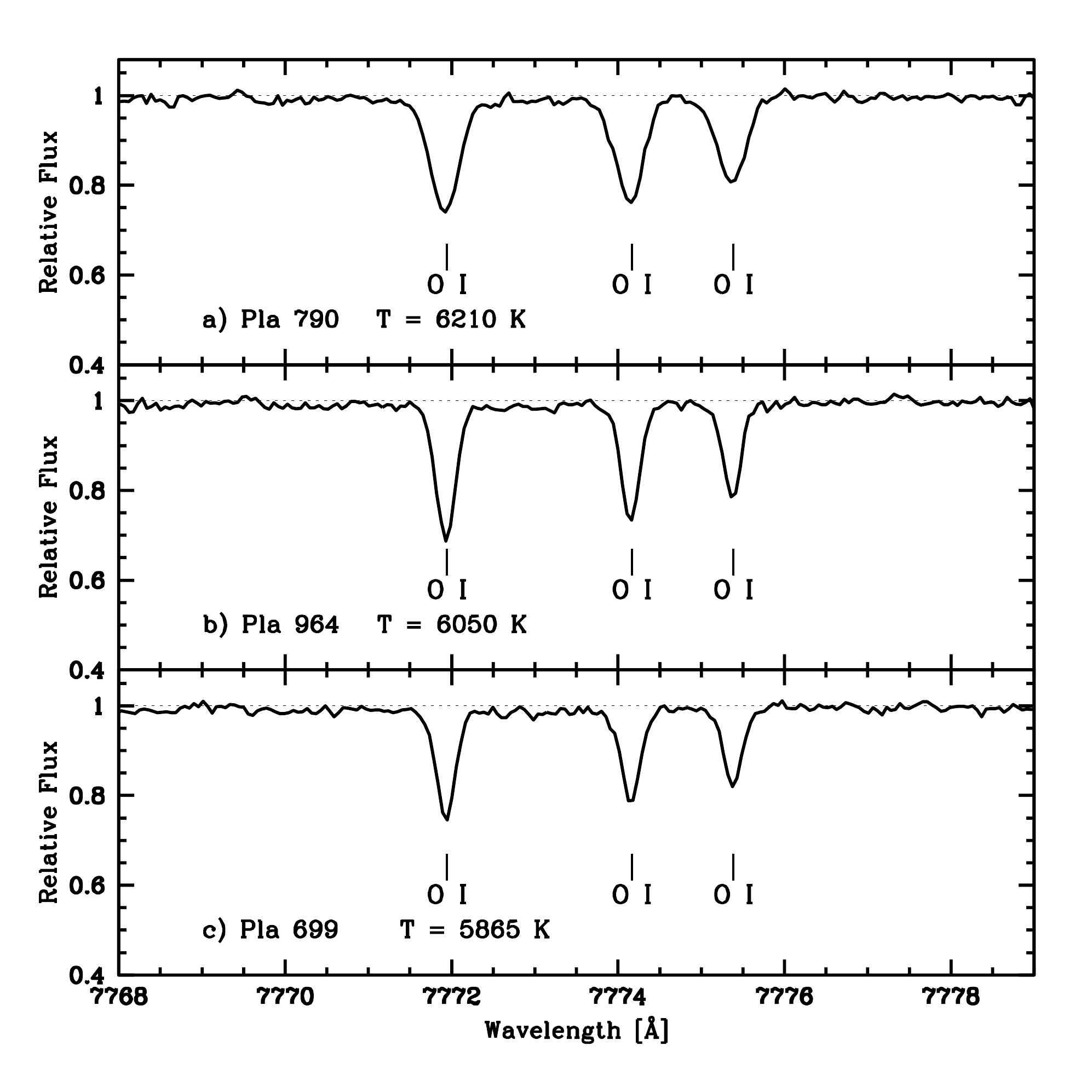}
    \caption{A sample of three spectra spanning the $T_{\rm{eff}}$ range for our sample showing the ``Oxygen triplet'' region.}   \label{fig:OSpecSN}
\end{figure}

\begin{figure}[H]
    \centering
	\includegraphics[width=0.80\textwidth]{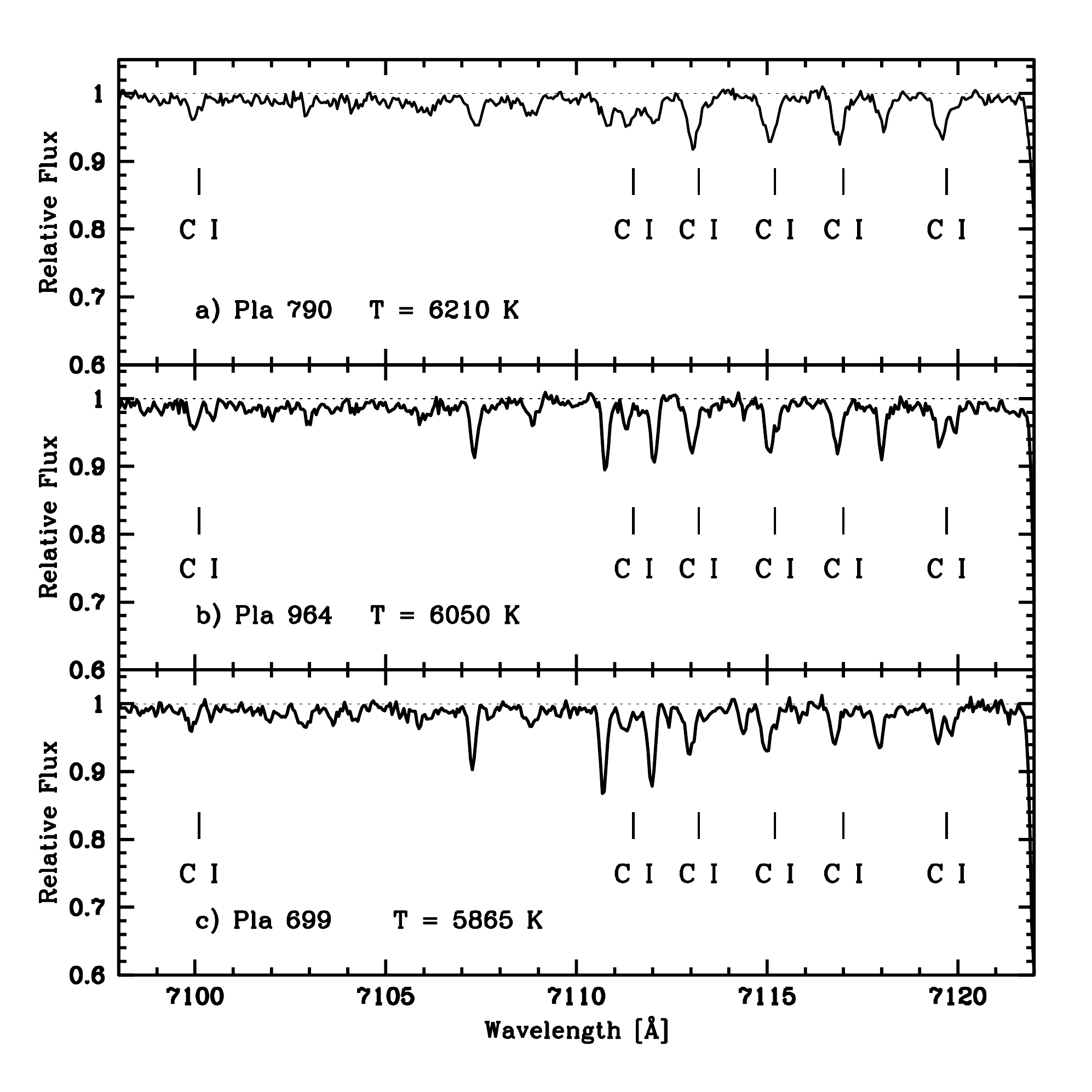}
    \caption{A sample of three spectra spanning the $T_{\rm{eff}}$ range for our sample showing the 7110-7120\AA region used for C\,\small{I} synthesis.}
 \label{fig:CSpecSN}
\end{figure}

\begin{figure}[H]
    \centering
    \includegraphics[width=0.80\textwidth, angle=0]{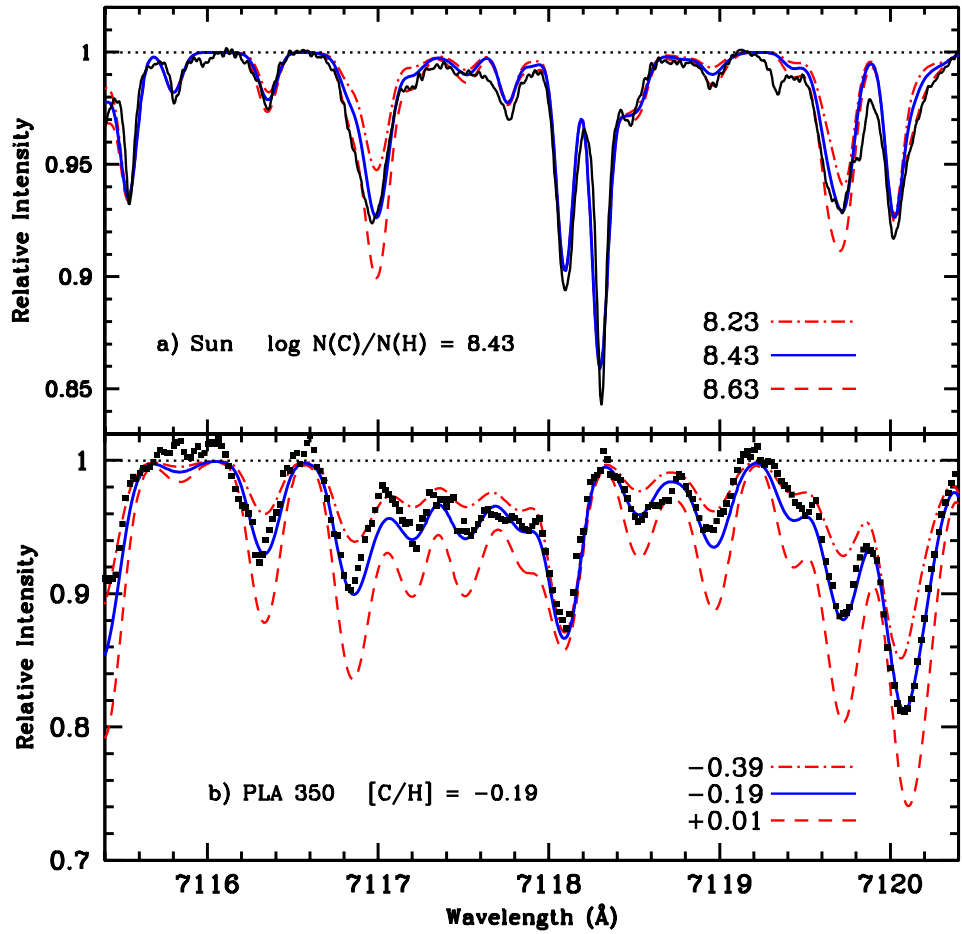}\vspace{-30mm}
    \caption{Comparison between the synthesis of a portion of the region used to measure C\,\small{I}. Note the CN molecular features between 7117 and 7118\,\AA~are visible in the giant, PLA-350 (bottom) spectrum, but not apparent in the dwarf (solar) spectrum.}
\label{fig:Synth}
\end{figure}

\begin{figure}[H]
    \centering
    \includegraphics[width=0.90\textwidth]{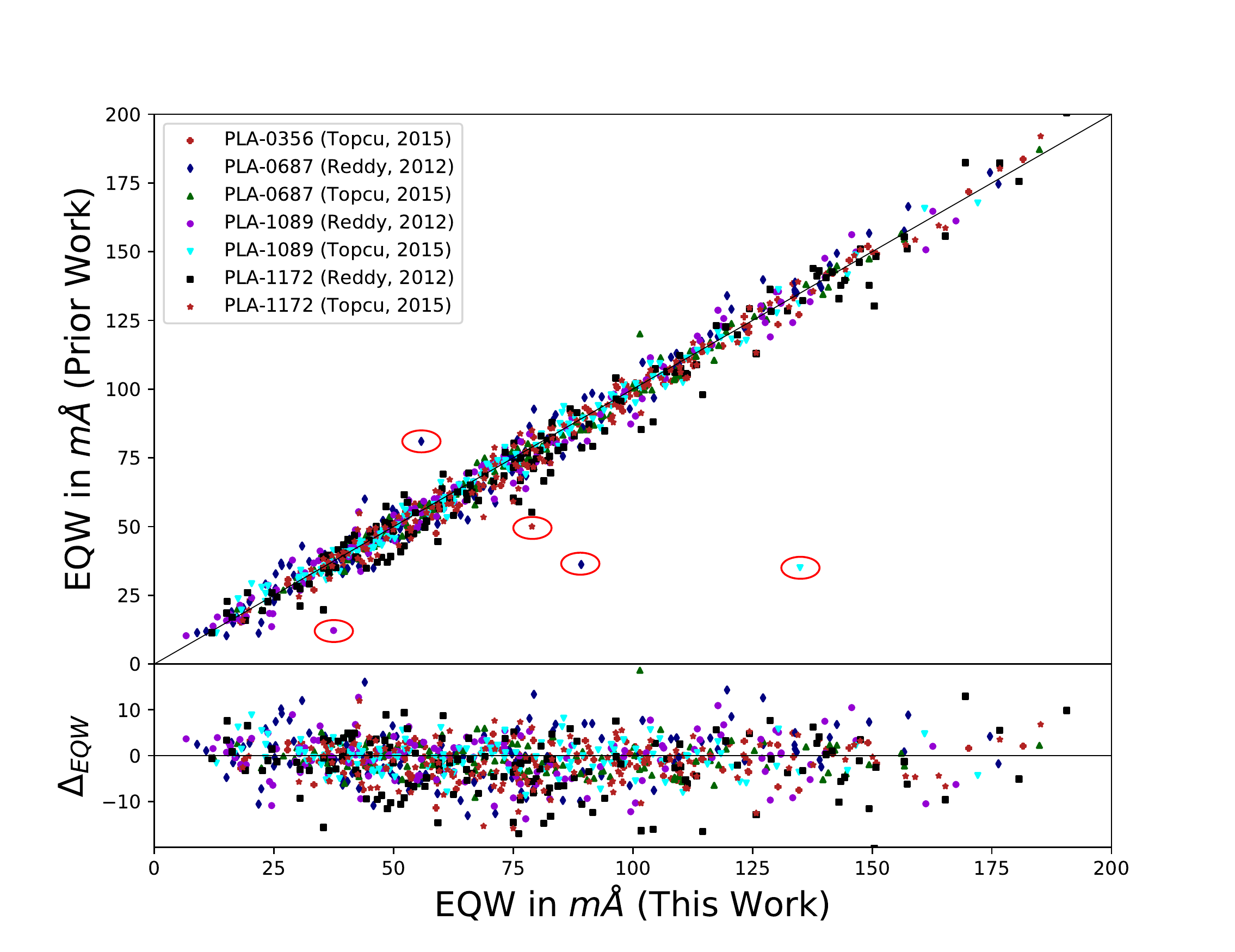}
    \caption{Comparison of 1211 Equivalent Width (EQW) measurements between \citet{Reddy:12}, \citet{Topcu:15}, and this work for common giant stars. The circled outliers are discussed in the text.}
 \label{fig:RelEQWs}
\end{figure}

\begin{figure}[H]
    \centering
    \subfloat[Dwarf Atmosphere Parameters]{%
        \includegraphics[width=0.49\textwidth]{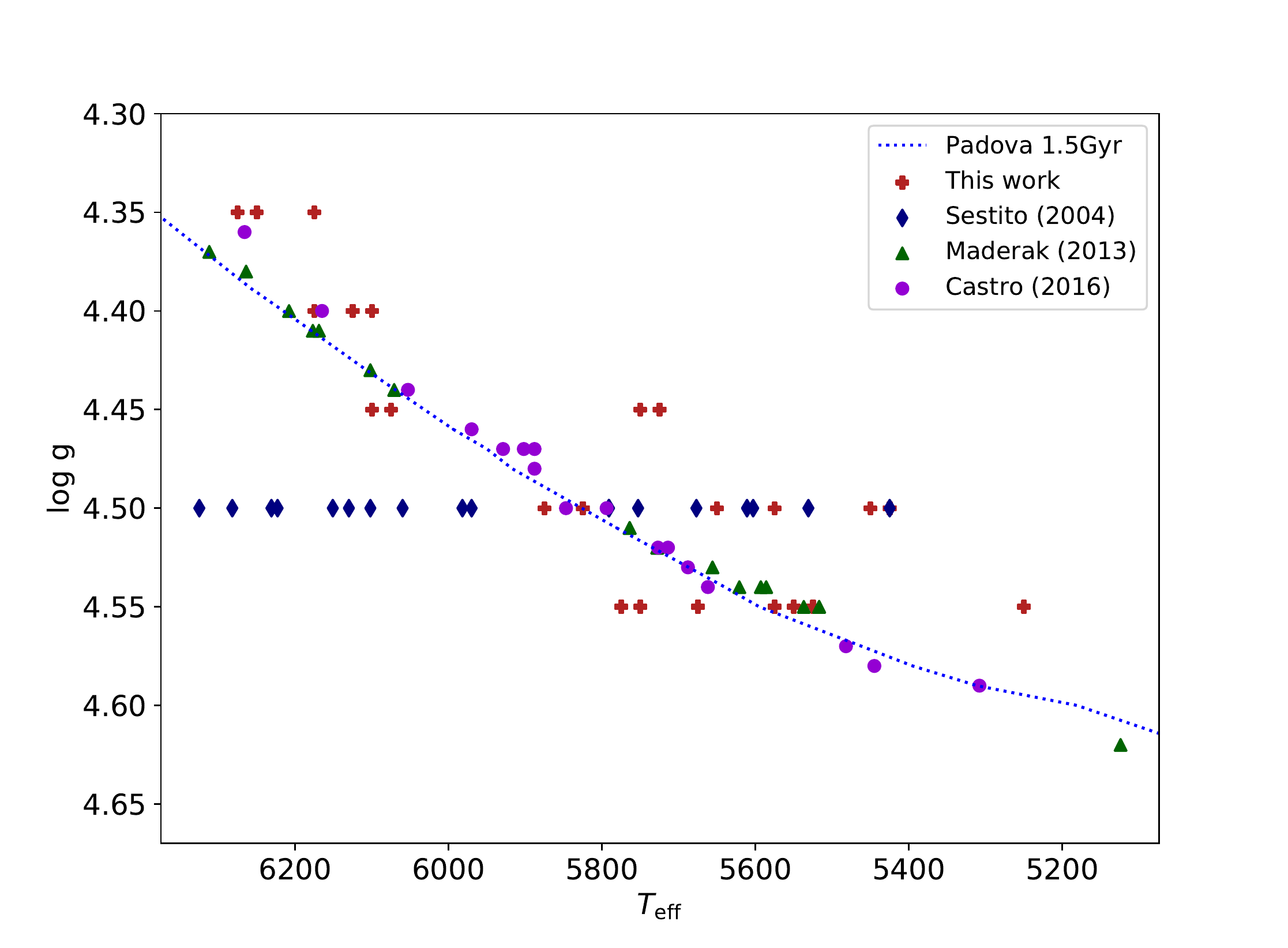}
        \label{fig:ParmComps:Dwarfs}}
    \subfloat[Giant Atmosphere Parameters]{%
        \includegraphics[width=0.49\textwidth]{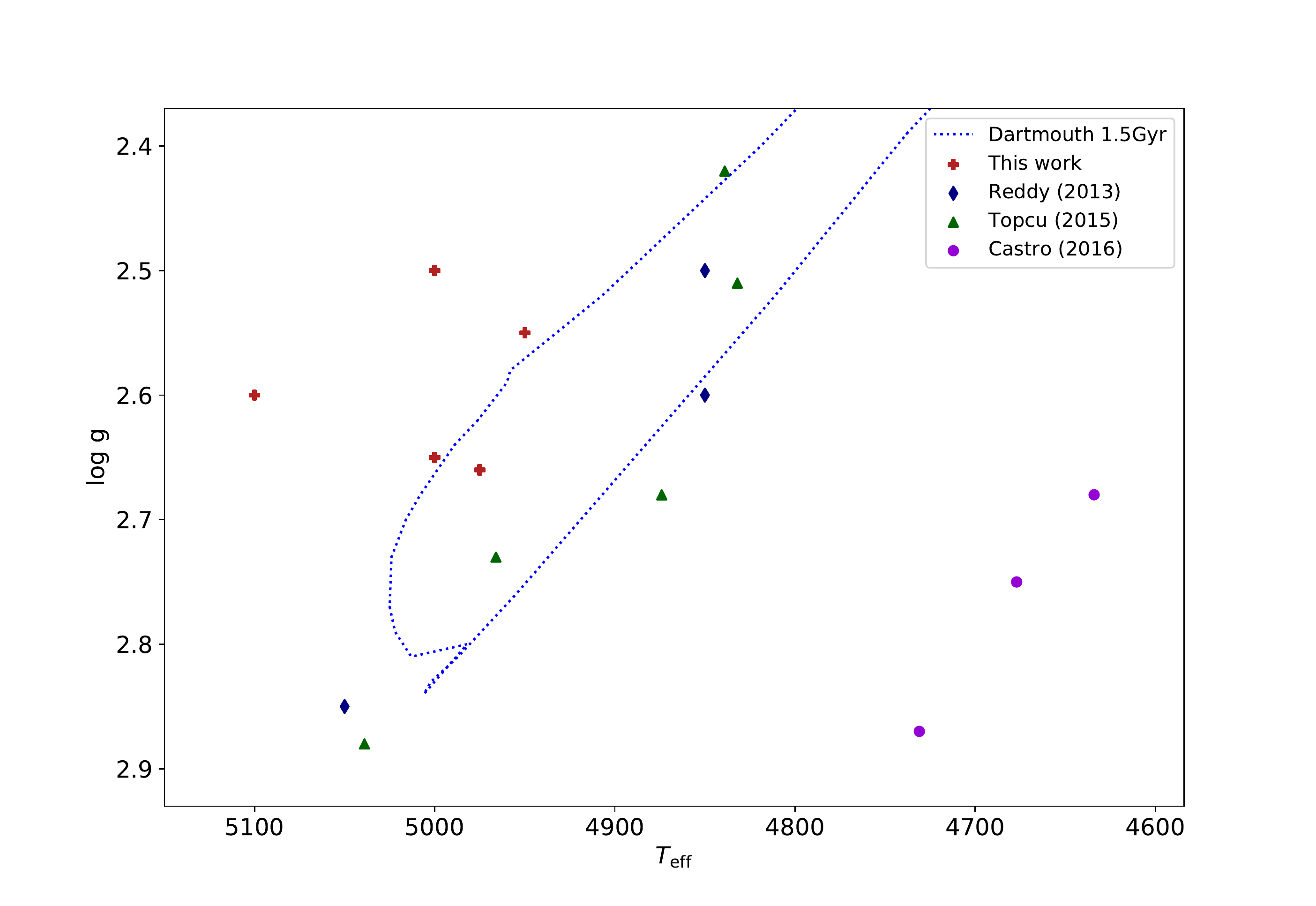}
        \label{fig:ParmComps:Giants}}
    \caption{$T_{eff}$-log\,g comparisons between this work and \citet{Sestito:04}, \citet{Reddy:13}, \citet{Maderak:13}, \citet{Topcu:15}, and \citet{Castro:16}. The 1.5 Gyr isochrones are shown as the blue dotted line in both figures. Note that \citeauthor{Sestito:04} assumed a constant log\,g of 4.50 for all dwarfs.}
\label{fig:ParmComps}
\end{figure}

\begin{figure}[H]
    \centering
    \subfloat[PLA-889 (Main-sequence)]{%
        \includegraphics[width=0.49\textwidth]{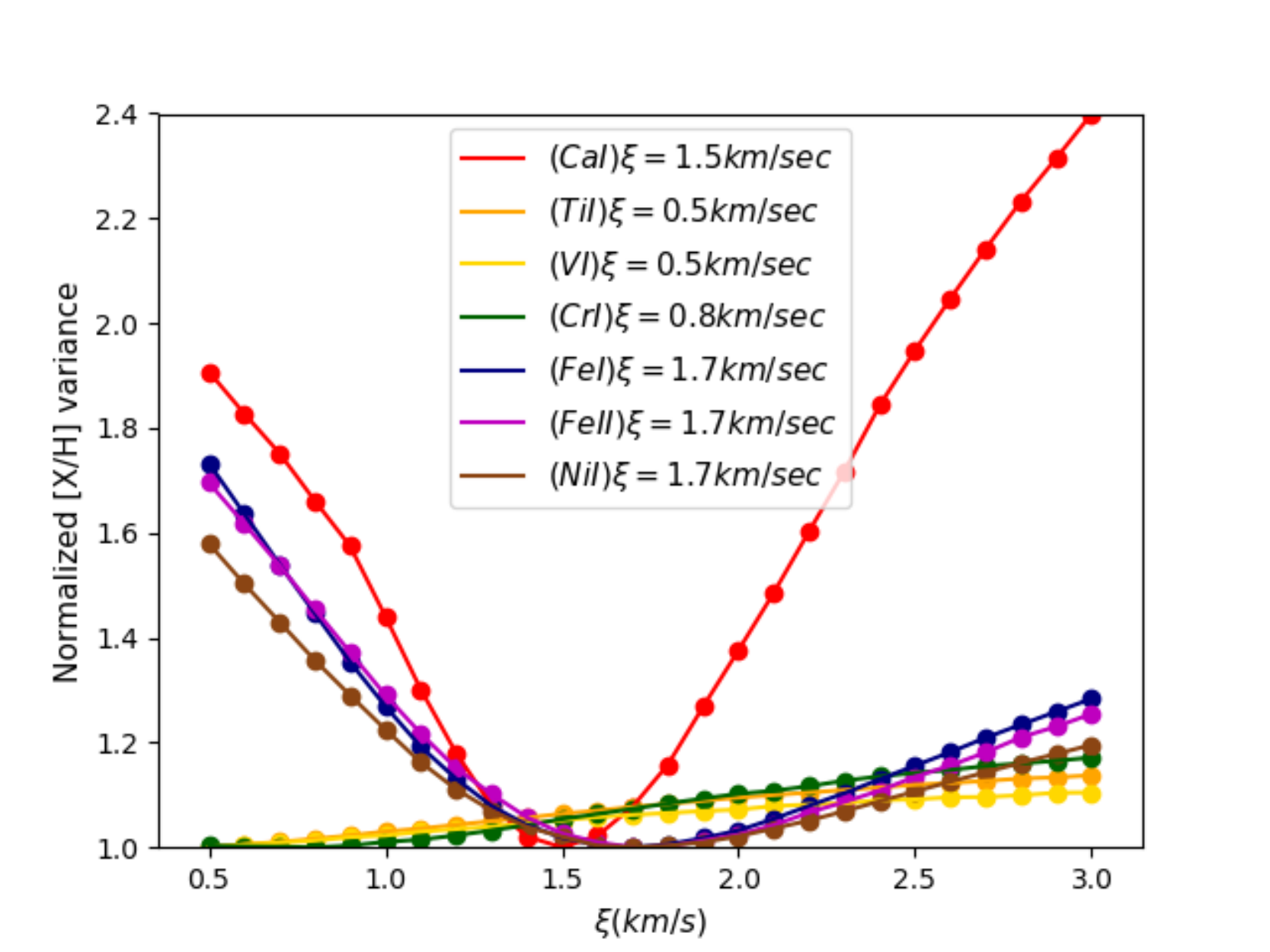}
        \label{fig:vTurbDetA}}
    \subfloat[PLA-1089 (Red giant clump)]{%
        \includegraphics[width=0.49\textwidth]{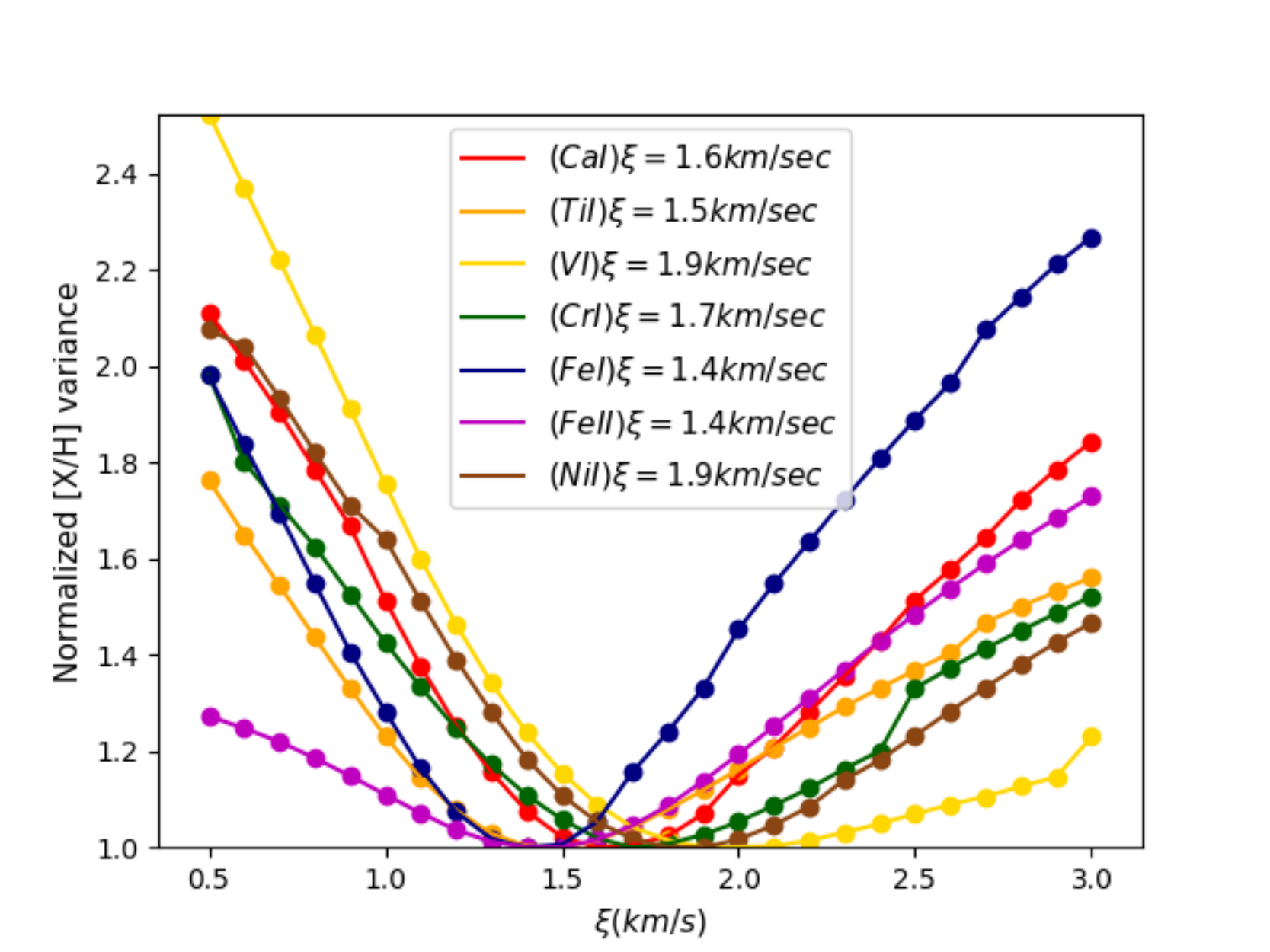}
        \label{fig:vturbDetB}}
    \caption{Normalized line-to-line abundance standard deviation ($\delta$) for various values of micro-turbulent velocity ($\xi$). The [Ca/H] variance best predicts the $\xi$ value corresponding to the minimum variance of the $\delta$-s for the suite of listed elements.}
 \label{fig:vturbDet}
\end{figure}

\begin{figure}[H]
    \centering
    \subfloat{%
        \includegraphics[width=0.49\textwidth]{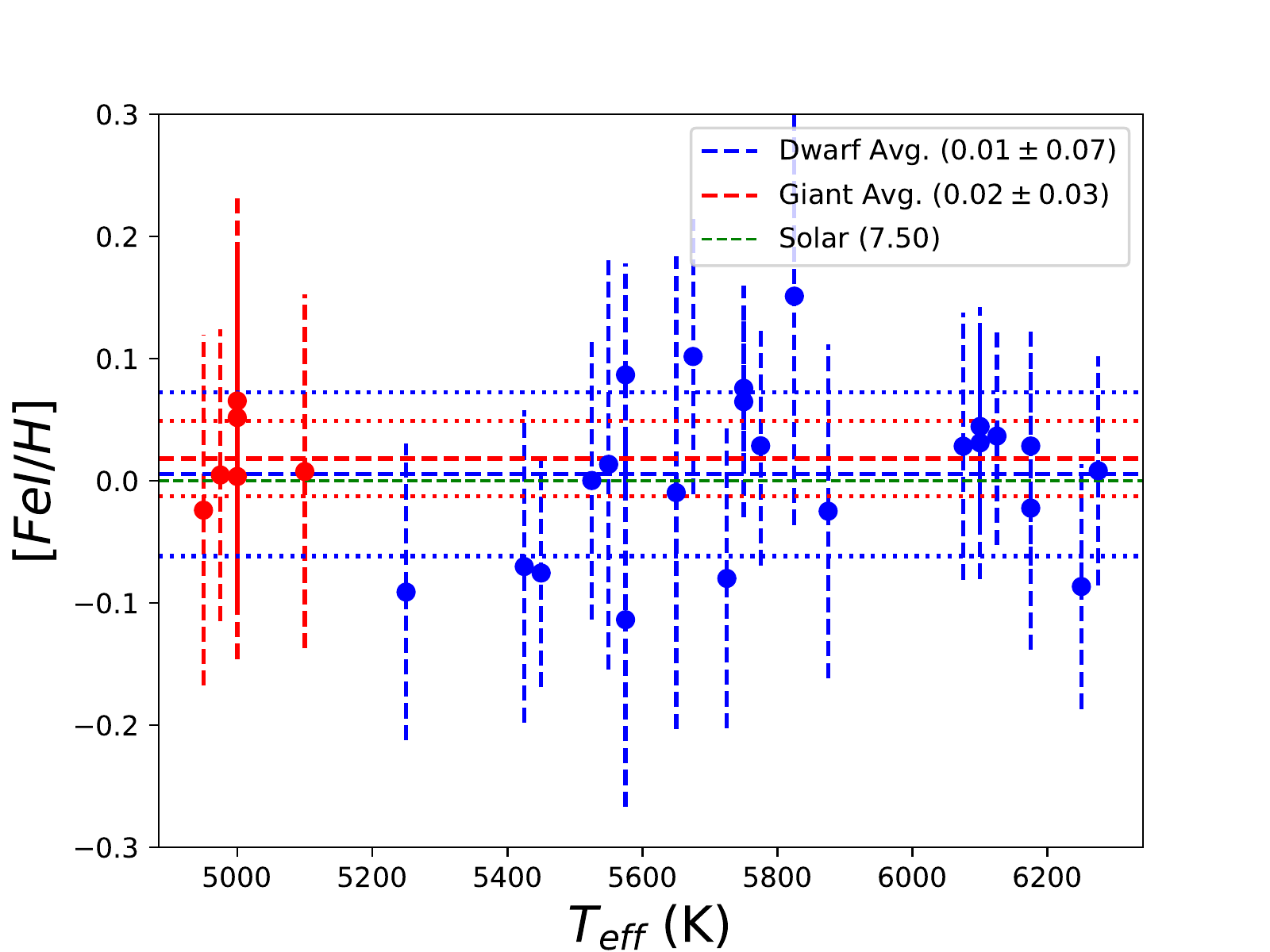}
        \label{fig:FeAbs:FeI}}
    \subfloat{%
        \includegraphics[width=0.49\textwidth]{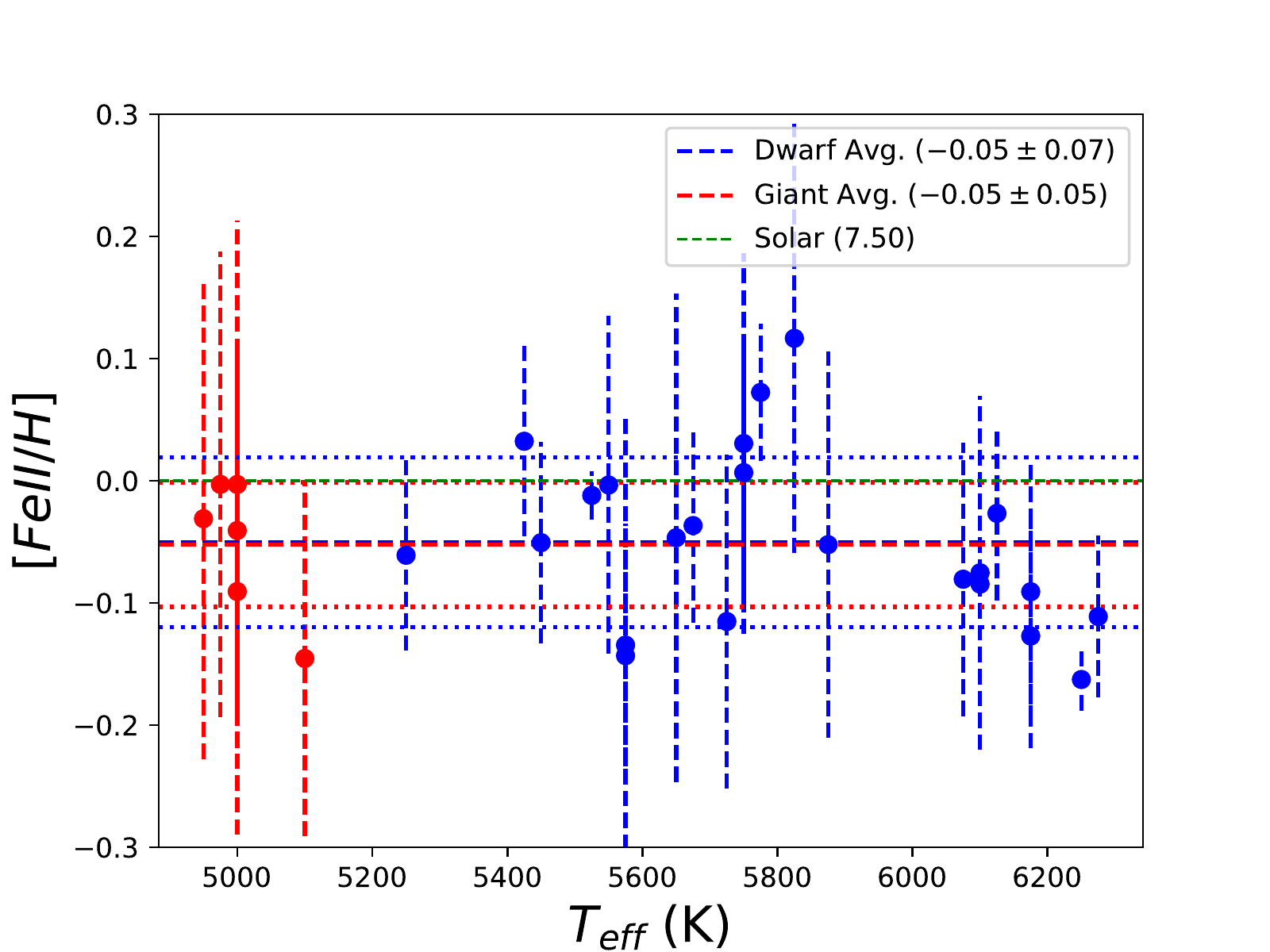}
        \label{fig:FeAbs:FeII}}
    \\
    \subfloat{%
        \includegraphics[width=0.49\textwidth]{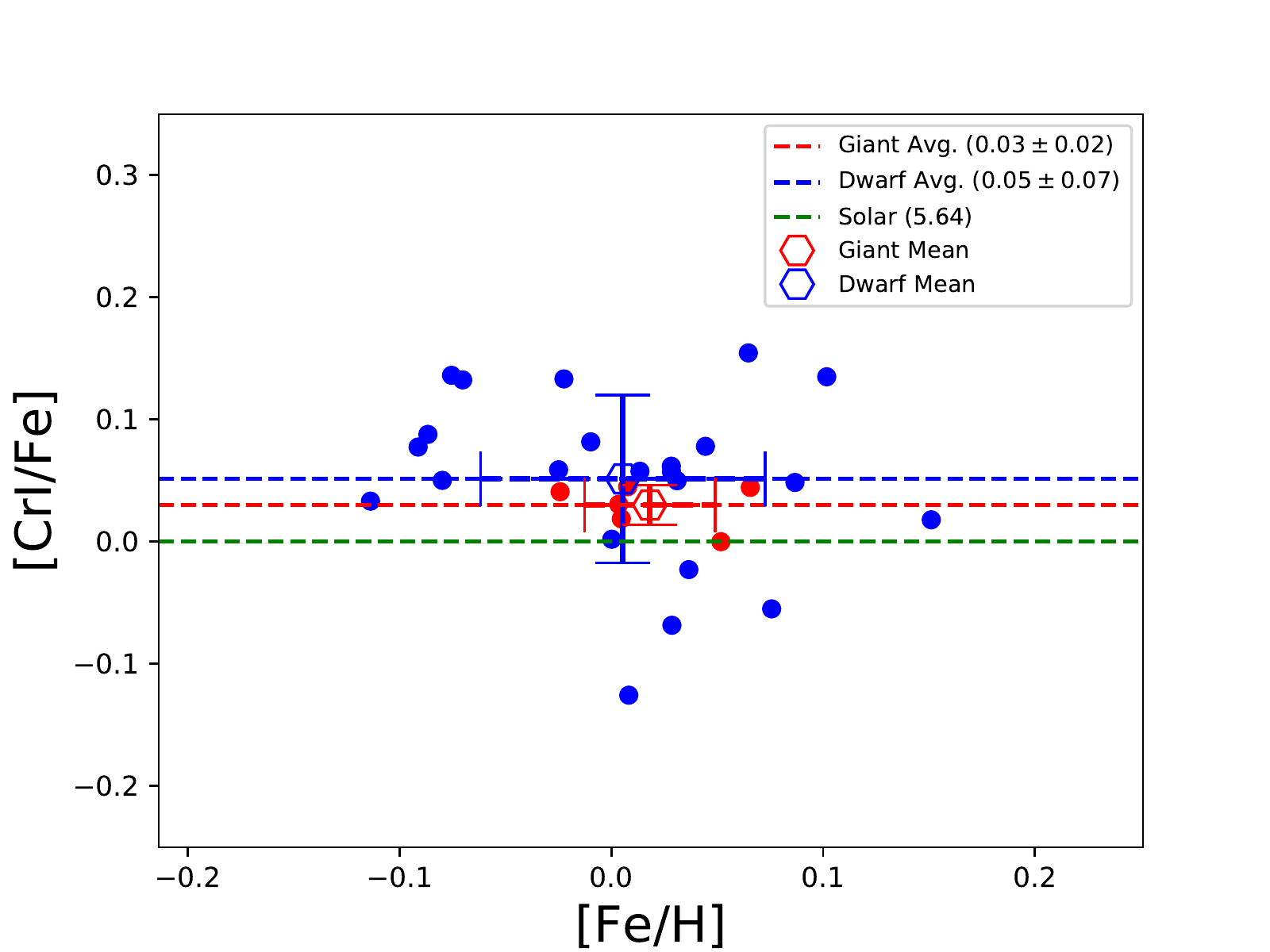}
        \label{fig:FeAbs:CrI}}
    \subfloat{%
        \includegraphics[width=0.49\textwidth]{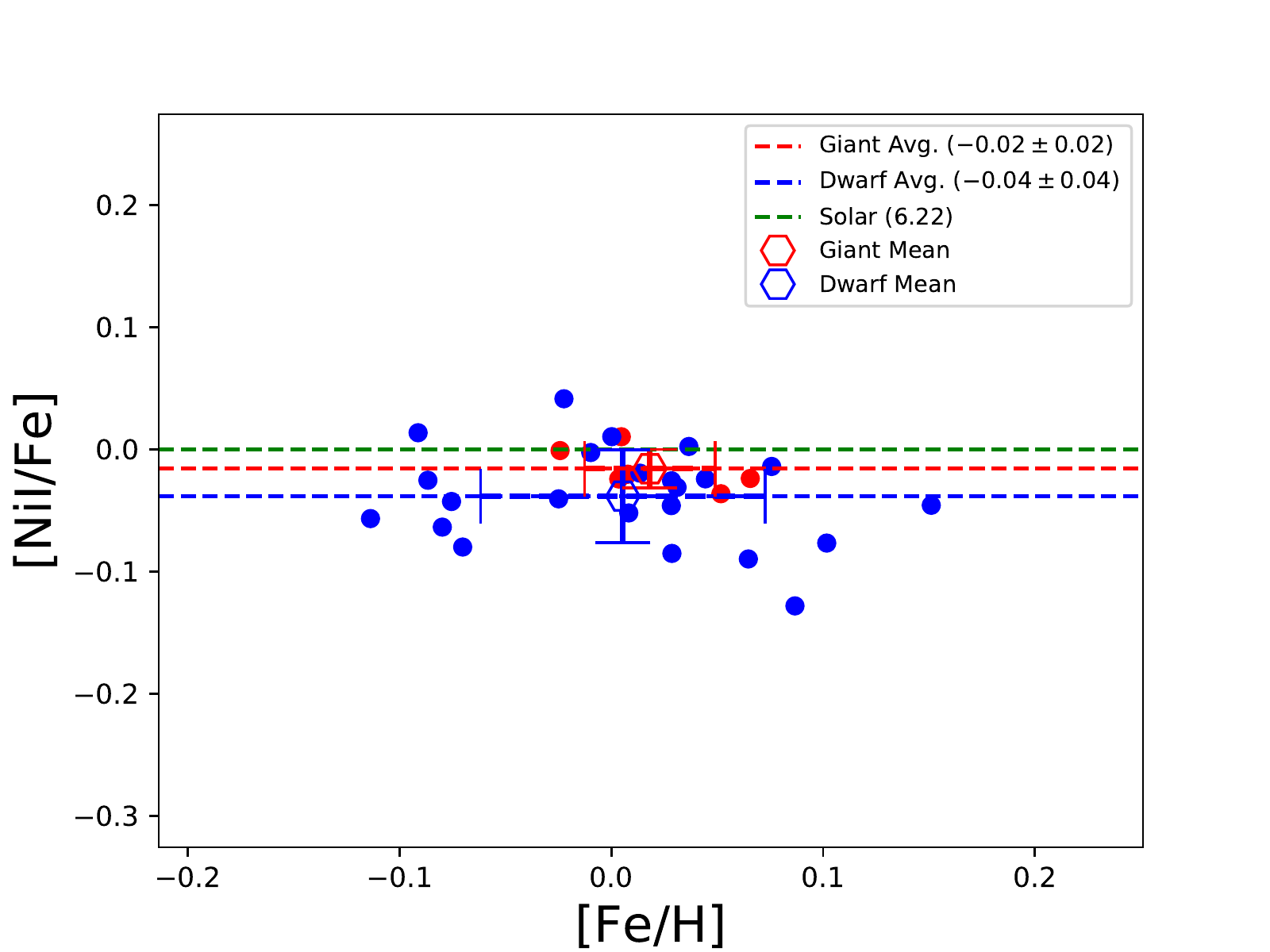}
        \label{fig:FeAbs:Ni}}
    \caption{Individual star abundances for Iron-peak elements: Fe\,\small{I}, Fe\,\small{II}, Cr\,\small{I}, and Ni\,\small{I}.  All three iron peak elements were measured at basically solar ([X/Fe]$=0.00$) levels, with no appreciable difference between the dwarf and giant populations. The abundance mean for the dwarf population is designated by the blue points and hexagon (group mean), and the giants in red.}
 \label{fig:FeAbs}
\end{figure}

\begin{figure}[H]
    \centering
    \includegraphics[width=0.69\textwidth]{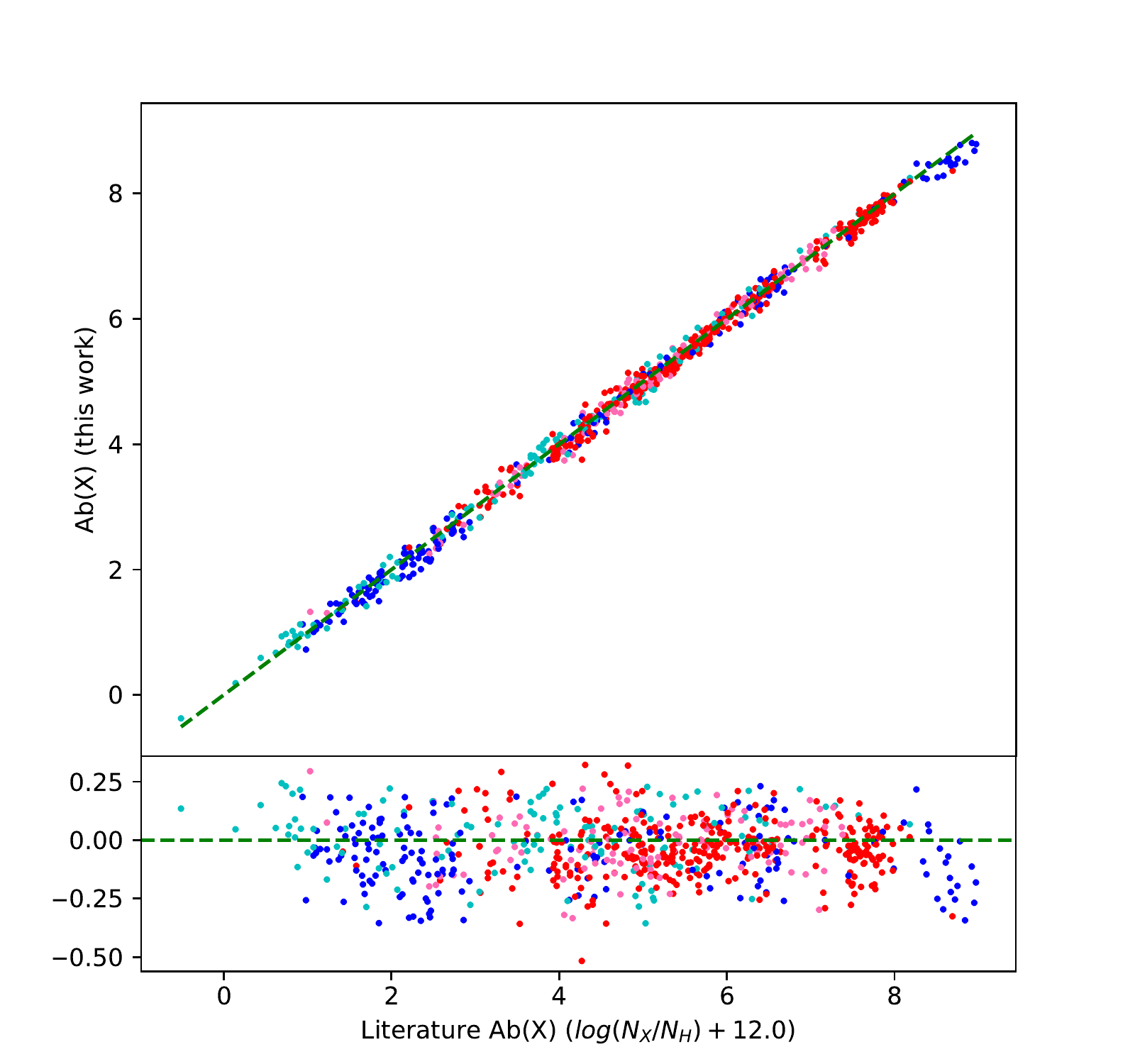}
    \caption{Comparison of elemental abundances as measured in literature \citep{Jofre:15} and through our process. Points are colored in red (red/pink) or blue (blue/cyan) hues to represent whether the literature value was directly measured (red) or if it was created from the literature value of [Fe/H] by assuming a solar ratio of [X/Fe] and extrapolating a ``literature" [X/H] (blue). The lighter saturated colored points (pink/cyan), represent low-metallicity stars ([Fe/H]~\textless~{-1.50}).}
 \label{fig:GAIA_Ab_Comp}
\end{figure}

\begin{figure}[H]
    \centering
    \subfloat[Carbon]{%
        \includegraphics[width=0.49\textwidth]{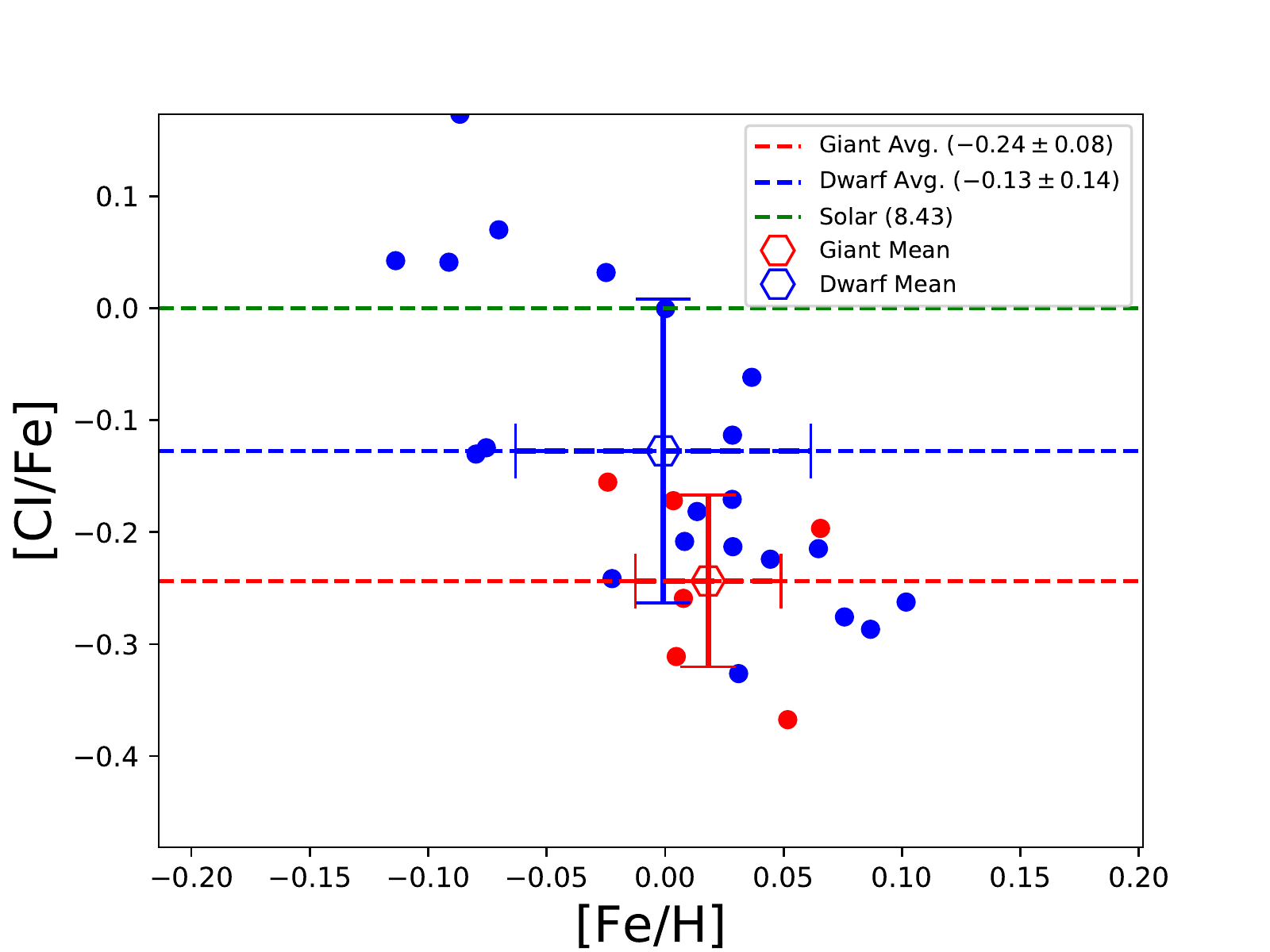}
        \label{fig:CNOAbs:C}}
    \\
    \subfloat[Nitrogen]{%
        \includegraphics[width=0.49\textwidth]{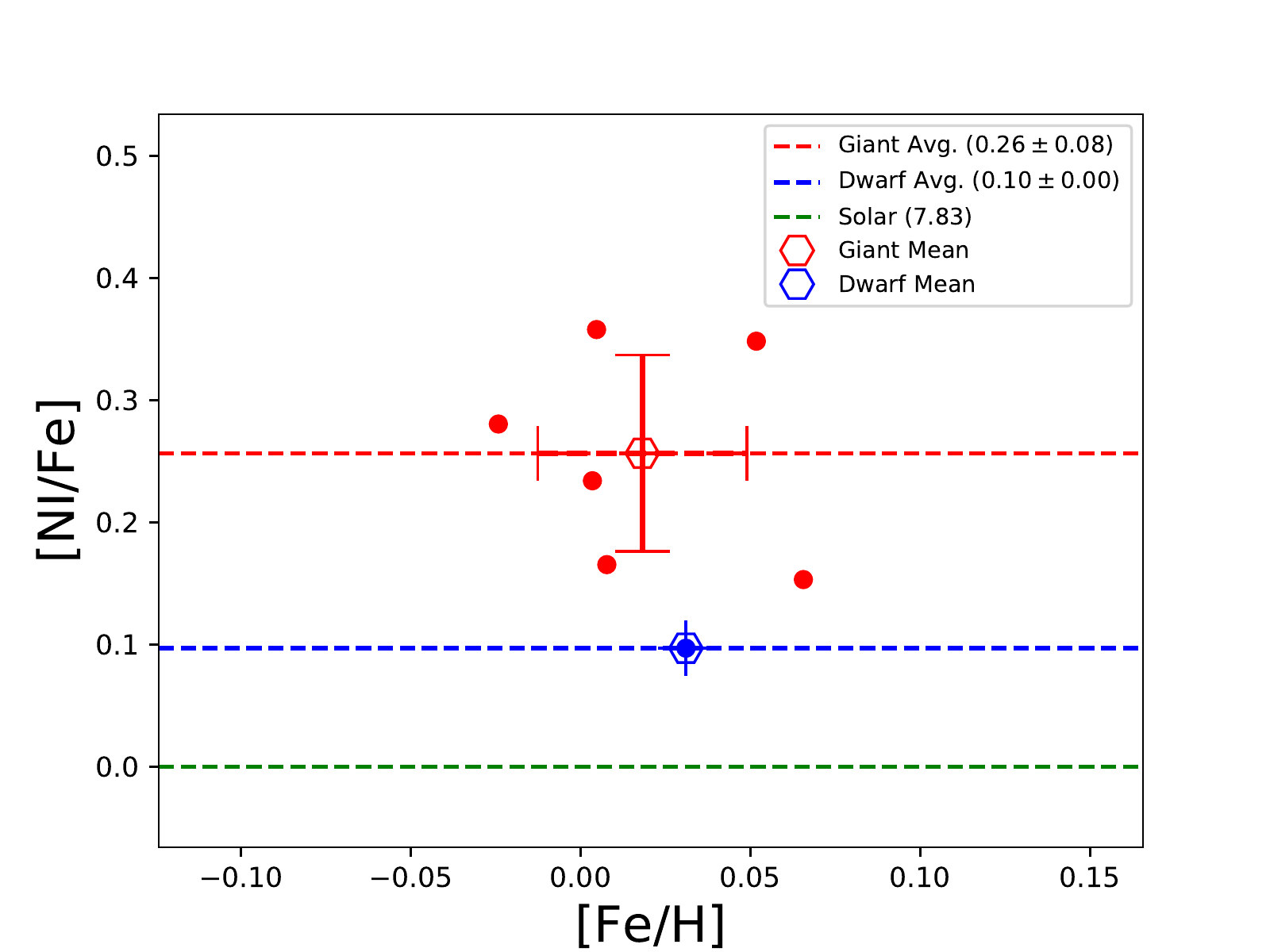}
        \label{fig:CNOAbs:N}}
    \subfloat[Oxygen]{%
        \includegraphics[width=0.49\textwidth]{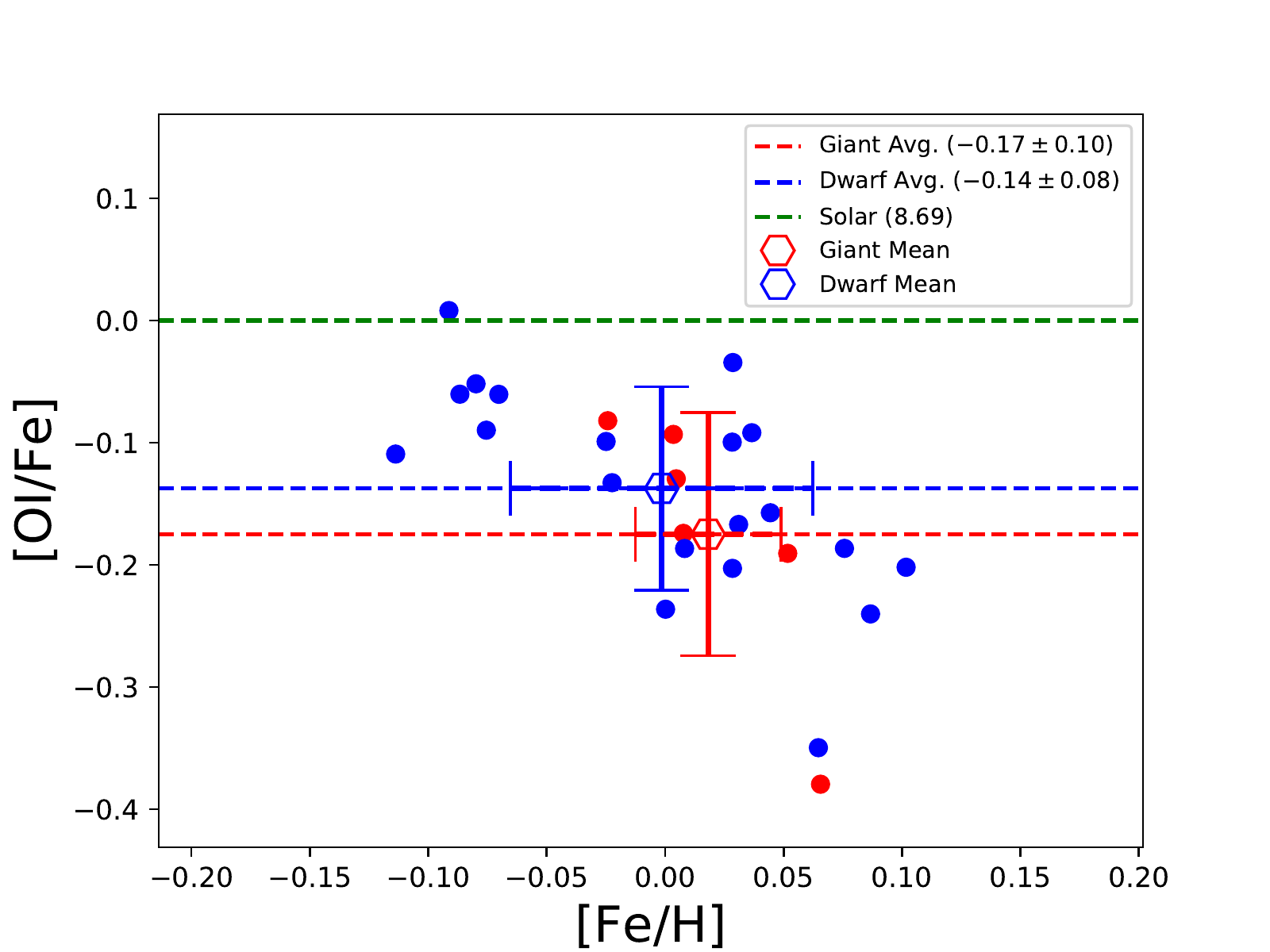}
        \label{fig:CNOAbs:O}}
    \caption{Individual star abundances for C\,\small{I}, N\,\small{I}, and O\,\small{I}. The abundance mean for the dwarf population is designated by the blue hexagon, and the giants in red.}
 \label{fig:CNOAbs}
\end{figure}

\begin{figure}[H]
    \centering
    \subfloat[Sodium]{%
        \includegraphics[width=0.49\textwidth]{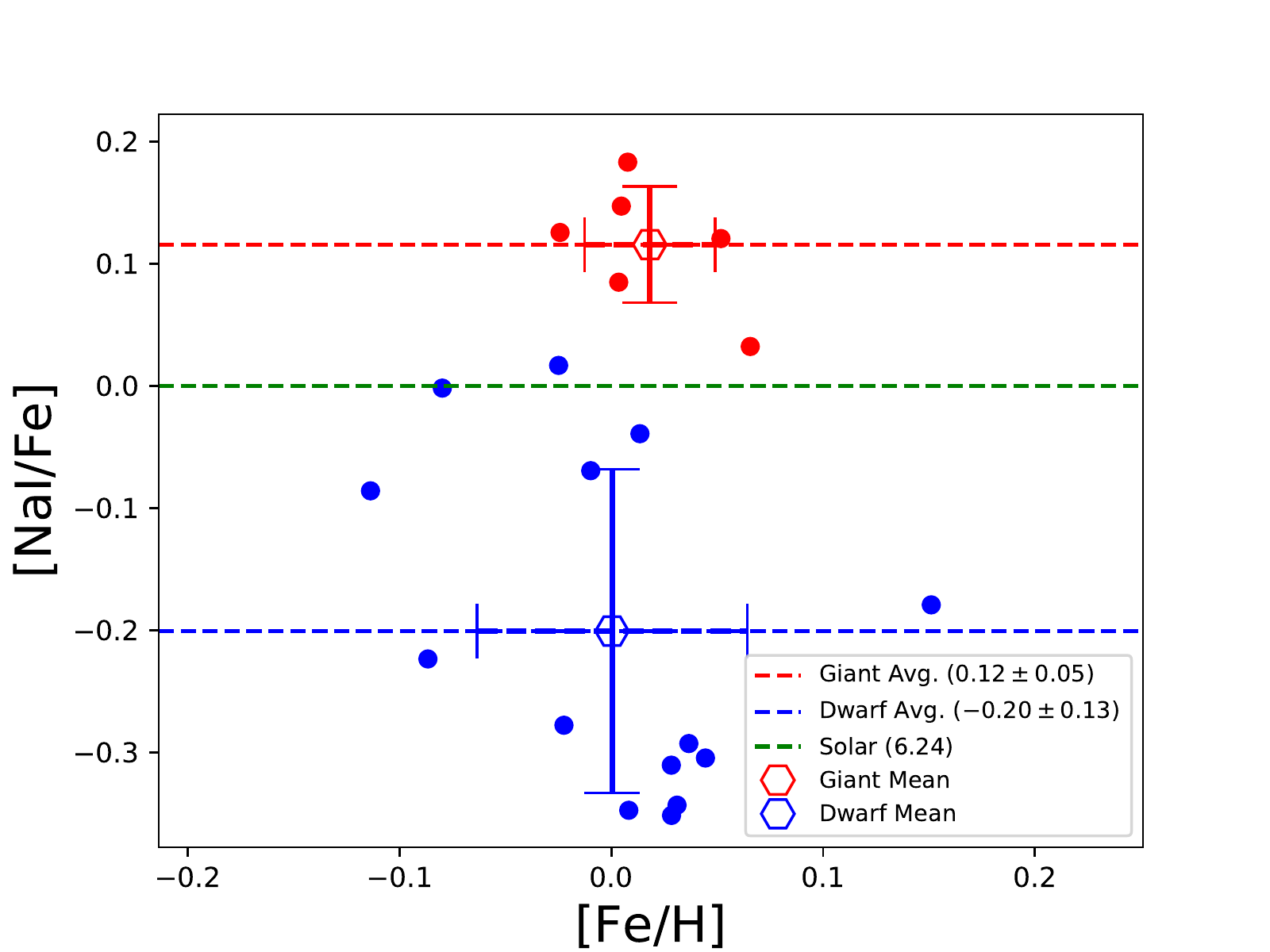}
        \label{fig:OddZAbs:Na}}
    \subfloat[Aluminium]{%
        \includegraphics[width=0.49\textwidth]{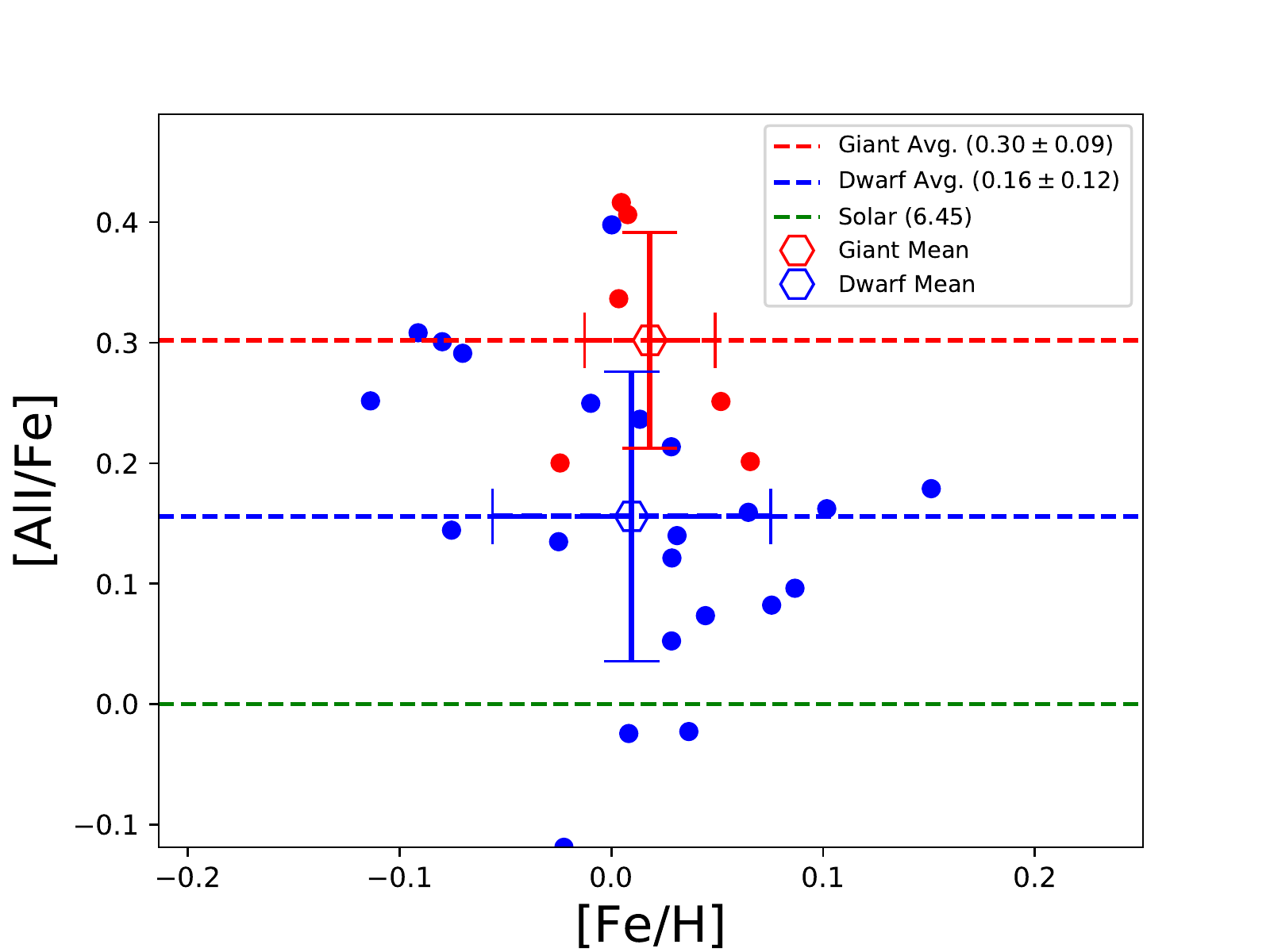}
        \label{fig:OddZAbs:Al}}
    \caption{Individual star abundances for the odd-Z light elements, Na\,\small{I}, and Al\,\small{I}. The abundance mean for the dwarf population is designated by the blue hexagon, and the giants in red.}
 \label{fig:OddZAbs}
\end{figure}

\begin{figure}[H]
    \centering
    \subfloat[Magnesium]{%
        \includegraphics[width=0.49\textwidth]{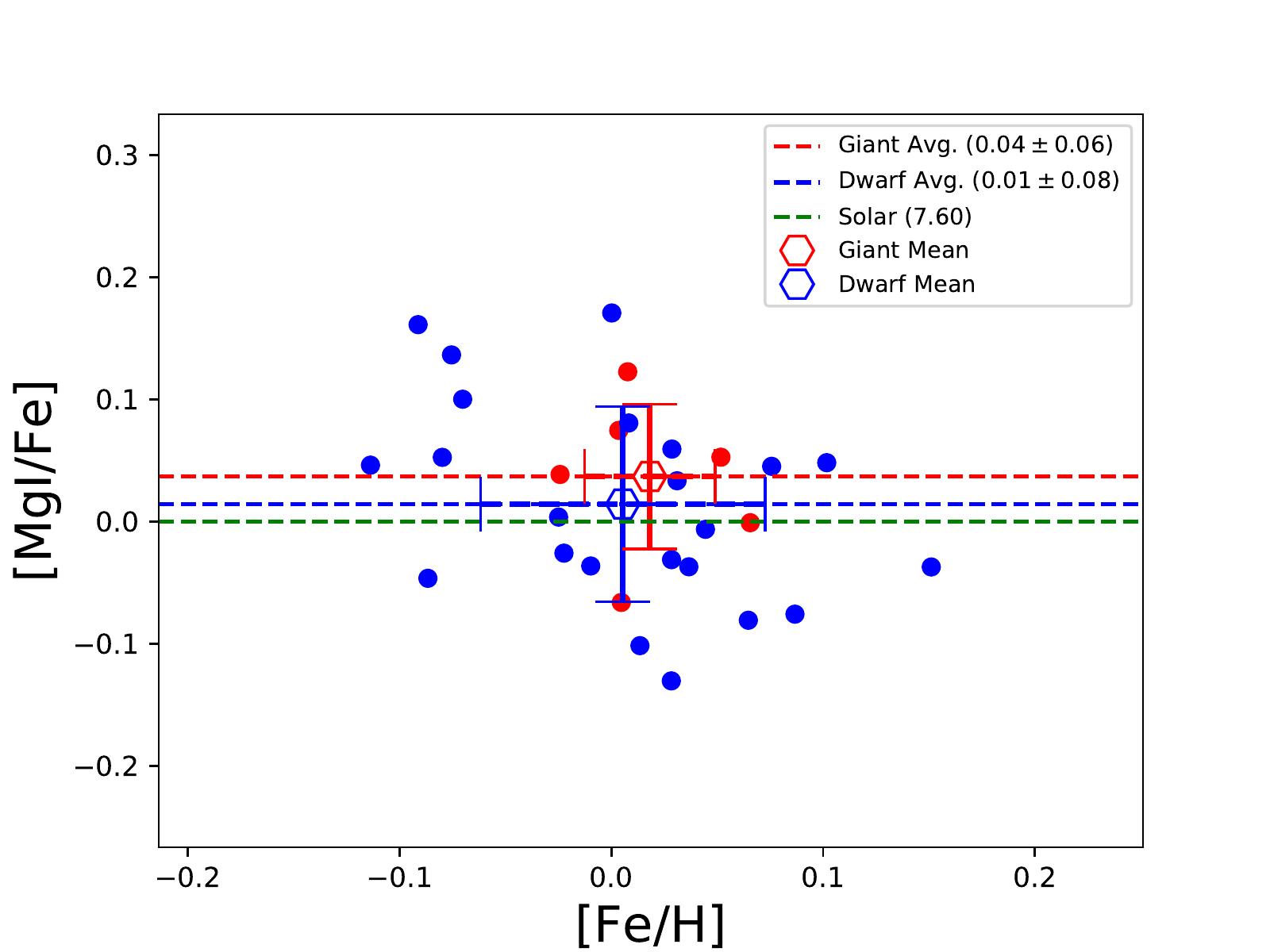}
        \label{fig:AlphaAbs:Mg}}
    \subfloat[Silicon]{%
        \includegraphics[width=0.49\textwidth]{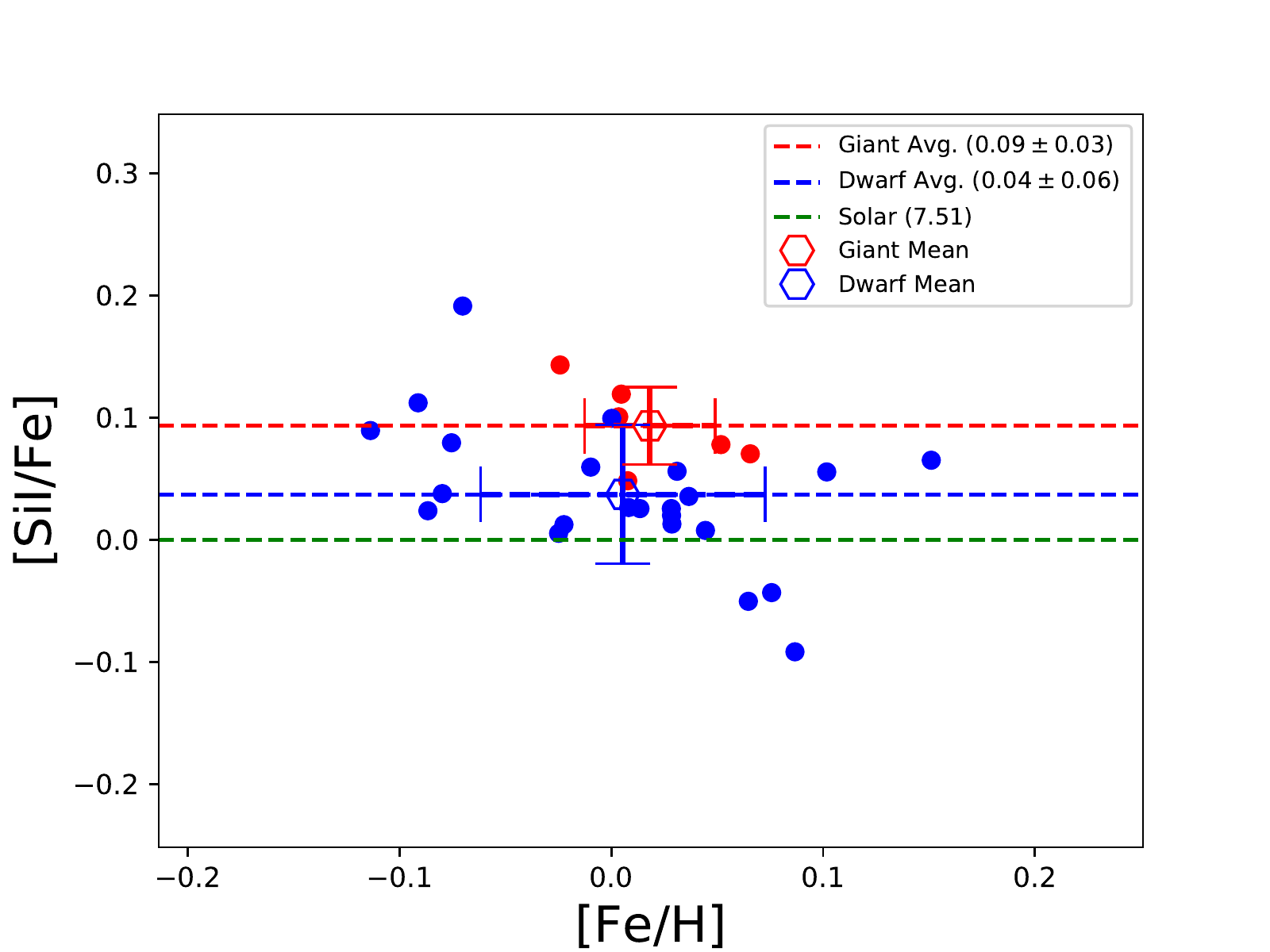}
        \label{fig:AlphaAbs:Si}}
    \\
    \subfloat[Calcium]{%
        \includegraphics[width=0.49\textwidth]{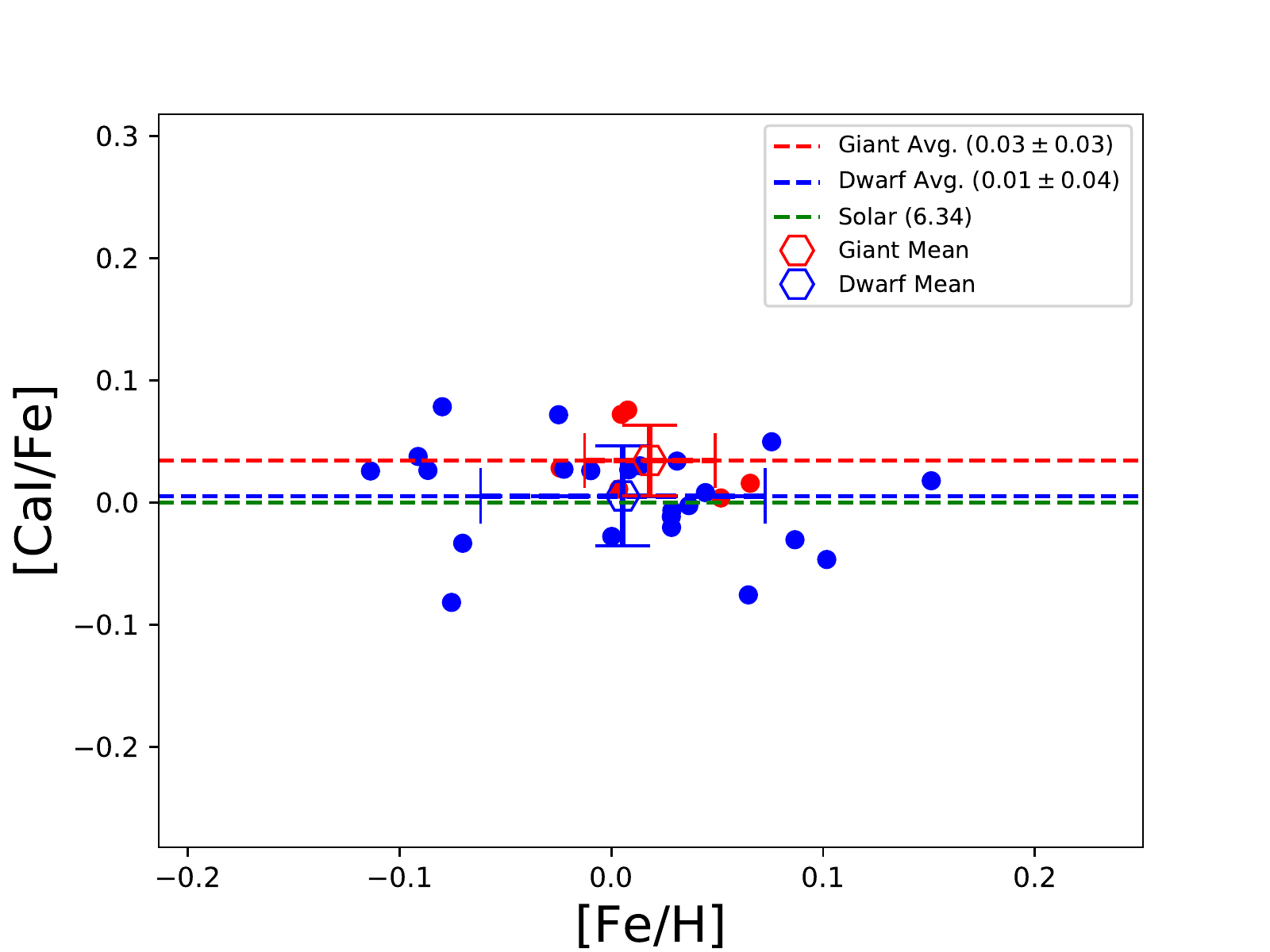}
        \label{fig:AlphaAbs:Ca}}
    \subfloat[Titanium]{%
        \includegraphics[width=0.49\textwidth]{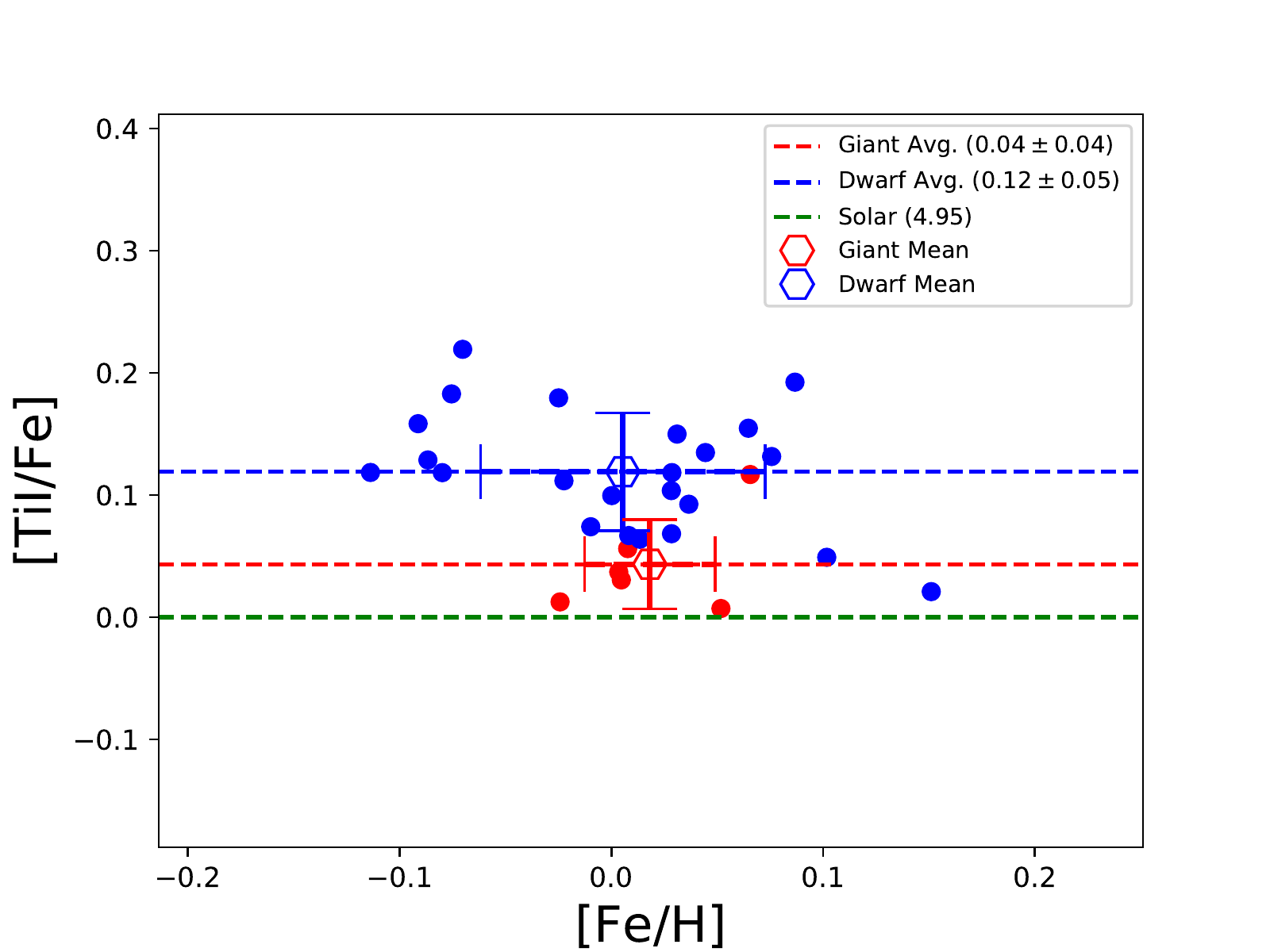}
        \label{fig:AlphaAbs:Ti}}
    \caption{Individual star abundances for the alpha-elements, Mg\,\small{I}, Si\,\small{I}, Ca\,\small{I}, and Ti\,\small{I}. The abundance mean for the dwarf population is designated by the blue hexagon, and the giants in red.}
 \label{fig:AlphaAbs}
\end{figure}

\end{document}